\documentclass{ws-ijmpd}
\usepackage[super,compress]{cite}
\usepackage{mathptmx}
\usepackage[perpage]{footmisc} 

\begin{document}


%
\catchline{}{}{}{}{}
%

\title{ON EXTENDED THERMODYNAMICS: FROM CLASSICAL TO THE RELATIVISTIC REGIME}

\author{J. FELIX SALAZAR and THOMAS ZANNIAS}

\address{Instituto de F\'isica y Matem\'aticas, Universidad Michoacana de San Nicol\'as de Hidalgo
Edificio C-3 , Ciudad Universitaria, 58040 Morelia, Michoac\'an, M\'exico.\\
jose.salazar@umich.mx, thomas.zannias@umich.mx}

\maketitle

\begin{history}
\received{Day Month Year}
\revised{Day Month Year}
\end{history}

\begin{abstract}
\noindent

The recent monumental detection 
 of gravitational waves by LIGO,
 the subsequent detection 
 by the  LIGO/VIRGO observatories 
 of a binary neutron star merger
 seen in the gravitational wave signal $GW170817$, 
 the first photo of the event horizon of the supermassive  black hole 
 at the center of the $M87$ galaxy
released by the EHT telescope
and the ongoing experiments
on 
Relativistic Heavy Ion Collisions
at the BNL and at the CERN,
demonstrate that we are witnessing  the second golden era of observational relativistic gravity.
These new observational breakthroughs,
although in the long run would influence our
views regarding this Kosmos, in the short run, they suggest
that 
relativistic dissipative fluids
(or magnetofluids)
and relativistic continuous media
play an important role on astrophysical-and also subnuclear-scales.
This realization brings into the frontiers of current research
theories of irreversible thermodynamics of relativistic continuous media.\\
Motivated by these
considerations,
in this paper, we summarize the progress that
has been made in
the last few decades in
the field of non equilibrium
thermodynamics 
of relativistic continuous media.
For coherence and completeness purposes, we begin with a brief 
description of the balance laws for classical (Newtonian) continuous media
and introduce
the classical irreversible thermodynamics (CIT) and 
the role of the local-equilibrium postulate within this theory. Tangentially,
we touch the program of rational thermodynamics (RT), the Clausius-Duhem
inequality, the theory of constitutive relations
and the emergence of the entropy principle and its role  in the description of continuous media.
We discuss at some length,
the theories of non equilibrium
thermodynamics that sprang out of 
a fundamental paper written by M\"uller in $1967$, with emphasis on the principles
of extended irreversible thermodynamics (EIT)
and the field of
rational 
extended irreversible thermodynamics (REIT).\\
Subsequently, after a brief introduction to relativistic fluids,  we 
discuss the Israel-Stewart transient (or causal) thermodynamics
and discuss its main features. Moreover, we introduce 
the Liu-M\"uller-Ruggeri  theory of irreversible thermodynamics
of relativistic fluids
and analyze and compare this theory to
the transient thermodynamics. We also discuss 
the 
 class of dissipative
relativistic fluids of divergence type developed in the late $1990$ by Pennisi, Geroch and Lindblom
and the emergence of symmetric-hyperbolic (or causal) system of dynamical equations
in the description of the such fluid states.\\
As we shall see in this review, the current efforts aiming
to develop viable theories 
of irreversible thermodynamics of continuous media 
is focused on theories whose dynamical equations constitute a symmetric-hyperbolic
and preferably causal set of dynamical equations. By design,
this class of theories eliminates
propagation of disturbances with
unbounded speed, a necessary condition for
the viability of the underlying theory.\\
Although it is 
fair to
state that substantial progress has been made in the field
of non equilibrium thermodynamics of classical media and many predictions 
of the extended theories (in the form of (EIT) or (REIT))
have been placed under experimental scrutiny, at the relativistic level the situation is different.
 Even though the time spent 
 aiming to the development of a sensible theory (or theories) of non equilibrium
 thermodynamics of relativistic fluids (or continuous media) is relatively short,
 enormous steps in the right direction
 have been taken.
 Still however, as we shall see in this review,
a successful theory of relativistic dissipation is lacking.\\

\end{abstract}


\ccode{PACS numbers:}


\section{Introduction}

The date $ 14th $ of September of $ 2015 $ will remain in the annals of gravitational physics as a landmark date. At that date, the upgraded Laser Interferometer Gravitational Wave Observatory (LIGO) captured gravitational waves from two black holes circling each other, the first detection of gravitational waves ever recorded in the human history.
 The event announced in a press conference on $ 11th $ of February in $ 2016$ and the reader 
is referred to the web page of the LIGO observatory \cite{LIGO} for a thorough analysis of this detection.\\ 
Two years later, specifically 
 on $17th$ of August $2017$, the LIGO/VIRGO gravitational wave observatory network
 recorded a gravitational wave signal, refereed as GW170817, which is consistent with a binary neutron star 
 inspiral and merger.
 For an analysis of this signal, its interpretation and consequences the
 reader is refereed to Refs. \refcite{NS1,NS2}.\\
On the $10th$ of April
$2019$, 
the Event Horizon Telescope (EHT) collaboration,
released the 
first ''photograph of the event horizon'' of the supermassive black hole at
the center of the $M87$ galaxy 
(for details, visit Ref. \refcite{EHT}).
This photograph
establishes the existence of an accretion disk around the black hole
and here is worth recalling that modeling of such a disk presuppose viscous accreting matter.\\
In another exciting development, ongoing experiments
at the  
Relativistic Heavy Ion Collider at BNL and at Large Hadron Collider at CERN,
show that in the relativistic heavy ion collision
the quark-gluon plasma formed in 
these
terrestrial
``mini bing-bangs''
is reliably described 
by  a viscous relativistic fluid.
For evidences supporting this 
unexpected connection the reader is referred
to recent reviews in Refs. \refcite{DKK,FHS,RR_1}.\\
These latest observational breakthroughs 
would have far reaching 
consequences upon our understanding of this 
Kosmos. 
At the fundamental level, once again,
 we reassured that 
Einstein's General Theory of Relativity, in the words of Penrose
 is a superb theory\footnote{We borrow the term 
 ''superb theory'' from the classification of fundamental theories
 proposed by Penrose in Ref. \refcite{Pen1}, (see discussion on pages $(225, 261)$ of 
 that ref.).} 
  where its minute predictions 
one after the other are observationally confirmed.
Moreover, in an 
 era where the LIGO/VIRGO observatories are operational and in the near future the advanced KAGRA observatory is expected to be operational,
accretion of matter into black holes and other compact objects, plasmas  in the early universe, 
supernovae explosions and core collapse, are going to be placed under observational scrutiny
and in these scenarios dissipative fluids or more generally relativistic continuous media
play an important role.
In this connection, it is worth mentioning that the current modeling of
neutron stars
suggest that viscosity and thermal conductivity 
play an important role in 
 their stability
 while results from relativistic heavy ion collisions suggest
that viscosity is also relevant at a subnuclear domain.\\
 These latest developments
challenges
relativists and high energy physicists
alike to develop
reliable theories of 
 irreversible thermodynamics
of relativistic continuous media in order to confront the new observational realities.\\
Interestingly, for relativistic fluids, the conventional 
first order theories of irreversible relativistic thermodynamics of Eckart
\cite{Eck} and  Landau-Lifshitz 
\cite{LaL}, predict instantaneous propagation of thermal and viscous effects which to use the words of Israel and Stewart \cite{Isr2} ``\textit{is an offense to the intuition, which expects propagation at about the mean molecular speed; in a consistent relativistic theory it ought to be completely prohibited}". Beyond this deficiency, these first order theories, suffer from other drawbacks: in Ref.\refcite{His2} it was proved that small-amplitude disturbances around equilibrium states diverge exponentially on a very short time scale.
These features of the conventional 
first order theories has been a source of concern
and many researchers are skeptical whether these theories could model accurately process characterized by rapid spatial and temporal
variations of heat fluxes and viscous stresses\footnote{A referee kindly pointed out to us
that actually the situation is even worst: \textit{They are (i.e. first order theories) ill-posed. The exponential modes grow without bound as the frequency increases. Thus, any high frequency perturbation, no matter how small we take it, can become as big as we want at any finite time just by increasing its frequency. Thus, these theories, do not have predictive power.}
We thank the referee for this comment.}. Fortunately however, in the last few decades, there 
has been a progress into the field of non equilibrium thermodynamics of continuous media both at 
the classical and the general relativistic domain. 
Theories of extended irreversible thermodynamics have been developed which
at least for states near equilibrium, lead to symmetric-hyperbolic system of equations 
predicting causal propagation of thermal and stress disturbances (see for instance Refs. 
\refcite{Isr2,Isr1,His1,Mul1,Mul2,JVL,Mul4,Mul6,Ger1,Ger2,Ger3}).
Motivated by these
developments,
this article introduces theories of extended irreversible thermodynamics
of continuous media (mostly Newtonian or relativistic fluids).
Our emphasis is on the structure 
of  theories
refereed as ``hyperbolic theories''
and in order to provide
a self contained introduction to their origin and the necessity
of introducing these  theories,
we begin by first offering 
a brief
account of the historical development of the subject.\\
Thermodynamics 
is an empirical science which has been developed after
persistent studies of the behavior of matter  under external stimuli
and 
 these studies culminated in the formulation of the four laws of thermodynamics
 (for an enjoyable reading of this historical development 
 see
 Ref. \refcite{Mul3}). Gradually, and beginning with the work of Onsager \cite{Ons1,Ons2}, states near equilibrium 
 began to
be incorporated into the field and this lead to the development of the Classical Irreversible Thermodynamics
 abbreviated here after as: (CIT). The cornerstone
that underlies this theory is the local thermodynamical equilibrium (often abbreviated as (LTE)) postulate
 which
 (when suitably applied)
assigns an entropy to non equilibrium states
and moreover 
implies that its 
evolution is deduced
from the Gibbs relation
combined with
the balance laws.
For a simple, heat conducting, viscous fluid,
(CIT) leads to the standard Fourier-Navier-Stokes theory and 
this success led to considerable amount of scientific work. 
Moreover, (CIT) 
had some experimental confirmation
and the reader is refereed 
to Refs.\refcite{JVL,Mei,Gro,Gya,Gla}
for an overview regarding these confirmations.\\
Despite these successes, (CIT)
suffers from some serious drawbacks. The 
(LTE) postulate 
implies that the entropy of non equilibrium states
 depends upon the same variables as  the entropy of equilibrium states.
It is conceivable however, that other variables may influence the thermodynamical behavior
 of non equilibrium states. Moreover, (CIT) predicts parabolic set of dynamical equations 
 which yield
 infinite speed of propagation of thermal and viscous signals.
Although these predictions are not really in contradiction with the 
classical (Newtonian) framework,
as we have already mentioned,
 they are  in contradiction with the spirit of the relativistic framework (and our intuition). The problem of the infinite propagation of thermal 
signals remained open until $ 1967 $, when M\"uller in an influential paper  \cite{Mul1}  has shown 
 that the paradox of Fourier's heat conduction (propagation of temperature disturbance with infinity velocity) is a consequence of an 
 insufficient description of the 
off-equilibrium 
thermodynamical state. For a simple fluid,
and for states near equilibrium,
he proposed 
a generalized entropy 
 that gets quadratic contributions
from 
the heat flux and stresses, a hypothesis that is in a glaring contradiction
to the spirit of the
(LTE) postulate as applied within the (CIT).
Via this bold hypothesis,
M\"uller arrived at a theory which removes the paradox of infinite propagation of heat conduction  
at least for fluids  with an appropriate equation of state.\\
M\"uller's hypothesis, received well within the scientific community
and his thesis extended 
in a variety of ways.
One popular extension lead
to the development of 
a theory of extended irreversible thermodynamics
of classical continuous media, designated here after by the acronym:
(EIT).
This theory
assigns a generalized entropy 
to arbitrary non equilibrium
states which
depends upon
dissipative fluxes (i.e. quantities that appear in the balance laws such as Cauchy's stress, heat flux, etc.).
This hypothesis, combined with a generalized 
Gibbs relation and 
the imposition of 
 the second law
 leads to
phenomenological equations
that describe 
the temporal and spatial variations of the
 fluxes
 and 
these equations generalize the
Maxwell-Cattaneo\cite{Cat1,Cat2} system.
The theory leads to a successful 
 treatment of heat transport in micro and nano systems, shock structure of waves 
 propagating on hydrodynamical systems, 
 etc. and 
for an overview of successes (and failures)
of (EIT),
the reader is refereed to the latest edition of
 the book\cite{JVL}, 
the $1984$ Barcelona conference proceedings \cite{new2},
and in Refs. \refcite{GC1,GC2,Sie1}.\\

M\"uller's original hypothesis even though
supported by the kinetic theory of dilute gases,
it was not entirely free of mathematical problems.
For a heat conducting, viscous fluid the resulting system of equations
 is generally non-hyperbolic
 and the characteristic speeds
 of disturbances are real provided some of the thermodynamical variables
 are suitably bounded.
 This property is however unsatisfactory since even if these bounds hold by 
 the  initial data,
 they may fail to the future of the initial surface (for a constructive criticism of
M\"uller's original approach consult Ref. \refcite{Rug1}). The quest for mathematical rigor lead
Liu, M\"uller, Ruggeri and collaborators 
to 
develop of a new theory, referred as
Rational 
Extended Irreversible Thermodynamics -abbreviated here after as  (REIT)-which
  is characterized by the following properties:\\

a) the dynamical equations of the theory are of balance type,\\

b) the set of constitutive equations are local in space and time,\\

c) the dynamical equations constitute a symmetric-hyperbolic system of field equations.\\

The tractable features of (REIT) is the restoration of finite propagation of
the heat and stress disturbances,
and 
the symmetric-hyperbolic nature
of the underlying field equations
(for an introduction to this  theory consult Refs. \refcite{Mul4}, \refcite{Mul5}, \refcite{MulW}).\\
The 
symmetric-hyperbolic nature
of the underlying field equations
of (REIT),
and the ensuing prediction of the
finite propagation of
the heat and stress disturbances
made relativists dreaming for an extension 
of (REIT) into the relativistic regime.
However, formulating a
theory describing the irreversible thermodynamics
of relativistic continuous media
and  obeying requirements $a)-c)$
it is not an easy task.\\ 
Besides the mathematical difficulties of extending
conditions a) to c) to the relativistic regime,
one 
confronts 
the problem related to the absence of
preferable rest frames\footnote{In more simple terms
this problem is associated with the inability within theories describing the  thermodynamics of relativistic fluids
 to uniquely single out the fluid's four velocity a notion that is the landmark of the non relativistic fluid dynamics.
 We will discuss that issue further ahead (see coment $e)$ on page $(27)$).} for
off-equilibrium states\footnote{We are assuming that within each theory, one would be able to
define
the class of equilibrium states.} . In general,
off-equilibrium states of 
fluids, 
or of elastic media, may 
single out more than one preferable rest frames.
The Eckart frame, the Landau-Lifshitz frame, the entropy frame\footnote{The energy frame
and particle frame will be defined further ahead.
The entropy frame is introduced by Carter in Ref. \refcite{Car0}.} etc.,
are examples of such
frames and in general it is not clear which one, if any of them
should be employed to express the 
 thermodynamical laws\footnote{One may argue
 for the development of a theory (or theories) that 
 is (are) ''frame-independent''.
 In fact, as we shall see further ahead, the formulation of
 the 
Liu-M\"uller-Ruggeri theory, the class of relativistic fluid theories of divergence type, etc
do not require a rest frame (or the fluid's four velocity) for their formulation. 
 However, at some stage, their interpretation
 becomes accessible to intuition
whenever some rest frame (or four velocity) in invoked.
 All so far experience of  the Fourier-Navier-Stokes fluids,
where  gradients of the velocity field enter into the description of stresses, shows
 that frames add to our intuition, and this trend also holds at the relativistic regime.}.
This subtle issue, has been settled by Israel who for long period of time
 has been stressing that 
 as long as considerations are restricted to states near equilibrium,
a theory 
of small deviations from equilibrium 
can be constructed
which is manifestly
invariant 
under first order changes
of the rest frame\footnote{\label{comment_2} This important property has been 
pointed out long ago in a M.Sc thesis written by Aitken \cite{Ait}, then student of Israel.}. 
For a simple fluid, it is shown,
in the Appendix A,
that
on the tangent space of any event within the region occupied by the fluid,  an invariant  
``cone" of opening pseudo-angle\footnote{The magnitude of the three velocity $ \vec v $ 
that appears in the definition of $\epsilon= \frac {v}{c}=O_{1}$,
stands for the relative velocity of the Eckart frame relative to the Landau frame
(the symbols $O_{1}, O_{2}....$ will be introduced further ahead).}
 $\epsilon= \frac {v}{c}$ can be defined
that
has the following property:
Any four velocity $u$ that falls within this ''cone''
can be used as a potential rest frame (or as a potential four velocity of the fluid)
so that observers at rest relative
to this frame deduce thermodynamical variables 
for the
non equilibrium state under consideration.
Even though, these thermodynamical variables
are frame dependent
and 
despite the plurality of such rest frames, as long as 
$\epsilon=\frac {v}{c}=O_{1}$ 
a consistent thermodynamical theory can
be developed that is
manifestly invariant under first-order changes of the rest-frame $u$, 
i.e. change\footnote{ 
Under such transformations, many thermodynamical variables
transform in a well defined manner
and these
transformation
laws are discussed in Appendix D.} in the rest frame described by 
\begin{equation}
	u^{\mu}\mapsto \hat u^{\mu}=u^{\mu}+\delta^{\mu},\quad \delta^{\mu}\leq O_{1}.
\end{equation}

Motivated by these considerations
and results from relativistic kinetic theory of gases, 
 Israel in 
 Ref.\cite{Isr1} proposed
 that for states near equilibrium of a simple fluid,
there exist a linear relationship
between the primary variables consisting of the energy momentum tensor $T^{\mu\nu}$
the particle current $J^{\mu}$ and the entropy four vector $S^{\mu}$ 
which however is
modified by a term $Q^{\mu}$
quadratic in the deviations $\delta J^{\mu}, \delta T^{\mu\nu}$
from the state of (local) thermodynamical equilibrium.
Based on this hypothesis, 
and independently
of
M\"uller's considerations,
he formulated in $ 1976 $ a  
 theory of irreversible thermodynamics
for fluid states near equilibrium
which elaborated further 
in Israel and Stewart
\cite{Isr2} and nowadays is known as transient
(or causal) thermodynamics (of relativistic fluids).
In Ref. \refcite{Isr2},
the fundamental
relation 
between the primary variables and 
the $Q^{\mu}$ term 
has been obtained in an elegant manner based on the covariant form of the Gibbs
relation and 
by invoking  ``\textit{the release of variation assumption}", a term that will be introduced
in section
$9$.
For a simple fluid,
and within the hydrodynamical approximation
advocated in 
\cite{Isr2} the  $ Q^\mu (u)$ term 
has the form:
\begin{equation}
Q^{\mu}(u)=
\frac {1}{2}u^{\mu}[\beta_{0}\pi ^{2}+\beta_{1}q^{\nu}q_{\nu}+\beta_{2}\pi^{\lambda\nu}\pi_{\lambda\nu}]-
\alpha_{0}\pi q^{\mu}-\alpha_{1}\pi^{\mu\nu}q_{\nu}+R^{\mu}(u),
\label{1}
\end{equation}
where
$\beta_{i}$, 
$i \in (0, 1, 2)$, $\alpha_{j}$, $j \in (0, 1)$
are undetermined  functions,
$q^{\mu}$ is
a frame invariant heat flux\footnote{This quantity is defined precisely
further ahead (see eq.
(\ref{IHF})). It is frame invariant in the sense that under a change of the velocity of the frame
changes by an $O_{2}$ term.}, $(\pi, \pi^{\mu\nu})$ 
are stresses
deduced by an observer with four velocity $u$ and $R^{\mu}(u)$ is a term whose specific nature will be 
discussed further ahead.
The phenomenological laws
resulting from the above choice of $Q^{\mu}(u)$
will be derived in section $(12)$
and 
as we shall see 
the functions $(\alpha_{j}$,$\beta_{i})$
are left unspecified.
By appealing to Boltzmann equation for a relativistic gas,
and within the Grad's approximation \cite{Gra}, 
Israel and Stewart\cite{Isr2}, 
have
evaluated the coefficients $ (\alpha_j, \beta_i) $ 
and have shown that 
their theory, at least for these values, predicts
that
the characteristic velocities of perturbations about equilibrium states are finite and sub-luminal i.e, they are
real and less than the speed of light.
The same conclusion has been also reached by 
Hiscock and Lindblom in \cite{His1}
who have shown 
within the framework of transient thermodynamics a close connection between
causality 
and stability of equilibrium states.\\
The results quoted above,
show that transient thermodynamics at least for the values of the coefficients 
$ (\alpha_i, \beta_i) $ predicted by the relativistic 
Boltzmann equation eliminates 
deficiencies
that plagues the theories
of Eckart and Landau-Lifshitz (or more generally the class of first order theories). 
Therefore transient thermodynamics,
at least when applied to fluid states near equilibrium,
seems to be a tractable theory with encouraging properties.
However, the
fact that the resulting phenomenological equations contain  
unspecified 
functions and the absence of a rigorous
proof demonstrating
that the dynamical equations
constitute a symmetric-hyperbolic system, lead
to the development 
of alternative theories.
The common feature of these alternative relativistic fluid theories is the symmetric-hyperbolic 
nature of their dynamical equations whose evolution respects causality.\\
One such theory,
describing states of relativistic fluids, proposed 
by Liu, M\"uller and Ruggeri in Ref. \refcite{Mul6},
and this theory extends the principles of 
 Rational 
Extended Irreversible Thermodynamics
(REIT) to
the relativistic regime.  
The
Liu-M\"uller-Ruggeri theory uses the particle current $ J^\mu $ and the conserved symmetric stress tensor $ T^{\mu \nu} $ as the basic
dynamical variables, but the dynamical equations $ \nabla_\mu J^\mu  = \nabla_\mu T^{\mu \nu} = 0 $ 
are enlarged by
an additional set of equations.
These additional equations contain the divergence of a third rank  totally symmetric tensor field $ A^{\mu \nu \lambda} $ and a second rank traceless tensor $ I^{\nu \lambda} $ refereed as the dissipation tensor.
Liu, M\"uller and Ruggeri\cite{Mul6} assumed that the tensors $ A^{\mu \nu \lambda} $, $ I^{\nu \lambda} $ as well as an entropy vector $ S^\mu $ are constitutive functions i.e.
are functions of the dynamical variables $  J^\mu  $ and $ T^{\mu \nu} $.
The functional form of 
 these constitutive 
 relations 
 are restricted by appealing to the following three fundamental principles:\\
  
  a) Entropy principle,\\ 
 
 b) Relativity principle,\\
 
 c) Symmetric-hyperbolic nature of the dynamical equations.\\
 
In section \ref{R_REIT}, we shall discuss the
 implications of these principles
 and shall see  that for states near equilibrium, the 
theory leads to phenomenological eqs. for the heat flux and shear stresses
that contain three unknown functions.\\
 The approach of  Liu-M\"uller-Ruggeri  has been extended
further by 
Pennisi \cite{Pen} and
by Geroch and Lindblom \cite{Ger1}, and these extensions
lead to the development
of a class of relativistic fluid theories 
having the mathematically tractable property 
that their dynamical equations are determined from the knowledge of a scalar generating function
depending however upon $14$ suitably defined variables and
from the knowledge of the traceless dissipation tensor $ I^{\nu \lambda} $.
The advantage of this class of theories lies in the
flexible structure they possesses.
It has been demonstrated
that there exist families of generating functions that 
lead
to causal dissipative fluid theories and in section  \ref{DRFDT}, we shall discuss 
these dissipative fluid theories  in a more details. \\

This review is organized as follows: In sections (\ref{Continuous thermodynamics}-\ref{The second law}), we remind the reader 
of a few properties of classical (Newtonian) continuous media with 
emphasis on the structure of 
 the balance laws for mass, linear momentum and total energy.
  Since these laws in general, fail to yield a closed system of equations,
 we provide a brief introduction to the theory of constitutive relations
 and point out the importance of the second law of thermodynamics
 in the description of continuous media.
 In particularly, within the 
 the program of Rational Thermodynamics (RT), we discuss  
 the implementation of the second law via the Clausius-Duhem 
 inequality
 and thus set the scene for the introduction of the entropy principle
 and its role as a selection rule for specifying appropriate constitutive relations.\\
In section \ref{S_CIT}, we introduce (CIT) i.e. the irreversible thermodynamics of classical continuous media
and discuss 
the local-equilibrium hypothesis and the implementation of the second law within this theory. In 
section \ref{EIT}, we 
 introduce theories of extended irreversible thermodynamics
 and define in particularly 
 two classes of such theories: the first one
 is the Extended Irreversible Thermodynamics 
refereed as 
  (EIT) and the second one as 
Rational   
 Extended Irreversible thermodynamics 
 indicated as (REIT). Section \ref{S_CRM}
discusses continuous media within the relativistic regime
and introduces primary and auxiliary variables
describing  arbitrary (non equilibrium) states of such media. From section \ref{GES}
onward, we restrict attention mostly to the description of relativistic fluids.
As a precursor to the development of transient thermodynamics, we identify a class of fluid states 
in
global (or local) thermodynamical 
equilibrium
and derive
the covariant form of the Gibbs relation for this class of states. In section \ref{NES},
we introduce
the 
Israel-Stewart transient thermodynamics.
First order theories and the phenomenological equations
for the theory of Eckart and Landau-Lifshitz are
derived as ''limits'' of the transient thermodynamics,
and these theories are discussed in 
section \ref{Eckart}. In section \ref{IST}, the phenomenological equations
 for the second order theories describing fluid states near equilibrium are derived. In section $12$, we discuss the
 Liu-M\"uller-Ruggeri theory
 describing relativistic fluids
 and we compare the predictions of this theory
  to transient thermodynamics.
 In section $13$, we discuss 
 the thermodynamics of the class of relativistic fluid theories 
 of divergence type introduced by Pennisi and Geroch-Lindblom.
 In the conclusion section,
 we discuss the current state of irreversible
 thermodynamics of relativistic fluids and open problems related to the thermodynamics of such media.\\
  The paper contains  four Appendixes. In the Appendix A,
 and within transient thermodynamics, 
we discuss in a coordinate free manner the identification of fluid states near equilibrium.
 Appendix B, deals with some mathematical aspects 
 regarding the implementation of the entropy principle
 and provides an introduction to Liu's procedure
 and to an alternative procedure
 introduced by Boillat, Ruggeri and coworkers for implementing
 the entropy principle.
  In Appendix C, we remind the reader of
 the first, second law 
 and the identification of equilibrium states for
relativistic continuous media.
In Appendix D, intermediate calculations regarding
 the transformations properties of thermodynamical variables under change of the rest frame
 within the context of transient thermodynamics, are
outlined.

\section{Balance Laws for Continuous Classical Media} \label{Continuous thermodynamics}

As a prelude,
in the next three sections, we discuss
a few relevant properties of classical continuous  media.
For our purposes, it is sufficient to consider an electrically neutral, continuous medium which at $t=0$ occupies
  a smooth bounded region $U$ of Euclidean $\mathbb{R}^{3}$ with smooth boundary $\partial U$.
The kinematics of the medium is described by one parameter family of orientation preserving diffeomorphisms\footnote{We assume the motion to 
be smooth enough
so that all mathematical operations defined
in the next sections are well defined.
In Ref. \refcite{MH}, $C^{r}$ regular motions are defined, but in this work
we do not enter in such fine
mathematical details.}
 $\phi_{t}, t \geq 0$ defined according to:
  
\begin{equation}
	\phi_{t}: U \to \phi_{t}(U):=U_{t} \subset R^{3}: \vec X \to \phi_{t}(\vec X):=\phi(t,\vec X),
\end{equation}  
so that the family of the images $\phi_{t}(U):=U_{t}, t>0$
describe the evolution of the 
initial distribution $U:=\phi_{0}(U)$.  Local coordinates  $\vec X:=(X^{1}, X^{2}, X^{3})$ over $U$ 
serve as Lagrangian labels for the elements of the medium.
Thus, for  a fixed $\vec X$,
the set: $\{ {\phi (t, \vec X), t\geq 0}\} $ 
describes a smooth trajectory of the element of the medium
labelled by $\vec X$.
For this fluid element, its velocity $\vec V(\vec X, t)$ and acceleration $g(\vec X, t)$ are defined by:
\begin{equation}
	\vec V(\vec X, t)=\frac {\partial \phi(\vec X, t)}{\partial t},\quad \vec g(\vec X,t)= \frac {\partial^{2} \phi(\vec X, t)}{\partial t^{2}}=\frac {\partial {\vec V}(\vec X, t)}{\partial t}.
\end{equation}
The Eulerian\footnote{An $\vec x \in \mathbb{R}^{3}$ 
is the Eulerian coordinate of a point on the trajectory 
$\{ {\phi(\vec{X},t), t\geq 0}\} $ provided
$\vec x=\phi(\vec{X},t)$ for some $t \geq0$.} 
velocity field $\vec u(\vec x,t)$
is defined via
\begin{equation}
	\vec u(\vec x,t)=\vec V(\vec X, t)\Big\vert_{\vec X=\phi^{-1}(t,x)},
\end{equation}
and by the chain rule it follows that
\begin{equation}
	\frac {\partial {\vec V}(\vec X, t)}{\partial t}=\frac {\partial {\vec u}(\vec x, t)}{\partial t} +\vec u \cdot \nabla \vec u,
\end{equation}
which means that the ``Lagrangian and Eulerian accelerations" are related via: 
\begin{equation}
	\vec g(\vec X,t)= \frac {\partial {\vec u}(\vec x, t)}{\partial t} +\vec u \cdot \nabla \vec u.
\end{equation}
For any function $Q(\vec X, t)$,
we associate its Eulerian counterpart
$q(\vec x , t)$
via 
\begin{equation}
	q(\vec x , t) = Q(\vec X, t)\Big\vert_{X=\phi^{-1}(t,x)},
\end{equation}
and again the chain rule implies
\begin{equation}
	\frac {\partial {Q}(\vec X, t)}{\partial t}=\frac {Dq}{dt}:=\frac {\partial {q}(\vec x, t)}{\partial t} +\vec u \cdot \nabla q(\vec x,t),
\end{equation}
where the Lagrangian coordinates $(\vec X, t)$ and Eulerians $(\vec x,t)$ are related via 
$(\vec x, t)=(\phi(\vec X,t), t)$.
The operator
\begin{equation}
\frac {D}{dt}=\frac {\partial}{\partial t}+\vec{u} \cdot \nabla,
\end{equation}
acting upon scalars, differentiates  along the flow lines of the velocity field and often 
it is denoted by on overdot, for instance for a sufficiently smooth $f$, we write;
$$\dot f:=\frac {Df}{dt}=\frac {\partial f}{\partial t}+\vec{u} \cdot \nabla f.$$
\\
We recall that for continuous media, the notion of the stress describes the internal forces generated by the medium itself
and
according to the Stress principle of Cauchy\footnote{This principle
and its history is discussed  in more details on page $132$ of Ref. \refcite{MH}.}
for any oriented surface element at $\vec x$ with normal vector $\vec n$
that finds itself within the medium, the force
$\vec t( \vec x, t, \vec n)$ 
that the medium generates at $(t, \vec x)$
depends upon
$t$, the spatial point $\vec x=\phi(\vec{X},t)$ and the normal vector $\vec n$. 
The so defined ''vector field" $\vec t(\vec{x}, t, \vec n)$ is
the Cauchy stress vector field and plays an important role in the structure of the dynamical laws describing the 
evolution of the medium.\\
Another important attribute of continuous  media
 is their property
 of permiting the phenomenon of the heat conduction to take place.
This phenomenon, is described by the heat flux function $h(\vec x, t, \vec n)$
which determines the rate of heat conduction at time $t$ across any oriented surface element at $\vec x$ with unit normal $\vec n$. Further ahead, we shall see, that the first law of 
thermodynamics
requires 
the existence of a vector field  $\vec q(\vec{x},t)$, 
refereed as the
heat flux vector, related to
the heat flux function $h(\vec x, t, \vec n)$
via
 \begin{equation}
	h(\vec{x},t,\vec n)=\vec q(\vec{x},t)\cdot \vec n.
\label{CT_HFV}	
\end{equation} 
\\
With
the introduction of the Cauchy stress and the heat flux function, we 
now turn our attencion to the 
description of the balance laws\footnote{For the development of the balance laws,
the transport theorem is very helpful: If $R(\vec x, t)$ is any sufficiently smooth scalar function
and  $\phi_{t}(\hat U)$ is any smooth region of the fluid at time $t$
transported from a subregion $\hat U$ of the initial $U$,
then the following identity holds:

\begin{equation}
	\frac {d}{dt}\int_{\phi_{t}(\hat U)}R(\vec x,t)dV(t)=
\int_{\phi_{t}(\hat U)}
\left[\frac {\partial R}{\partial t} +\nabla \cdot (R\vec{u})\right] dV(t).
\label{INF}
\end{equation}
The derivation of this identity
can be found in \cite{MH} (or in any text of advanced calculus).}. In the classical framework, these laws incorporate the following principles:\\
  
a) mass is neither created nor destroyed.\\

b) Newton's second law is valid in the sense that  the rate of change of linear momentum of any part of the medium equals to the total external and internal forces acting on this part of the medium.\\

c) Energy neither is created nor is destroyed.\\

d) Entropy never decreases in the forward time direction.\\

The first principle expresses mass conservation. For any continuous medium is 
assigned the mass density function $\rho(\vec x, t)$ 
with the property that the integral $M(V)=\int_{V}\rho d\vec x $ is the mass contained
within $V$ (assuming here that both $\rho(\vec x, t)$ and $V$ are sufficiently smooth for the integral to exist).
This $\rho(\vec x, t)$ incorporates the conservation of mass principle, provided\footnote{Here after 
and in order to avoid repetitions, the 
symbol $ \phi_t(\hat{U}) $
that appears in integrals stands for the evolution under $\phi_{t}$ of any arbitrary
open subset $\hat U$ of U.}

\begin{equation}
	\frac {d}{dt}\int_{\phi_{t}(\hat U)}\rho(\vec x, t)dV(t)=0.
\label{DMC}
\end{equation}
This condition, coupled with 
the transport theorem leads to the following easily verifiable Lemma:\\

\begin{lemma}
Let $\phi_{t}, t >0$ a motion 
and $\rho(\vec x, t)$ a smooth density function, then the following are equivalent:\\

1) Conservation of mass holds,\\

2) $\rho(\vec x, t) J(\vec {X},t) = \rho_{0}(\vec X)$, where $\vec  x =\phi( \vec {X},t)$,\\

3) the equation of continuity holds:
$\frac {D\rho}{Dt} +\rho\nabla \cdot \vec u = 0\quad \Longleftrightarrow \quad\frac {\partial \rho(t, x)}{\partial t}+\nabla \cdot(\rho \vec{u})=0$,\\

where $\rho_{0}(\vec X)$ is the density of mass
at the $t=0$ and 
$J(\vec {X},t)=det(\frac {\partial \phi^{i}}{\partial X^{j}})$ is the Jacobian of the family of maps:
$\phi(\vec X,t)=
(\phi^{1}(\vec X,t),\phi^{2}(\vec X,t),\phi^{3}(\vec X,t))$. 
\end{lemma}
The following definition expresses the balance of linear momentum: 
  
\begin{definition}
For any motion $\phi_{t}, t\geq 0$ of a continuous medium characterized by a mass density $\rho(\vec x, t)$, Cauchy's stress vector
 field $\vec t(\vec x, t, n)$ and finds itself in an external force field $\vec b(\vec x, t)$, the balance of linear momentum\footnote{
Just for completeness we 
also mention the balance of angular momentum.
 \begin{definition}
Under the same assumptions as in the Definition $1$, we  
 are saying that the balance of angular momentum is satisfied provided 
 \begin{equation}
	\frac {d}{dt}\int_{\phi_t(\hat{U})}{\vec x}\times \rho\vec u d {\vec x}=\int_{{\partial \phi_t(\hat{U})}_{t}}{\vec x}\times\vec t(\vec x, t, \vec n) d a +\int_{\phi_t(\hat{U})}{\vec x}\times\rho\vec b(\vec x, t)d {\vec x}
\end{equation}
\end{definition}
where $\vec x$ stands for the position vector relative to same origin
and $\vec x \times \vec u$ is the operation of the standard cross product of the Euclidean
$R^{3}$.}
 is satisfied provided for every $\phi_{t}(\hat U)$ the following relation holds:    
\begin{equation}
	\frac {d}{dt}\int_{\phi_{t}(\hat U)}\rho(\vec x, t)\vec u(\vec x,t) d {\vec x}=\int_{{\partial \phi_{t}(\hat U)}}\vec t(\vec x, t,\vec n)da +\int_{\phi_{t}(\hat U)} \rho(\vec x, t) \vec b(\vec x, t)d {\vec x},
\label{BLM}
\end{equation}
\end{definition}
where
${\partial \phi_{t}(\hat U)}$ stands for the boundary 
of $\phi_{t}(\hat U)$
and $da$ is the surface element 
of ${\partial \phi_{t}(\hat U)}$.\\	

This integral relation expresses linear momentum balance, but
as it stands, it does not lead to a local conservation law of linear momentum. However, 
as long as 
 (\ref{BLM}) holds,
 and under some
mild restrictions upon the smoothness of $\phi_{t}, t\geq 0$ and the Cauchy's stress vector field 
$\vec t(\vec x, t,\vec n)$, it can be shown that there exist a unique
second rank tensor field\footnote{
For the derivation of this important property
see Theorem $2.2$ on page $134$ in Ref. \refcite{MH}.} $\sigma=\sigma(\vec x, t)$ referred as the Cauchy's stress tensor, so that  the 
components of $\vec t(\vec x, t,\vec n)$ can be written in the form:
\begin{equation}
	t^{a}(\vec x, t,\vec n)=\sigma^{ab}(\vec x, t)g_{bc}n^{c}=\sigma^{a}{}_{c}n^{c},\quad a,b, c \in (1,2,3),
\end{equation}
where 
$\sigma^{ab}(\vec x, t)$ are the components of the 
Cauchy's stress tensor, $g_{ab}$ are the components of the Euclidean metric of $\mathbb{R}^{3}$ and as always $n^{a}$ 
stand for the components of the outward pointing normal vector field of the
${{\partial \phi_{t}(\hat U)}}$.
Using this representation of 
$\vec t(\vec x,t, \vec n)$ in 
(\ref{BLM}),
and assuming mass conservation,
then the divergence theorem 
yields the local law\footnote{Similar analysis holds for the balance law of the angular momentum.
Here a local conservation law for angular momentum requires 
the Cauchy's stress tensor to be symmetric.
For a derivation of the local angular momentum conservation law
see Theorem $2.10$ on page $138$ of Ref. \refcite{MH}.}:

\begin{equation}
	\rho {\dot {\vec u}}(\vec x, t)=\nabla \cdot \sigma +\rho \vec b,\quad (\nabla \cdot \sigma)^{i}=\frac {\partial \sigma^{ij}}{\partial x^{j}},\quad i, j \in (1,2,3).
\label{CoLM}
\end{equation}
 This law in Cartesian coordinates takes the form{\footnote{It is should be mentioned, that both the balance of linear momentum and angular momentum
discussed here, use the linear structure of the underlying Euclidean 
$\mathbb{R}^{3}$ and in particularly the global existence of 
Cartesian coordinate systems.
Therefore special care is required when these laws 
are required to be written down relative to arbitrary coordinates systems.}

\begin{equation}
	\rho \left(\frac{\partial u^i}{\partial t} + u^j\frac{\partial u^i}{\partial x^j}\right) = \frac{\partial \sigma^{ij}}{\partial x^j} + \rho b^i, \qquad i,j \in (1,2,3),
\end{equation}
and as derived here,
these equations 
have a formal character. They acquire a well defined 
meaning provided first the 
dependence of $\sigma(\vec x, t)$ upon the
motion of the medium itself is spelled out. We will come back to this point further ahead.

\section{The First Law for Continuous Classical Media}  \label{The first law}

So far, the thermodynamical state of the medium is described by
the mass density $\rho(\vec x, t)$, the stress tensor $\sigma(\vec x, t)$, the external force field/unit mass $\vec{b}(\vec x, t)$. However these variables alone do not specify the medium and they are enlarged by the inclusion of:\\

a) the internal energy/unit mass $e(\vec x, t)$,\\ 

b) the heat flux vector $\vec q(\vec x, t)$ so that
$h(\vec x, t, \vec n)=-\vec q(\vec x, t)\cdot \vec n=-g_{ab}q^{a}(\vec x, t)n^{b}$,\\

c) the heat supply/unit mass $Q(\vec x, t)$.\\
 
Also for  any
$t > 0$ and any $\phi_{t}(\hat U)$, we introduce the total kinetic energy $T(t)$ and total internal energy $U(t)$
 given by

\begin{equation}
	T(t)=\int_{\phi_{t}(\hat U)}\frac {u^{2}}{2}dm,\quad U(t)=\int_{\phi_{t}(\hat U)}edm, \quad dm=\rho 
dV(t),
\end{equation}
and in terms of these variables we have the following definition:
\begin{definition}
For a motion $\phi_{t}, t >0 $,
a state of the medium defined by $\rho(\vec x, t)$, $\vec  u(\vec x, t)$, $e(\vec x, t)$, $ h(\vec x, t, \vec n)$, $Q(\vec x, t)$ obeys the balance of energy principle 
provided:    
\begin{equation}
	\frac {d}{dt}\int_{\phi_{t}(\hat U)}\left(\frac {u^{2}}{2}+e\right) dm=\int_{\phi_{t}(\hat U)}(\vec u \cdot \vec b+Q)dm+\int_{\partial \phi_{t}
(\hat U)}(\vec u \cdot \vec t+ h)da.
\label{ECO}
\end{equation}

\end{definition}

This energy principle 
is a restatement of the first law of thermodynamics as applied to continuous media.
It asserts that
within 
any $\phi_{t}(\hat U)$, 
the rate of change of the total energy (= kinetic energy + internal energy (including potential energy)) 
is due to the work done by the combined
external force $\vec b$ and the stress force $\vec t$ 
on this 
 $\phi_{t}(\hat U)$
 augmented by 
the amount of the total heat crossing
$\partial\phi_{t}(\hat U)$
or to the total heat supplied
to $\phi_{t}(\hat U)$ 
by an external agency.\\

As for the case of linear momentum, this 
 integral 
law yields a local conservation
law provided the heat flux function $h(\vec x, t, \vec n)$ and the Cauchy's 
stress field  $\vec t(\vec x, t, \vec n)$ are written in terms of the
heat vector\footnote{The proof that the local law of the energy conservation,
then requires the representation 
$h(\vec x, t, \vec n)=-\vec q(\vec x, t) \cdot \vec n $
see proposition $(3.2)$ in Ref.\cite{MH}.}
$\vec q(\vec x, t)$ 
and Cauchy's
stress tensor $\sigma(\vec x, t)$.  Under
these conditions and by applying  the divergence theorem
to 
(\ref{ECO})
one obtains a 
local law:

\begin{equation}
	\rho\frac {D}{dt}\left(\frac {u^{2}}{2}+e\right)=\rho(\vec u \cdot \vec b+Q)+\nabla \cdot(\sigma  \cdot \vec u-\vec q), \quad
\nabla \cdot(\sigma  \cdot \vec u):=\frac {\partial (\sigma^{i}{}_{j}u^{j})}{\partial x^{i}}.	
\label{ConE}
\end{equation}
Combining 
this law
with the local law 
in (\ref{CoLM})
expressing linear momentum conservation 
and some algebra,
yields: 
\begin{equation}
	\rho\frac {De}{dt}=\rho Q -\nabla \cdot \vec q +\sigma: \nabla \vec u,\quad\quad \sigma: \nabla \vec u =
\sigma^{ij}u_{i,j}
\label{FL_CM_1}
\end{equation}
which describes the manner that the internal energy $e $ varies as it transported along along the flow lines
of the fluid's velocity field.\\

To get insights into the structure of the 
so far derived balance laws,
 let us assume that
for a particular medium the Cauchy tensor $\sigma(t, \vec x)$
 has the special form:
\begin{equation}
\sigma^{ij}(\vec x, t)=-P(\vec x, t)g^{ij},\quad i,j \in(1,2,3).	
\label{PF}
\end{equation}
For such medium,  the law of linear momentum	
(\ref{CoLM}) yields
\begin{equation}
	\rho {\dot {\vec u}}(\vec x, t)=-\nabla P(\vec x, t)+\rho \vec b(\vec x, t),
\label{EE}
\end{equation}
which is recognized as the Euler's equation
with $P(\vec x, t)$ interpreted as the (thermodynamical) pressure for an ideal fluid.\\
However, for an arbitrary medium the 
Cauchy's stress tensor $\sigma(\vec x, t)$ may exhibit a 
much more 
complicated structure
than the one exhibited in 
(\ref{EE}).\\
In order to describe this more general case, it customarily to 
 decomposed it according to:
\begin{equation}
	\sigma^{ij}=-Pg^{ij}-\hat \sigma^{ij},\quad \hat \sigma^{ij}=P^{v}g^{ij}+\sigma_{(v)}{}^{ij}, \quad \sigma_{(v)}{}^{ij}g_{ij}=0,
\label{SRD}	
\end{equation}
where $P$ is the thermodynamical pressure,
$P^{v}$ is the bulk viscous stress,
while  the traceless part $\sigma_{(v)}^{ij}$
 is the shear viscous stress\footnote{This terminology is motivated by 
 the theory of Navier-Stokes fluids
 and in such theory often 
 $\sigma^{ij}$ are also refereed as a the components of the pressure tensor.}. 
 In view of this splitting,  the  evolution of the internal energy $e$
 in (\ref{FL_CM_1})  takes the equivalent  form:
\begin{equation}
\rho \dot e=\rho Q-\nabla_{\mu}q^{\mu}-(P+P^{v})\nabla_{\mu}{u}^{\mu}-\sigma_{(v)}^{ij}u_{(t)ij},
\label{EvIN}
\end{equation}
where 
$u_{(t)ij}$ stands for the symmetric trace free part of $u_{i,j}$ defined via the decomposition:
$u_{{i,j}}=\frac {1}{2}(u_{i,j}+u_{j,i})
+\frac {1}{2}(u_{i,j}-u_{j,i})=u_{(ij)}+u_{[ij]}$ so that
\begin{equation}
	u_{(i,j)}=\left(\frac {1}{3}\nabla^{a}u_{a}\right)g_{ij}+
\frac {1}{2}\left(u_{i,j}+u_{j,i}-\left[\frac {2}{3}(\nabla ^{a}u_{a})g_{ij}\right]\right)=\left(\frac {1}{3}\nabla^{a}u_{a}\right)g_{ij}+u_{(t)}{}_{ij}.
\end{equation}
For latter use, we introduce here
the specific density 
$v(\vec x, t)=\rho (\vec x, t)^{-1}$.
This 
$v(\vec x, t)$
and as a consequence of the continuity equation satisfies
\begin{equation}
\rho \dot v=\nabla_{\mu}u^{\mu},
\label{ESD}
\end{equation}
a formula that will be helpful further ahead.\\

The balance laws derived in the
last two sections
apply to a large class of (classical) continuous media.
It is a peculiarity of the nature of continuous media that these laws 
alone fail to yield a closed system of equations.
Closure
is accomplished
by the specifications of constitutive relations. 
These relations amount to specifying the functional dependance of the Cauchy's stress tensor $\sigma(\vec x, t)$,
  the heat flux vector $\vec q(\vec x, t)$ and 
internal energy $e(\vec x, t)$ upon suitable   
variables describing the medium
and 
 often these variables are refereed as basic variables.
 For instance, for a heat conducting, viscous fluid 
 a set of basic
 variables consists of 
 the density, the velocity field,
 temperature (and may be their derivatives).
 For the specification
of the constitutive\footnote{For an update on issues regarding the theory of 
constitutive relations,  
the reader 
is refereed to Refs. \refcite{Mul4}, \refcite{MH}, \refcite{Trues1}, \refcite{Trues2}, \refcite{Trues3}, \refcite{Rug4},
for further details.}
relations describing
a particular medium, certain selection rules apply 
and at this stage 
the entropy 
and the second law play a central role.
In the next section, we shall discuss the interplay between balance laws,
 constitutive
relations and thermodynamic
and it should be mentioned that 
the focus of the rest of the paper is to delineate the frontiers where
hydrodynamics ends and thermodynamics begins.

\section{The Second Law and Continuous Media-The Entropy Principle} \label{The second law}

Although the derivation of the balance laws discussed in the last two sections
were free of ambiguities, maters complicate considerably once we pass
to the subtle task of assigning entropy to arbitrary states of a continuous media.
In fact, it is fair to claim that except for states describing thermodynamical equilibrium, there exist no universally
accepted recipes 
to assign a meaningful entropy to non equilibrium states
of continuous media.
The possible choices of this non equilibrium entropy, the implementation 
of the second law and its consequences 
will be the central themes for the rest of this paper.
In the next section, we shall introduce classical irreversible thermodynamics (CIT)
where the 
local thermodynamical equilibrium postulate determines the entropy of off-equilibrium states. In 
section$-6$, we introduce the Extended Irreversible Thermodynamics (EIT), where
M\"uller's postulate determines the entropy of states near equilibrium, while within the Rational Extended Irreversible Thermodynamics
(REIT), constitutive relations determines the entropy of off-equilibrium states.\\
 Below and 
 motivated by reasons of
presentations, 
 we do not follow the 
 historical development, 
 but start discussing first
 the issue of assigning an entropy 
to off-equilibrium states
and the implementation of the second  law
 within the
framework of Rational Thermodynamics\footnote{
For an introduction to the principles of Rational Thermodynamics, the reader is refereed to Refs. \refcite{Trues1,Trues2,Trues3}, see also discussion in Refs. \refcite{JVL,MH}.} abbreviated here after as (RT).\\

Rational thermodynamics, leaves the balance laws 
for continuous media derived in the previous sections intact, but 
introduces two additional functions:\\

 the entropy/unit mass $s(\vec x, t)$,\\

 the local temperature $T(\vec x, t)$.\\

The local temperature 
$T(\vec x, t)$ is considered to be an absolute element
and the  theory
implements the second law of thermodynamics via the 
Clausius-Duhem 
inequality according to the definition:

\begin{definition}
For any motion $\phi_{t}, t >0$, 
a state of a continuous media described by
$\rho(\vec x, t)$,
$e(\vec x, t)$, $ h(\vec x, t, \vec n)$, $Q(\vec x, t)$, $s(\vec x, t)$, $T(\vec x, t)$ satisfies the second law of thermodynamics, if the rate of entropy production within any $ \phi_t(\hat{U}) $, i.e.
\begin{equation}
	\frac {d}{dt}\int_{\phi_{t}(\hat U)}s(\vec x, t)dm,
\end{equation} 
satisfies the Clausius-Duhem integral inequality: 
\begin{equation}
	\frac {d}{dt}\int_{\phi_t(\hat{U})} sdm \geq \int_{\phi_{t}(\hat U)}\frac {Q}{T}dm+
	\int_{\partial \phi_{t}(\hat U)}\frac {h}{T }da.
\label{THO}
\end{equation}
\end{definition}

Thus within (RT) the rate of entropy increase within $\phi_{t} (\hat U)$, is greater (or at best equal) 
to 
the entropy generated by the 
heat supplied reversibly to 
$ \phi_{t}(\hat U)$
and the entropy generated by heat flux through the boundary
${\partial \phi_{t}}(\hat U)$.
 For states subject to: $ Q(\vec x, t)=h(\vec x, t)=0 $,
the inequality 
implies 
\begin{equation}
\frac {d}{dt}\int_{\phi_{t}(\hat U)}s(\vec x, t)dm= \frac {d}{dt}\int_{\phi_{t}(\hat U)}\rho(\vec x, t)s(\vec x, t)dV \geq0,
\label{EP1}
\end{equation}
i.e. the entropy cannot decrease in the forward time direction.\\

The Clausius-Duhem inequality
can be expressed in a local 
form provided that one employs the heat flux vector $ \vec{q}(\vec x, t) $ via
 $h(\vec x, t,\vec n)=-\vec q(\vec x, t)\cdot \vec n=-g_{ab}q^{a}(\vec x, t)n^{b}$ 
 where as before $\vec n$ is the outward pointing normal vector of $\partial \phi_{t}(\hat U) $. In that 
event, 
 (\ref{THO}) yields 
the point wise 
form of the Clausius-Duhem
inequality:
\begin{equation}
\rho \frac {Ds}{dt} \geq \frac {Q \rho}{T}-\nabla \cdot \left(\frac {\vec q}{T}\right)\quad\Longleftrightarrow\quad 
\frac {\partial (\rho s)}{\partial t}+\nabla \cdot(\rho s \vec u+\frac {\vec q}{T})\geq  \frac {Q \rho}{T},
\label{CDI}
\end{equation}
which is the local formulation of the second law within the theory.
Within (RT), 
the entropy density $s$ is
considered to be a constitutive 
function i.e.
$s$ dependents
upon 
suitably defined set of basic variables describing the state of the medium
and the explicit dependance of this 
$s$ 
upon the basic variables is a thorny issue.\\

Coleman and Noll\cite{ColN1} in the $1960s$,
were the first 
to suggest 
 a limited version of what nowadays is refereed as the entropy principle.
They suggested that the 
dependance of $s(\vec x, t)$ upon the basic variables must be assigned in such a manner so that the second law of thermodynamics is satisfied for any 
arbitrary thermodynamic processes\footnote{In the language of thermodynamists, a thermodynamic process
 is any solution of the balance equations.}.
 M\"uller\cite{MulN1, MulN2} in the $1970s$ refined
the  Coleman-Noll 
approach and this refinement cemented the 
modern version of the entropy principle.
Specifically 
M\"uller at first postulated that for any medium,
there exists  a scalar additive quantity the entropy density $s$,
and 
an entropy flux vector $\vec J$,
such that 
in the absence of any heat supply $Q$ these
 $s$,
and $\vec J$,
satisfy the entropy inequality:
\begin{equation} 
\frac {\partial (\rho s)}{\partial t}+\nabla \cdot(\rho s \vec u+\vec J)\geq 0.
\label{MCDI}
\end{equation}
Moreover, he postulated that
both $s$ and $\vec J$ are
considered to be constitutive functions
and this introduces an additional degree of flexibility 
in the formulation of the second law\footnote{
Compare
(\ref{CDI})
to the form of 
 the entropy inequality
in 
(\ref{MCDI}). The latter,
does not make 
any reference to the heat flux
nor makes 
no reference to 
the notion of temperature.
Away from equilibrium states, 
the notion of temperature is a subtle concept.}.
In this modern version, the entropy principle 
asserts that 
the dependance of both i.e $s$
and $\vec J$
(or more generally of any constitutive function)
upon the basic variables should be chosen in such a manner
so that every solution of the balance laws
satisfy the above form of the entropy inequality.\\
Nowadays, the entropy inequality and entropy principle 
as refined by 
M\"uller,  lie at the center of the thermodynamics of 
classical continuous media
(and as we shall see further ahead they remain so in the 
theory of relativistic continuous media).
However implementing the entropy principle is a delicate matter.
Does there exists an algorithmic procedure to pick up
the right form of the constitutive relations
so that the entropy principle holds?\\
Since the late $1960$ until recently, there have been
invented various procedures
to implement the entropy principle.
The Coleman-Noll procedure
(see Refs. \refcite{ColN1}, \refcite {ColN2}),
the 
M\"uller procedure (see Refs. \refcite{MulN1}, \refcite{MulN2}), the 
 Liu-procedure (see Ref.\refcite{Liu}) 
 and the procedure developed by  Boillat, Ruggeri 
 and coworkers (see Refs. \refcite{RugF1}, \refcite{RugF2},
\refcite{RugF3}, \refcite{Boi}, \refcite{RugStr}).
In order to facilitate matters, in the Appendix $B$,
we outline 
the basic features of the  Liu's and Boillat-Ruggeri's procedures
and the manner that these procedures implement the principle. \\ 
We finish this section by mentioning
that 
the entropy inequality (\ref{MCDI}) can be cast in the equivalent form
\begin{equation} 
\rho\dot s+\nabla \cdot {\vec J}=\sigma,
\qquad \dot{s} = \frac{\partial s}{\partial t} + u^a\nabla_a s
\label{New}
\end{equation}
where the scalar  $\sigma$
is interpreted as the the density of entropy production
per unit volume and unit time
and the  restriction 
 $\sigma \ge 0$ implements the second law.
 The entropy inequality
in the form shown in (\ref{New}) will be used  further ahead.\\

\section{Classical Irreversible Thermodynamics}\label{S_CIT}
In this section,
we introduce
the first theory of irreversible thermodynamics
describing classical continuous media. The theory
 initiated long ago by Onsager\footnote{Besides Onsager, the works by Eckart, Meixner, Progogine, are also associated with
 the developments of (CIT). For a historical account consult
  Ref.\refcite{JVL}, and also the critical review in 
   Ref.\refcite{MulW}} and will be abbreviated  here after as:
(CIT).
The theory
 leaves intact the 
balance laws
 for mass, linear momentum and total energy introduced earlier on 
 but
uses the local thermodynamical equilibrium postulate (local-equilibrium in short)
to assign an entropy to non equilibrium states.\\
According to (CIT)  a medium finds itself in a state of a local-equilibrium, if at any point $ (\vec x, t) $
a sufficiently small cell can be introduced so that the cell by itself is considered to be a 
thermodynamical subsystem satisfying
the following property:
\textit{within this cell, the state variables are well defined (i.e. do not exhibit wild fluctuations)
and obey the same thermodynamical relations as if this subsystem was in a state of a global thermodynamical equilibrium}.
Accepting this hypothesis, then 
it is relatively straihtforward to assign the equilibrium entropy 
$s(t, \vec x) $ at that cell\footnote{ This can be done for instance 
by appealing to Clausious, Caratheodory's or to Gibbs axiomatic approach
to the equilibrium entropy. Notice however, that (CIT) postulates something more: it postulates
that the functional form of the this local equilibrium
entropy can be taken as describing the physical entropy of the underlying state.}.
It is worthwhile
to briefly describe states of a simple, heat conducting, viscous fluid
within the framework of (CIT).\\
Under the hypothesis 
that the fluid finds itself in a state of local-equilibrium,
 at any $ (\vec x, t) $
the entropy density $s(\vec x, t)$ depends only upon $e(\vec x, t)$ and 
specific volume $v(\vec x,t)=\rho^{-1}(\vec x, t)$
and moreover 
$s(\vec x, t), e(\vec x, t)$ and 
 $v(\vec x,t)=\rho^{-1}(\vec x, t)$
 satisfy : 
\begin{equation}
	T(\vec x, t)ds(\vec x, t)  = de(\vec x, t) + P(\vec x, t)d\left(\frac{1}{\rho(\vec x, t)}\right)
\end{equation}
which is the familiar form of the 
Gibbs equilibrium
relation (except from the crucial dependance of the thermodynamical variables upon 
$(\vec x, t)$).\\
It follows from this relation that 
 the evolution of 
 $s(\vec x, t)$
 along the flow lines obeys: 
 \begin{equation}
	T\frac{Ds}{dt} = \frac{De}{dt} + P\frac{D}{dt}\left(\frac{1}{\rho}\right), \qquad \frac{D}{dt}:= \frac{\partial}{\partial t} + \vec{u}(\vec x, t) \cdot \nabla, 
\label{XXX}
\end{equation}
while for any two nearby fluids element at $ (\vec{x},t) $ and $ (\vec{x} + d\vec{x},t) $,
the density $s(\vec x, t)$
 satisfy

\begin{equation}
	T \nabla s(\vec x, t) = \nabla e(\vec x, t) + P \nabla \left(\frac{1}{\rho(\vec x, t)}\right),
\end{equation}
where $\nabla$ stands for the gradient operator on the Euclidean $\mathbb{R}^{3}$.\\
Now (\ref{XXX}) implies
\begin{equation}
	\rho\dot{s} = \frac{\rho}{T}\dot{e} + \frac{P\rho}{T}\dot{v}, \qquad \dot{s} = \frac{\partial s}{\partial t} + u^a\nabla_a s,
\quad v = \frac{1}{\rho},
\label{CITSL}
\end{equation} 
and using to
(\ref{EvIN}-\ref{ESD}),
yields
\begin{equation}
	\rho\dot{s} = \frac{\rho}{T}Q - \nabla \cdot \left(\frac{\vec{q}}{T}\right) - \frac{\vec{q}\cdot \nabla T}{T^2} 
	- \frac{P^{v}}{T}\nabla_a u^a - \frac{1}{T}\sigma_{(v)}^{ij}u{_{(t)}}_{ij},
\label{EPS}	
\end{equation}	 
which can be cast in the form:
 \begin{equation}
	\rho\dot{s} +\nabla \cdot \vec J_{E}= \sigma,
\label{EPS}	
\end{equation}	   
\begin{equation}
\vec J_{E}=\frac {\vec q}{T},\quad \sigma= \frac{\rho}{T}Q  - \frac{\vec{q}\cdot \nabla T}{T^2} 
	- \frac{P^{v}}{T}\nabla_a u^a - \frac{1}{T}\sigma_{(v)}^{ij}u^{(t)}_{ij},
\label{EPFX}	
\end{equation}	  
where $\vec J_{E}$ is the entropy flux while  $\sigma$ is
the entropy production per unit volume and unit time\footnote{It is a common notation
 to refer to $ \vec{q}, P^v, \sigma_v^{ij} $ in 
 (\ref{EPFX}),  as dissipative fluxes while the corresponding
$$
	\frac{\nabla T}{T^2},\quad \frac{\nabla_a u^a}{T} \quad \text{and} \quad \frac{u_{(t)ij}}{T}
$$
as thermodynamical forces.}.\\ 
Whenever
$Q=0$, then (\ref{EPS}-\ref{EPFX}) 
show that 
the time evolution of fluid states
characterized 
by
$ \vec{q}=P^v=\sigma_v^{ij}=0 $  do not generate entropy.
For this case, 
the 
balance laws 
show that we are dealing with Eulerian hydrodynamics
and 
these Eulerian states may be viewed as the equilibrium states
within the space of all states describing heat conducting, viscous fluids.

However, for states characterized by 
non vanishing heat flux or (and) Cauchy stress,
formulas (\ref{EPFX}) shows that their evolution
generates entropy.
A  look at  
(\ref{EPS})
shows that
the second law is satisfied 
provided 
$\sigma$ in 
(\ref{EPFX})	
	is semi-positive definite
	and this holds 
  provided
  the following linear constitutive relations hold:

\begin{equation}
	\vec{q} = -\mu_1 \frac{\nabla T}{T^2},
\label{A1}
\end{equation}  
\begin{equation}
	P^{v} = -\mu_0 \frac{\nabla_a u^a}{T},
\label{A2}
\end{equation} 
\begin{equation}
	\sigma^{ij}_{(v)} = -\mu_2 \frac{u_{(t)ij}}{T},
\label{A3}
\end{equation} 
 
where the coefficients $ \mu_1, \mu_0, \mu_2 $ are in general
temperature dependent and subject to the restrictions: $ \mu_1 \geq 0 $, $ \mu_0 \geq 0 $ and $ \mu_2 \geq 0 $. 
By introducing the coefficients of the thermal conductivity $k$, bulk viscosity $\zeta$ and 
shear viscosity $\eta$ via
\begin{equation}
	\mu_{1}=kT^{2},\quad \mu_{0}=\zeta T,\quad\mu_{2}=2\eta T,
\end{equation} 
then (\ref {A1}- \ref {A3}) take the form

\begin{equation}
 \vec q=-k\nabla T,\quad P^{v}=-\zeta\nabla_{a}u^{a},\quad \sigma^{ij}_{(v)}=-2\eta u_{(t)ij},
\label{NSE}	
\end{equation}
which are the standard forms of the Fourier and Navier-Stokes linear constitutive
relations known long time ago
\footnote{It
is worth noting that the constitutive
relations in 
(\ref{NSE}) have 
deduced by implementing the second law within (CIT).}.\\
The Fourier-Navier-Stokes theory of heat conducting, viscous fluid
is the standard theory 
describing laboratory and astrophysical fluids and it is 
 reliable\footnote{Although in this section, for illustration purposes, we applied the principles of (CIT) to
states of a heat conducting, viscous fluid, nevertheless,
(CIT) can be employed to describe
states for a fluid mixture or states of other physical systems
and 
the reader is referred to the Refs. \refcite{JVL,Mei,Gro,Gya,Gla} for further applications.} 
as long as restrictions as restricted to length scales greater than certain microscopic distances
(for instance mean free path for the case of a gas).\\
On the other hand, it was recognized long ago,
that as a consequence of the structure of the constitutive relations
in 
(\ref{NSE}), the Fourier-Navier-Stokes system 
predicts 
that 
thermal and viscous signals
propagate with unbounded speed 
and this unsatisfactory property motivated the search for
alternative theories describing
 the  irreversible thermodynamics of a heat conducting, viscous fluid. 
A large number of such alternative theories have been put forward
after M\"uller's\cite{Mul1} in $1967$ introduced a key postulate regarding
the entropy of non equilibrium states and in the next section we discuss 
this postulate in details.

\section{Extended Irreversible Thermodynamics}\label{EIT}

M\"uller\cite{Mul1} in 
 $1967$, suggested that 
for a simple, 
heat conducting, viscous  fluid, 
 the entropy 
 of non equilibrium states differs drastically from the entropy resulting by invoking
 the local equilibrium postulate within the framework of (CIT).
 For states near equilibrium, 
 M\"uller\cite{Mul1} postulated  
 that their entropy 
 receives quadratic contributions\footnote{According to a historical account on Ref \refcite{Jou1},
it appears that other workers before
M\"uller's $1967$ paper were contemplating the enlargement of the
entropy function by the inclusion of dissipative fluxes, but  it seems that 
it was M\"uller's paper\cite{Mul1}
that trigger the development of new
theories.}
from the fluxes $(\vec q, P^{v}, \sigma_{(v)}{}^{ij})$ that appear in the balance laws.
Specifically, 
he assigned
to these near equilibrium  states a generalized entropy $s_{ge}(\vec x,t)$
that has the form:
\begin{equation}
 s_{ge}(\vec x, t)=s_{lep}(e, v)+a \vec q \cdot {\vec q}+b({P^v})^{2}+c \sigma_{(v)}{}^{ij}\sigma_{(v)}{}_{ij}
\label{EEIT}
\end{equation}
where 
$s_{lep}(e, v)$ is the entropy density assigned  by the local equilibrium postulate within (CIT),
and $(a,b,c)$ are smooth functions of $(e,v)$.
Remarkably, M\"uller's hypothesis
yields a theory that cures 
the problems of the 
Fourier-Navier-Stokes mentioned in the last section.
By appealing 
to (\ref{EEIT}) and imposing the second law,
a new set of constitutive relations 
are derived which are structurally different than those predicted by the (CIT).
In turn, these new constitutive relations are the crucial elements  in arriving 
at a set of dynamical equations for the 
Fourier-Navier-Stokes
system predicting finite propagation 
for the heat and shear waves.\\
Due to the close connection of 
M\"uller's postulate to the development
of transient  thermodynamics of relativistic fluids, below, we 
treat 
briefly non equilibrium states of a simple, heat conducting, viscous fluid
within the
Extended Irreversible Thermodynamics\footnote{We 
remind again the reader, that in this paper, the term (EIT), stands 
for the theory 
that abandons
the recipe of assigning entropy to non equilibrium states
as within (CIT) by appealing local equilibrium postulate.
The (EIT)
assigns an entropy to non equilibrium states that depends explicitly
upon fluxes appearing in the balance laws
and whose evolutions are obtained by imposing the second law combined
 with an extended version of the Gibbs relation.
(EIT) is
discussed at length
in the book of Ref.\refcite{JVL}, see also Ref.\refcite{last}.
However often in the literature, the	
 acronym (EIT) stands for alternative
 theories of 
non equilibrium thermodynamics of continuous media
which in one or another way
are based on a form of non equilibrium entropy
which is
motivated largely by
M\"uller's original idea.
For
an overview of such theories, consult for instance Refs. \refcite{Rug2,Rug3}.} 
 (abbreviated as (EIT))
which is a theory that adopts
M\"uller's $1967$ postulate.
According to this theory, for an arbitrary non equilibrium state
of this fluid,
one assigns a generalized entropy
$s_{ge}(\vec x, t)$ 
which is a smooth function of 
 $(e, v, \vec q, P^{v}, \sigma_{(v)}{}^{ij})$ i.e. 
 $s_{ge}=s_{ge}(e, v, \vec q, P^{v}, \sigma_{(v)}{}^{ij})$.
 For states near equilibrium, a series expansion of 
this $s_{ge}(e, v, \vec q, P^{v}, \sigma_{(v)}{}^{ij})$
 around $( e, v, 0, 0,0)$ yields (\ref{EEIT})
 and the resulting theory leads to 
 M\"uller's conclusions
 while for states away from equilibrium, this $s_{ge}(e, v, \vec q, P^{v}, \sigma_{(v)}{}^{ij})$
 may be expanded
 around a background non equilibrium steady state
 (for the definition and treatment of such states,  see Ref.\refcite{last} section $9$).
\\
Within this theory, let
$(e, v, \vec q, P^{v}, \sigma_{(v)}{}^{ij})$
a background non equilibrium steady state
and let
$(e+de, v+dv, \vec q+d\vec q, P^{v}+dP^{v}, \sigma_{(v)}{}^{ij}+d \sigma_{(v)}{} ^{ij}$
a ''nearby non equilibrium state''.
The difference $ds_{ge}$ in the generalized entropies of these two states at least formally,
can be written in the form

\begin{equation}
	ds = \frac{\partial s}{\partial e}de + \frac{\partial s}{\partial v}dv+ \frac{\partial s}{\partial q^i}dq^i + \frac{\partial s}{\partial P^v}dP^v + \frac{\partial s}{\partial \sigma_{(v)}{}^{ij}}d\sigma_{(v)}{}^{ij},\quad i,j \in (1,2,3),
\label{EIT_Gibbs1}
\end{equation}
 where for typographical convenience we write $s$ intead of
 $s_{ge}(e, v, \vec q, P^{v}, \sigma_{(v)}{}^{ij})$,
 summation over the indices $(i,j)$ is understood and
 the partial derivative of $s$ with respect to $e$
 is taken by keeping $(v,  \vec{q} $, $ P^v $, $ \sigma_{(v)}^{ij} )$ fixed (similar restrictions for the other derivatives as well).
This relation is interpreted within (EIT) as a
``generalized Gibbs relation'' 
and introduces formally a non-equilibrium
 absolute temperature $\Theta$ via
\begin{equation}
	\Theta^{-1}(e,v,\vec{q}, P^v, \sigma_{(v)}^{ij}) = \left(\frac{\partial s}{\partial e}\right),
\label{EIT_temperature1}	 
\end{equation} 
and  a non-equilibrium thermodynamical pressure $\pi$  via 
 
 \begin{equation}
 	\frac {1}{\Theta} \pi(e,v,\vec{q}, P^v, \sigma_{(v)}^{ij}) = \left(\frac{\partial s}{\partial v}\right).
\label{EIT_P}	  
\end{equation}
The remaining partial derivatives in (\ref{EIT_Gibbs1}) 
are written\footnote{For a discussion regarding the physical significance
and observability of the non equilibrium temperature $\Theta$ 
defined in (\ref{EIT_temperature1})-
not to be confused with 
the local-equilibrium temperature $T^{-1}=\frac {\partial s(e, v)}{\partial e}$-as well as 
the significance of the  non equilibrium pressure $\pi$
defined in (\ref{EIT_P}), the reader is refereed to chapters $(2,3)$ of 
Ref. \refcite{JVL},
see also Refs. \refcite{last}, \refcite{new1}.}  in the form:

\begin{equation}
	\frac{\partial s}{\partial q^i} = -va_{10}q^i, \quad i \in (1,2,3),
\label{EIT_1a}	
\end{equation}  
\begin{equation}
	\frac{\partial s}{\partial P^v} = -va_{00}P^v, 
\label{EIT_2a}	
\end{equation}   
\begin{equation}
	\frac{\partial s}{\partial \sigma_{(u)} ^{ij}} = -va_{21}\sigma_{(v)} ^{ij}, \quad i,j \in (1,2,3),
\label{EIT_3a}	
\end{equation}   
where the scalar functions $(a_{10}, a_{00}, a_{21}) $
depend in general  upon 
$(e, v, \vec q, P^{v}, \sigma_{(v)}{}^{ij})$.	
With this notation, the generalized Gibbs relation (\ref{EIT_Gibbs1})
takes the form
\begin{equation}
ds = \frac{1}{\Theta}de + \frac{\pi}{\Theta}dv - va_{00}P^vdP^v - va_{10}\vec q \cdot d{\vec q}  - va_{21}\sigma_{(v)}{}^{ij}d\sigma_{(v)}{}^{ij},
\label{EIT_Gibbs2}
\end{equation} 
which implies that the evolution of the generalized entropy along the flow lines is 
governed by
\begin{equation}
\dot s = \frac{1}{\Theta}\dot e + \frac{\pi}{\Theta}\dot v - va_{00}P^v {\dot P}^v - va_{10}
\vec q \cdot \dot {\vec q} - va_{21}\sigma_{(v)}{}^{ij}{\dot \sigma}_{(v)}{}^{ij}
\label{EIT_Gibbs2}.
\end{equation} 
Multiplying this equation by $\rho$ and using
(\ref{EvIN}-\ref{ESD}),
one finds\footnote{Notice that
equation (\ref{EIT_Gibbs2}) in combination  to 
(\ref{EvIN}-\ref{ESD}) imply that
a term $\frac{\pi-P}{\Theta}\nabla_au^a$ 
ought to be included in the right hand side 
of (\ref{EIT_bl2}) and here we see the first conceptual problems of the (EIT) to present themselves.
It is assumed here that the thermodynamical pressure $P$ and 
the non-equilibrium thermodynamical pressure $\pi$
are related and in fact that they are equal. For states near to local equilibrium
that may be the case but for states away from equilibrium it  is far from clear whether this is the case.}	
\begin{equation}
	\rho \dot{s} = -\frac{1}{\Theta}\nabla_iq^i - \frac{P^v}{\Theta}\nabla_au^a 
 - \frac{1}{\Theta}\sigma^{ij}u_{(t)ij} -
a_{00}P^{v}\dot P^{v} - a_{10} \vec q \cdot \dot {\vec q}
- a_{21}\sigma^{ij}_{(v)}{\dot {\sigma}}^{ij}_{(v)},
\label{EIT_bl2}	
\end{equation} 
where 
$u_{(t)ij}$ stands for the symmetric traceless part of $u_{i,j}$ 
and as before 
$\sigma^{ij}_{(v)}$ is the symmetric traceless part
of the stress tensor $\sigma^{ij}$
(see the decomposition in 
(\ref{SRD})).\\	
Ignoring for the moment conceptual problems
related to the physical significance 
of the 
non-equilibrium
 absolute temperature $\Theta$, the non-equilibrium thermodynamical pressure $\pi$  
 as well as the meaning of the other partial derivatives in the generalized
Gibbs relation,
the evolution equation 
$\dot{s} $ in (\ref{EIT_bl2}) can be cast in the form: $ \rho \dot{s} + \nabla \cdot \vec J = \sigma $,
where the restriction $\sigma \geq 0$ enforces the second law.
The entropy flux $\vec J$ is considered to be a 
constitutive
function (see comments following eq.
(\ref{MCDI}))
and for isotropic states\footnote{For more details regarding the choice 
of (\ref{DefJ}) and the results of this section
consult sections $(2.3, 2.4)$ of Ref.\refcite{JVL}.}, 
$\vec J$
is postulated to have the structure:
\begin{equation}
\vec J=\frac {\vec q}{\Theta}
+\beta' P^{v} \vec q+\beta '' \sigma_{(v)}\cdot \vec q, \qquad \sigma_{(v)}\cdot \vec q=\sigma_{(v)}^{ij} q_{j},
\label{DefJ}	
\end{equation} 
where
$\beta '$ and $\beta ''$ are unspecified coefficients depending
upon $e$ and $v$. For this choice
of $\vec J$, the entropy production
$\sigma$ 
takes the form
\begin{equation}
\sigma=
\vec q \cdot \vec X_{1}+P^{v} X_{0}+\sigma_{(v)}{}^{ij}X_{(2)}{}_{ij},
\label{EntP}	
\end{equation} 
with	
\begin{equation}
\vec X_{1}=\nabla \Theta^{-1}+\nabla_{i}\cdot (\beta'' \sigma_{(v)}^{ij})+\nabla (\beta' P^{v})-a_{10}\dot {\vec q}, 
\label{Y_{1}}	
\end{equation} 
\begin{equation}
X_{0}=- \Theta^{-1}\nabla_{i}u^{i}-a_{00}\dot {P^{v}}+\beta' \nabla_{i}q^{i},
\label{X_{1}}	
\end{equation} 
\begin{equation}
{X_{(2)}}_{ij}=-\Theta^{-1}u_{(t)ij}-a_{21}\dot{\sigma}_{(v)ij}+\beta''(\partial_{i} q_{j})_{st},
\label{X_{2}^{ij}}	
\end{equation} 
where 
$(\partial_{i} q_{j})_{st}$ stands for the symmetric and traceless part of the tensor $\partial_{j}q_{i}$.\\
It follows from (\ref{EntP}),
 that 	
a simple manner to implement the second law (i.e.
to enforce a non negative $\sigma $)
is via the choices:
\begin{equation}
\vec X_{1}=\mu_{1}\vec q,\quad X_{0}=\mu_{0}P^{v},\quad 
{X_{(2)}}_{ij}=\mu_{2}\sigma_{(v){ij}}, 
\label{SECL}	
\end{equation} 
where 
$\mu_{1}\geq0, \mu_{0}\geq0, \mu_{2}\geq0$,
are new phenomenological coefficients that
may depend upon $(e, v)$.
By combining
(\ref{SECL})  with  
(\ref{Y_{1}}-\ref{X_{2}^{ij}})
one obtains a set of constitutive relations for this theory
that are structurally different to those
constructed within the framework of (CIT) (compare
(\ref{SECL}) to (\ref{NSE})).\\	

To get insights into these relations,
one neglects 
in (\ref{SECL})
quadratic terms in the fluxes and products of fluxes 
as well as time gradients of  $e$ and $u$.
Under these simplifications,  
(\ref{SECL}) 
combined with
(\ref{Y_{1}}-\ref{X_{2}^{ij}})
yields:
\begin{equation}
\nabla \Theta^{-1}+\beta'' \nabla_{i}\cdot ( \sigma_{(v)}^{ij})+\beta' \nabla (P^{v})-a_{10}\dot {\vec q}=\mu_{1}\vec q, 
\label{F_{1}}	
\end{equation} 
\begin{equation}
- \Theta^{-1}\nabla_{i}u^{i}-a_{00}\dot {P^{v}}+\beta' \nabla_{i}q^{i}=\mu_{0}P^{u},
\label{F_{2}}	
\end{equation} 
\begin{equation}
-\Theta^{-1}u_{(t)ij}-a_{21}\dot{\sigma}_{(v)ij}+\beta''(\partial_{i} q_{j})_{st}=\mu_{2}{\sigma}_{(v)ij}.
\label{F_{3}}	
\end{equation} 
For stationary and spatially homogeneous 
states, the spatial and temporal gradients of the fluxes are zero
and thus
\begin{equation}
\mu_{1}\vec q=\nabla \Theta^{-1},\quad \mu_{0}P^{v}=- \Theta^{-1}\nabla_{i}u^{i},\quad
\mu_{2}{\sigma}_{(v)ij}=
-\Theta^{-1}u_{(t)ij}.
\label{Z_{1}}	
\end{equation} 
Comparing these relations with the standard 
Navier-Stokes formulas in (\ref{NSE}) fixes the parameters $\mu_{0}, \mu_{1},  \mu_{2}$
to the values	
 \begin{equation}
 \mu_{0}=(kT)^{-1},\quad \mu_{1}=(\zeta T)^{-1},\quad
 \mu_{2}=(2 \eta T)^{-1},
\label{NSEE}	
\end{equation}
 where 
$(k, \zeta, \eta)$ stand for the coefficients of thermal conductivity, bulk viscosity and 
 shear viscosity,
 and we identified $\Theta$ with the local equilibrium temperature $T$.\\
 Leaving aside intermediate details (see Ref. \refcite{JVL})
 and introducing relaxation times
 $(\tau_{0}, \tau_{1},\tau_{2})$
via
\begin{equation}
	a_{10} = \tau_1(\kappa T^2)^{-1},
\end{equation}
\begin{equation}
	a_{00} = \tau_0(\zeta T)^{-1},
\end{equation}
\begin{equation}
	a_{21} = \tau_2(2\eta T)^{-1}
\end{equation}
the linearized evolution equations
(\ref{F_{1}}-\ref{F_{3}})
take the form:

\begin{equation}
	\tau_1 \dot{\vec{q}} = -(\vec{q} + \kappa \nabla T) + \beta'' \kappa T^2 \nabla \cdot \sigma_{(v)ij} + \beta' \kappa T^2 \nabla P^v,
\label{G1}
\end{equation}
\begin{equation}
	\tau_0 \dot{P}^v = -(P^v + \zeta \nabla \cdot \vec{u}) + \beta' \zeta T \nabla \cdot \vec{q},
\label{G2}
\end{equation}
\begin{equation}
	\tau_2 \dot{\sigma}_{(v)ij} = -(\sigma_{(v)ij} + 2\eta u_{(t)ij}) +2\beta'' \eta T (\nabla \vec{q})_{st},
\label{G3}
\end{equation}
and these equations  in the limit where
$\beta '$ and $\beta ''$ are vanishing reduce to 
the Maxwell-Cattaneo laws \cite{JVL,Cat1,Cat2}
\begin{equation}
	\tau_1 \dot{\vec{q}} + \vec{q} = -\kappa \nabla T,
\end{equation}
\begin{equation}
	\tau_0 \dot{P}^v + P^v = -\zeta \nabla \cdot \vec{u},
\end{equation}
\begin{equation}
	\tau_2 \dot{\sigma}_{(t)ij} + \sigma_{(t)ij} = -2\eta u_{(t)ij}.
\end{equation}

The time evolution equations
for the fluxes $(\vec q, P^{v}, \sigma_{(v)}{}^{ij})$
shown
in $(\ref{G1}-\ref{G3})$
 are one of the striking implications of (EIT)
when this theory is applied to 
spatially homogeneous states.
These evolution equations combined with the balance yields a closed system 
of equations which are very different than those predicted by (CIT).\\
One of the issues concerning this new system is 
 whether it predicts finite propagation for heat and viscous disturbances.
According to Ref. \refcite{JVL} page (85),
it appears that the (EIT) passes that test 
provided 
the generalized entropy 
$s_{ge}=s_{ge}(e, v, \vec q, P^{v}, \sigma_{(v)}{}^{ij})$
is chosen to be a concave function of its arguments i.e.
 the  second variation $ \delta^2 s_{ge} $ of $ s_{ge} $
 evaluated on the background state is
negative definite. Moreover, 
it is stated in \cite{JVL}, that as long as
the generalized entropy $s_{ge}$ is chosen
to be a concave function of its arguments, then this property of 
$s_{ge}$ is  
equivalent to the symmetric-hyperbolic\footnote{Since in this theory the background state is arbitrary,
it would be nice if this property 
reexamined further. The structure of the dynamical equations needed to have (or transformed into)  the form of conservation laws.} nature of the
underlying equation
which in turn ensures that the  characteristic propagation speeds are real and finite
and in  the same reference document these assertions through specific examples.\\
We conclude this brief introduction of (EIT) 
by mentioning
that this theory has been developed enormously in the last few decades and the reader is 
referred to the vast literature 
(see for instance \cite{JVL,last, new2,GC1,GC2,Sie1})
for further discussion and open problems within (EIT).\\
 
M\"uller's original hypothesis, 
regarding the notion of generalized entropy $s_{ge}(\vec x,t)$
discussed at the beginning of this section,
as well as many of the theories arising from his thesis,
have been placed
 under intense scrutiny.
A comparison
of the predictions of these
extended theories to
the standard theory of Fourier-Navier-Stokes fluids
reveal signs of concern.
In Refs. \refcite{Anil1,Maj1}
the shock structure of M\"uller's original theory was investigated and compared
to the structure predicted by the
Fourier-Navier-Stokes theory.
It is found that regular shock structure exist only 
for sufficiently low Mach numbers and this prediction lead
to questioning the concept of an extended entropy.\\
This crisis, as well as the quest to have a satisfactory 
resolution of the velocity propagation of thermal and shear waves,
lead 
 Liu, M\"uller, Ruggeri and collaborators
to develop a new theory dealing with classical fluids, refereed as
Rational Extended Irreversible Thermodynamics (abbreviated here after as (REIT)). This theory 
takes for granted the moments of the non relativistic Boltzmann
equation for a monoatomic classical dilute gas truncated at some large integer $N$.
Using these moments, and employing Grads $13$-moment distribution function,
 M\"uller and Liu in 
Ref. \refcite{Mul5}
derived equations for the heat flux $q^{i}$ and the components of Cauchy stress $\sigma^{ij}$
for a classical dilute monoatomic gas.
They noticed that these equations take the form of equations of balance form, i.e.
equations of the form
\begin{equation}
\partial_{\alpha} {\vec F}^{\alpha }(\vec u)=\vec f(\vec u),\quad \alpha \in (0,1,2,3), 
\label{MT1}
\end{equation} 
where
$$\vec F^{0}:=(F_{1}^{0},
F_{2}^{0},..........,F_{n}^{0})^{T},\quad
\vec F^{1}:=(F_{1}^{1},
F_{2}^{1},..........,F_{1}^{0})^{T},etc,$$
$$
\vec u(t, \vec x):=(u_{1}(t, \vec x),
u_{2}(t, \vec x),..........,u_{n}(t, \vec x))^{T}
$$ 
and $T$ signifies transpose (the structure of
such balance laws  and some of their basic properties are discussed in the Appendix 2).\\
These results
of 
M\"uller and Liu in 
Ref. \refcite{Mul5} lead to the foundation of (REIT).
Within (REIT), it  is postulated
that  
the moments of the Boltzmann equation ought to be treated as phenomenological equations
describing heat conducting, viscous fluids
so that 
the heat flux $q^{i}$ and the components of Cauchy stress $\sigma^{ij}$
for such fluids
satisfy equations analogous to
(\ref{MT1}). 
This is the central hypothesis
underlying (REIT) and thus at a first side it appears 
that this theory is dealing with dilute gases. However,
that is not the case. Ought to be kept in mind
that the problem of the closure of the $N$ moments 
within (REIT) is 
dealt via the entropy principle
and other methods of classical continuous media. In particular
an entropy law
 is incorporated as an additional balance law
in (\ref{MT1}) and under appropriate restrictions
a symmetric - hyperbolic system of field equations
is emerging (for a brief discussion
on that issue see Appendix B).
The symmetric- hyperbolic nature 
of the dynamical equations within (REIT) is to be compared with 
the parabolic nature of the dynamical equations characterizing
for instance the
Fourier-Navier-Stokes system.
Thus in summary, 
M\"uller's original idea of an extended entropy lead eventually to
the emergence of 
 a symmetric- hyperbolic system of field equations
 as the dynamical equations describing states of classical fluids.
 Due to space limitations, we shall not discuss any further
 the principles of (REIT). For a detailed introduction to this theory, the reader is refereed to
Ref.  \refcite{Mul5},
the monograph 
entitled: \textit{Rational Extended Thermodynamics}, Ref. \refcite{Mul4}
and the relatively recent review 
 in Ref. \refcite{MulW}.\\
The current status of (REIT)
is best described by quoting a passage\footnote{There is
a comments regarding the title of the book of Ref. \refcite{Mul4}. The term rational thermodynamics
should not confused with the theory of rational thermodynamics (RT) 
introduced in Refs. \refcite{Trues1,Trues2,Trues3}
 and
mentioned briefly in section $4$. The authors of Ref. \refcite{Mul4} explain 
 the title of their book in the introduction section as follows:
 \textit{......Rational Extended Thermodynamics. The literature is full of papers referring to extended thermodynamics which, however, are devoid of rational methodology and mathematical cohesion. The epithet rational in the present title is chosen so as  to emphasize the systematic procedure which the book espouses, a procedure typical for a deductive science. }\\
 }
  from the introduction section
of 
this monograph:\\
\textit{.... the shock wave structure calculated in extended thermodynamics....is worse than the shock wave structure in ordinary thermodynamics; and again: many moments are needed to put things right.....
 When enough moments are used to describe the state, (REIT) leads to perfect agreement of theory and experiment.}\\
 We are not going to pursue any further the analysis of 
 irreversible thermodynamics of classical (Newtonian)
media either in the form of (EIT) or (REIT).
We hope that this brief exposition
highlights the potentialities
(and challenging open problems) of non equilibrium thermodynamics
of continuous media.\\
The rest of the paper is focused
on 
aspects of the  thermodynamics
of relativistic continuous media and confronts
 the subtleties arising from the blending
of thermodynamics with the principle of general covariance.

\section{Continuous  media in a relativistic setting}\label{S_CRM}

From this section onward, we discuss
thermodynamical aspects
of
continuous media propagating  on an
arbitrary smooth four dim. spacetime $(M,g)$.
Thermodynamical properties of such media has been the subject of many past investigations,
see for instance
\cite{Eck,LaL,Tol1,Kle,Klu,Syn,Car1,Car2,Dix,Ehl},
and these investigations  cover
the nature of relativistic equilibrium, 
the formulation of the second law, aspects of irreversible
thermodynamics of relativistic fluids, elastic solids etc.
However, the aim of the following sections 
is to  
discuss the progress
that has been made in the last few decades regarding the development of theories of irreversible thermodynamics
of relativistic media that predict finite propagation of disturbances and admitting stable equilibrium states.
To put matters in mathematically precise terms, the aim is to introduce
theories 
of irreversible thermodynamics
of relativistic media 
that are described by dynamical equations 
that constitute a 
symmetric-hyperbolic and causal system of equations. Such systems of equations guarantee 
  the well posedness of the Cauchy problem 
 and moreover
 general perturbations of a background solution propagate within the light cone.\\
 
 As for the case of theories of non equilibrium thermodynamics of classical (Newtonian)
media
outlined in the previous sections,
similarly the field of irreversible
thermodynamics of relativistic media has evolved.
As early as $1940$ Eckart \cite{Eck} and later in $1950s$ Landau and Lifshitz \cite{LaL}, introduced the first theories of 
dissipative relativistic fluids that now days are refereed as conventional theories.
In the late 
$1970$,
 Israel\cite{Isr1}
and 
Israel and Steward\cite{Isr2}
introduced the theory of transient thermodynamics.
In the late $1980$'s 
 Liu,  M\"uller and Ruggeri, \cite{Mul6} (see also
\cite{Mul4})
developed 
a
version
of irreversible
thermodynamics 
for relativistic fluids that extends the principles of 
Rational Extended Irreversible Thermodynamics
(REIT)
to the general relativistic regime.
Motivated by the 
 approach of Liu, M\"uller and Ruggeri, in the early $1990$, Pennisi\cite{Pen}, 
 Geroch and Lindblom\cite{{Ger1}, {Ger2}}, developed 
the theory of relativistic dissipative fluids of divergence type.
In the following sections, we discuss these theories
 and whenever convenient, we illustrate their principles
 by considering states of either simple fluids or fluid mixtures.
In order to pave the way towards to these developments,
in the Appendix C, we remind the reader of a few basic aspects of the
thermodynamics of relativistic media.\\

We begin by setting
the scene for the development of Israel-Stewart's transient thermodynamics\footnote{The theories of 
Eckart \cite{Eck} 
Landau and Lifshitz \cite{Eck}
will emerge as a special limit of the transient thermodynamics.}
and shall illustrate its  principles by specializing the theory  for the moment to arbitrary fluid states\footnote{As it will become clear further ahead
transient thermodynamics describes states near equilibrium, however
for this initial setting up of the theory, this point is irrelevant.
Also, part of the discussion that follows remains valid
if instead of 
fluids, more general continuous media are
considered, like relativistic elastic media, polarized media etc.
For the description of off-equilibrium states 
of such
media, see  for instance Refs. \refcite{Kr1}, \refcite{Kr2}, \refcite{Isr3}.}.
Within this theory, states of a simple fluid
propagating on a smooth spacetime $(M,g)$,
are described by a set of primary variables\footnote{Besides these primary variables, 
a complete specification of arbitrary fluid states requires 
the specification of additional auxiliary variables and these variables will be introduced further ahead.}
consisting of the conserved and symmetric\footnote{It will be assumed here
after
that the fluid is isolated and interacts only with a background gravitational field.}
energy momentum tensor $T$, a conserved timelike particle current $J$ and the entropy
 four vector $S$ obeying:
\begin{equation}
 \nabla_{\mu} T^{\mu\nu}=0,\quad 
 \nabla_{\mu}J^{\mu}=0,\quad \nabla_{\mu}S^{\mu}\geq 0,
\label{Pr1}
\end{equation}
where the inequality satisfied by $S^{\mu}$ is dictated by the second law
and at  this point
we do not impose any restriction upon
the dependance of $S^{\mu}$ upon other basic variables, this dependance will enter the scene gradually.\\
For a fluid mixture, 
an arbitrary state involves 
$n-$particle currents
described by $n-$timelike vector fields $J_{A}$ with $A\in (1,2,.......n)$.
 In the absence
of chemical or nuclear reactions, the primary variables  $(T,  S, J_{A})$ for this fluid mixture, satisfy
 \begin{equation} \nabla_{\mu} T^{\mu\nu}=0,\quad 
 \nabla_{\mu}S^{\mu}\geq 0,\quad \nabla_{\mu}J_{A}^{\mu}=0,\quad A\in (1,2,....n),
 \label{Pr2}
\end{equation}
while in the presence of chemical reactions  the $n-$particle currents  $J_{A}$ satisfy:  
 
\begin{equation}
	\nabla_{\mu}J_{A}^{\mu}= \sum_iC^{i}_{A}r_{i},\quad i \in {(1,2,.....,k)}, 
 \label{Pr3}
\end{equation}
here $k$ is the number of reactions
that involve the species of type $A$,
$r_{i}$ is the $i_{th}$ reaction rate and $C^{i}_{A}$ are the stoichiometric coefficients
(for properties of these coefficients see for instance Refs. \refcite{Isr1,Gro}).\\
For classical fluids, it  is common to assume 
that  the energy momentum tensor satisfies  the weak energy condition i.e.
 $T_{\mu\nu}u^{\mu}u^{\nu}\geq0$ for all future directed timelike vectors $u$
 and thus by Synge's theorem Ref.\refcite{Syn},  
$T$ admits a  unique timelike eigenvector $u_{E}$,
$g(u_{E},u_{E})=-1$ that defines the Landau-Lifshitz or energy frame.\\
On the other hand, every particle current $J_{A}$ defines a unique timelike future directed vector field $u_{A}$ via $J_{A}{}^{\mu}=n_{A}{u_{A}}^{\mu}$ with $g(u_{A}, u_{A})=-1$
and each one of these 
$n$-fields $u_{A}$ define
their own rest frame.
For 
the case of a simple fluid, the 
unique $u_{N}$ parallel to $J$
 defines the Eckart or particle frame.
Thus the primary variables assigned to an arbitray state of a  relativistic fluid (away from the case of relativistic perfect fluids), 
offers the possibility  to introduce 
 more than one rest frame and in general
no fundamental reason exists to choose one versus the other\footnote{To view matters from different perspective:
 the four velocity associated with states of 
 a dissipative fluid 
 ceases to be a well defined element. For instance, within the context of a simple dissipative fluid, should one identify the fluids four velocity with the flux of energy i.e. assign the fluid's for velocity to $u_{E}$ or 
identify the fluid's four velocity with $u_{N}$ i.e. with the flux of particles or may be take
the fluid's velocity to be a combination of the two?
There is not a satisfactory resolutions of this dilemma. 
In this work, this dilemma appear as 
an issue 
regarding the 
specification of 
suitable classes of fields of rest frames.
Transient thermodynamics deals satisfactory
with the absence 
of what we are intuitively accustomed i.e. the fluids four velocity
by building a framework
where the possible choice of fluid's four velocity is restricted within a suitable cone
specified further below.}.\\
It is a common practice amongst relativists
to express the thermodynamical properties of
the fluid either relative to the Eckart frame or relative the Landau-Lifshitz frame. 
This tendency 
gives the impression
 that the laws of irreversible thermodynamics
of relativistic fluids are tied to
a particular frame, even though 
the primary variables do not single out
such a frame.
As we have mentioned in the introduction,
a suggestion by Israel asserts
that as long as considerations are restricted
to states close to thermal equilibrium, there is some sort of ''gauge freedom" 
regarding the choice of the rest frame (or according to the comment $e$ of the previous page
a  ''gauge freedom" in the choice of the fluid's four velocity). A consistent thermodynamical theory can
be developed  that is
manifestly invariant under changes of the rest-frame $u$, 
as long  as $u$ remains within the ``cone" of opening angle 
$\epsilon$
defined
by $u_{E}$ and  $u_{N}$ (for further discussion on this
cone see Appendix A)
and this theory is the
Israel-Stewart transient thermodynamics that we are aiming to develop in the following sections.\\

\section{Global thermodynamical equilibrium}\label{GES}

Of importance for the development
of transient thermodynamics, is the identification
of a class of fluid states refereed as 
equilibrium states.
The primary variables 
for such class of states define a unique 
rest frame (and thus a unique fluid four velocity)
and in this case 
naturally 
the thermodynamical laws
are expressed relative
to this special rest frame\footnote{We ought to be aware however,
that there exist systems that do not admit a rest frame, in the sense that the frame moves with the speed of light.
This for instance occurs for 
the Hawking radiation field on the event horizon of a black
hole.}
(for more discussion on this point see Appendix C or Ref. \refcite{MTW}).
As we shall show with more details further ahead, 
 transient thermodynamics deals with
 the description of states
that are considered to be perturbations of equilibrium states and
thus this  later family 
of states needs to be
specified precisely\footnote{ Hiscock and Lindblom in 
Ref.\refcite{His1}, identified the class of equilibrium states
within transient thermodynamics by a different root than the one that we are going to follow. They first postulated
a particular functional form for the entropy $S^{\mu}$ 
for fluid states and subsequently required that equilibrium states are those
subject to: $\nabla_{\mu}S^{\mu}=0$. They arrived at the same conditions as those postulated
by Israel in 
Ref.\refcite{Isr1} and conditions $1)-5)$ that we are discussing in this section.}.\\

 For a fluid mixture, the primary variables describing
equilibrium states within transient thermodynamics
satisfy the following 
five conditions
(for a detailed discussion regarding
the choice of these conditions, consult Ref.\refcite{Isr1}):\\

$1$) The entropy production vanishes i.e. 
\begin{equation}
	\nabla_{\mu}S^{\mu}=0.
\end{equation}
\\

$2$) There exists a unique hydrodynamical $ 4$ velocity $ u $, $ g(u,u) = -1 $ such that
the primary variables $ (T^{\mu \nu}, J_A^\mu, S^\mu) $
for all $ A\in (1,2,....n)$,
take the form: 
\begin{equation}
	T^{\mu \nu} = \rho u^\mu u^\nu + P {\Delta}^{\mu \nu}(u), \quad \Delta^{\mu \nu}(u)=g^{\mu\nu}+u^{\mu}u^{\nu},\quad
	J^\mu_{A}=n_{A}u^{\mu},\quad S^{\mu}=su^{\mu}
	\label{tensor energia momento_1}	
\end{equation}
where 
$(\rho, P, n_A)$ stand for the energy density, thermodynamical pressure and particle densities
measured 
by an observer comoving with the flow
defined by $u$,
while $s=-S^{\mu}u_{\mu}$ is the entropy density perceived by that observer.\\

$3$) There exist an equation of state of the form $ s=s(\rho,n_{A}) $ from 
which the equilibrium pressure $ P(\rho,n_{A}) $ can be derived from the relation
\begin{equation}
s=\frac {\rho +P}{T}-
\sum_{A=1}^n \Theta_{A}n_{A},
\label{Entro_py}
\end{equation}
with the temperature $ T $ and the thermal potentials $ \Theta_{A} $, $ A \in (1,..,n) $ defined 
from the Gibbs equilibrium relation
\begin{equation}
ds=\frac {d\rho}{T}-
 \sum_{A=1}^n\Theta_{A}dn_{A},
\label{GibbsR}
\end{equation}
(the origin of the last two fundamental relations 
are discussed in the Appendix C).\\

$ 4 $) The thermal potentials $\Theta_{A}$ and the reaction rates $ r_i $ obey
\begin{equation}
\Theta_{A}C^{i}_{A}=0, \quad r_{i}=0,  \quad \Theta_{A}=\text{const.}, \quad A \in (1,2,...,n).
\label{TherP}
\end{equation}
\\

$ 5 $) The motion is rigid
in the sense of Born i.e. satisfies
\begin{equation}
\Delta^{\mu a}\Delta^{\nu b}(\nabla_{a}u_{b}+\nabla_{b}u_{a})=0,
\label{BorM}
\end{equation}
(for properties of this type of fluid motions, see for instance discussion in Ref. \refcite{Syn}).\\

When these five conditions hold 
simultaneously, then states
defined by the
primary variables
$(T^{\mu\nu}, J_{A}^{\mu}, S^{\mu})$
in 
(\ref{tensor energia momento_1}),	
describes global thermodynamical equilibrium.
To 
get insights into these states, let us define the inverse temperature vector $b^{\mu}$
via 
\begin{equation}
	b^\mu = \frac{u^\mu}{T},
\label{InT}
\end{equation}
and note that this definition, in combination to $\Theta_{A}=\text{const.}$ and
(\ref{BorM}), implies
that $ b^\mu $ is a timelike Killing field\footnote{We may add here that conformally invariant fluid
theories (see discussion and references at the end of section $(13)$)
have equilibrium states where $ b^\mu $ is a conformal Killing vector field.
We expect that this is due to the radiation equation of state
$P=\frac {1}{3}\rho$ which is dictated by conformal invariance, although this claim needs to be checked
in details.}
i.e. obeys 
\begin{equation}
	\nabla_\mu b_\nu + \nabla_\nu b_\mu = 0.
\end{equation}
This equation implies
\begin{equation}
	T(-b_\mu b^\mu)^{1/2} = \text{const.},
\end{equation}
as well as the Tolman-Klein law:
\begin{equation}
	(T\Theta_A)(-b_\mu b^\mu)^{1/2} = \text{const.} 
\end{equation}
Thus states 
in global thermodynamical equilibrium are special.
Besides the vanishing
of the entropy production
$\nabla_{\mu}S^{\mu}=0$, 
and the restrictive nature of the primary variables 
shown in (\ref{tensor energia momento_1}),	 
they require
 stationarity of the background
 spacetime $(M,g)$ 
 and in addition the four velocity $u^{\mu}$ should be parallel to
 the timelike Killing field $b^{\mu}$
 and these conditions are very restrictive.\\
 On the other hand, states
 that obey conditions $1)$ through $3)$
 describe states in a local thermodynamical equilibrium
  (as opposed to states describing global thermodynamical equilibrium). 
  These states
  are characterized by the vanishing of entropy production i.e.
  $\nabla_{\mu}S^{\mu}=0$ and this property follows as a consequence of 
  (\ref{tensor energia momento_1}) coupled to	  
 (\ref{Entro_py}) (see Appendix A for details regarding this point).
  Moreover, for these states, the fluid flow it is not 
 any longer  required to be rigid 
  since condition $5)$
it is not required to hold
 and
 thus the background $(M,g)$ it is not required to be stationary\footnote{There exist plenty of systems
 admitting states
 in local equilibrium. For instance, any state describing a simple perfect fluid propagating in 
 an arbitrary $(M,g)$ satisfies  conditions 
 $1)$ to $3)$ 
 but not in general conditions $4)$ to $5)$.
 Relativistic kinetic theory offers other states in local thermodynamical equilibrium.
 For a simple relativistic gas,  a local Maxwellian distribution makes the collision integral in the relativistic 
 Boltzmann equation to vanish and thus locally the entropy production vanishes.
 Moreover a local Maxwellian distribution, introduces a natural rest frame and the corresponding energy momentum
tensor and particle currents have the form as in 
(\ref{tensor energia momento_1}). 
Small deviations from a local 
Maxwellian distribution satisfy  
relation (\ref{Entropy_Off_E}) derived further ahead.
The interest reader is referred 
to Refs. \refcite{Ehl}, \refcite{O1,O2,O3,An1}
for an introduction to this theory
while for a relation of kinetic theory to phenomenology see sections $(3-7)$ of Ref. \refcite{Isr2}.}.	  
 States in local (or global)  thermodynamical equilibrium
 can be thought as comprising a $ (n + 4) $ dimensional space $ \hat E $ parametrized by the $n-$thermal potentials $ \Theta_A $, $ A \in (1,...,n) $ and an inverse temperature vector $ b^\mu $ defined
 in (\ref{InT}) and this observation will be useful further ahead. \\
 In the above formulation of conditions $1)-5)$,
  the rest frame 
  $u$ or the unique hydrodynamical four velocity defined
   in (\ref{tensor energia momento_1}),
 played a prominent role.
  However, this prominence appears to subside 
  by passing to a 
  covariant form of the equilibrium Gibbs relation.
  To derive
  this version, note that the standard form of the Gibbs relation (\ref{GibbsR}) in combination 
  to (\ref{Entro_py}) imply the following identities
  derived first by Israel\cite{Isr1}

\begin{equation}
	T^{-1}dP + (\rho + P)dT^{-1} = \sum n_Ad\Theta_A,
\label{SL_CM_GE}	 
\end{equation}

\begin{equation}
	Td(sX) = d(\rho X) + PdX - T\sum \Theta_Ad(n_A X),
\label{SL_CM_GE1}	
\end{equation}

where $X$ stands for an arbitrary function.
Choosing $ X=1 $ 
in (\ref{SL_CM_GE1})	
yields (\ref{GibbsR})
while 
for a simple fluid
the choice $ X = V := \frac{1}{n} $ yields

\begin{equation}
	d\hat s = T^{-1}(de + PdV), 
\label{GSF}
\end{equation}

where
$\hat s=sn^{-1}$ is the entropy per particle and
 $e $ is the internal energy per particle defined according to $ \rho = n(m +e) $ (we are employing
units so that $ k = c = 1 $).
However, the most important
relation 
hidden in 
(\ref{SL_CM_GE1})
arises by choosing:
$X=u^\mu$.
Remembering that we are dealing with states 
obeying $1)$ to $3)$
(or possibly $1)$ to $5)$), then 
for the choice 
$X=u^\mu$,
the identity
(\ref{SL_CM_GE1})
implies\footnote{Notice that in the following three relations,
we denoted the primary variables that describe states in (local or global) equilibrium by $ (T_0^{\mu \nu}, J_{0A}^\mu, S_{(0)}^\mu) $ and the reason for introducing this notation will become clear shortly.}
\begin{equation}
dS_{(0)}^\mu = - \sum_{A=1}^n \Theta_AdJ^\mu_{(0)A} - b_\lambda dT_{(0)}^{\lambda \mu},
\label{Entropy_ne0}
\end{equation}
while (\ref{Entro_py}) yields
\begin{equation}
S_{(0)}^\mu = Pb^\mu - \sum_{A=1}^n \Theta_AJ^\mu_{(0)A} - b_\lambda T_{(0)}^{\lambda \mu}.
\label{Entropy_ne1}
\end{equation} 
Moreover, multiplying 
(\ref{SL_CM_GE}) by $u^{\mu}$ and after some algebra, we obtain 
the useful relation	 
\begin{equation}
d(Pb^{\mu})=\sum_{A=1}^n J^{\mu}_{(0){A}}d\Theta_{A}+T_{(0)}^{\mu\nu}db_{\nu}.
\label{FuG}
\end{equation} 
Relation 
(\ref{Entropy_ne0})
is the covariant version of (the equilibrium) Gibbs relation
which involves
only covariant objects and thus 
eliminates any quantity defined relative to a specific frame. This
 covariant version
describes reversible transformations
from the (global or local) equilibrium state parametrized by 
$(\Theta_{A}, b^{\mu})$ 
to a nearby (global or local) equilibrium state\footnote{Within the mathematical framework introduced in the Appendix C,
the $d$-variations
that appear in 
(\ref{Entropy_ne0})
are variations along a fiber $Y$ over a point on $(M,g)$
but leave us within
the equilibrium manifold $\hat E$, see Appendix C.}
parametrized by $(\Theta_{A}+d\Theta_{A}, b^{\mu}+db^{\mu})$.
It is an  aesthetically pleasing
formula and 
as we shall see in the next section,
plays an important
role in the formulation of the transient thermodynamics.

\section{The Principles of Transient Thermodynamics}\label{NES}

 In section \ref{EIT},
we have seen 
that the deficiencies in the Fourier-Navier-Stokes system appear to be cured
by abandoning the 
formula for the entropy of off-equilibrium states arising by appealing to (CIT)
and introducing 
instead the notion of the generalized entropy 
that receives quadratic contributions from
dissipative fluxes.\\
The development of the transient (or causal) thermodynamics
parallels similar route.
 Israel in Ref.\refcite{Isr1}, 
 motivated from the relativistic kinetic theory of gases, 
 put forward the hypothesis 
that an arbitrary non equilibrium fluid state,
beyond the primary variables $T^{\mu\nu},
J_{A}^{\mu}$ and $S^{\mu}$ 
is characterized 
by an additional (perhaps an infinite) set of auxiliary variables denoted collectively by 
$X_{(i)}^{\mu\nu\lambda...}$ where $i\in (1,2,....).$
Furthermore, he suggested that an equation of state (EOS) should exist 
of the form 
\begin{equation}
S^{\mu}=F^{\mu}(T, J_{A}, X_{(i)}^{\mu\nu\lambda...}),\quad i\in (1,2,3.....),
\label{EOS1}
\end{equation}
having the following properties:\\

a) For any equilibrium state (where all
$ X_{(i)}^{\mu\nu\lambda...}$ vanish)
this (EOS) reduces to 
the linear 
relation between
$(T_{(0)}{}^{\mu\nu}$, $J{_{A}}_{(0)}^{\mu}$, $S_{(0)}^{\mu})$ 
shown in
(\ref{Entropy_ne1}).\\

b) Away from equilibrium states,
and as long as consideration are restricted to states near equilibrium,
then 
$S^{\mu}$ 
in 
(\ref{EOS1}),
can  be expanded in a Taylor series around a background  equilibrium state
and in such expansion
contributions higher than quadratic, at a first instance, can be neglected.\\

The assumption that an (EOS) 
of the form
(\ref{EOS1})
exists that obeys conditions a) and b) constitutes the backbone of transient thermodynamics.
Assuming validity of a) and b), let
$(\Theta_{A}, b^{\mu})$ 
 be an arbitrary point
 on the space of equilibrium states $\hat E$ 
so that 
 $(T_{(0)}{}^{\mu\nu}, J{_{A}}_{(0)}^{\mu}, S_{(0)}^{\mu})$
stand for the primary variables describing this particular equilibrium state. 
Let now $(dS^{\mu}, dT^{\mu\nu}, dJ_{A}{}^{\mu})$
be an infinitesimal perturbations of this equilibrium state
 so that 
 \begin{equation}
S^{\mu}=S_{(0)}^{\mu}+dS^{\mu},\quad T^{\mu\nu}=T_{(0)}{}^{\mu\nu}+dT^{\mu\nu},\quad  J_{A}^{\mu}=J{_{A}}_{(0)}^{\mu}+dJ_{A}{}^{\mu},
\label{PPP}
\end{equation}
 define a new state that remains
 ``infinitesimally near'' to 
$(T_{(0)}{}^{\mu\nu}, J{_{A}}_{(0)}^{\mu}, S_{(0)}^{\mu})$
but lies\footnote{In the Appendix C, a mathematical framework is outlined
where the notion of states close to the equilibrium manifold $\hat E$ acquires 
a well defined mathematical meaning.} off-$\hat E$.\\
 ``\textit{The release of variation postulate}" introduced
in \cite{Isr1,Isr2},
states 
 that  the infinitesimal perturbations
 $(dS^{\mu}, dT^{\mu\nu}, dJ_{A}{}^{\mu})$
in (\ref{PPP}) are not independent but satisfy  
 a (non equilibrium) covariant Gibbs relation
of the form
  \begin{equation}
dS_{}^\mu = - \sum_{A=1}^n \Theta_AdJ^\mu_{A} - b_\lambda dT_{}^{\lambda \mu},
\label{ad_1}	
\end{equation}  
i.e. 
a relation
that has the
same functional form as the 
  equilibrium Gibbs relation in 
(\ref{Entropy_ne0}) 
 except 
 that presently  
 the perturbations\footnote{Notice
 that the $d$-variations that appears in 
 (\ref{ad_1}) are considered to be fiber variations
 i.e. variations 
 that move us along the fiber $Y$ over a basis point
 and ''away'' from the equilibrium manifold $\hat E$.
 These variations are distinct to the $d$-variations appearing
 in the equilibrium Gibbs relation
(\ref{Entropy_ne0}).
The underlying  mathematical framework is briefly outlined in Appendix C.}	 
 $(dS^{\mu}, dT^{\mu\nu}, dJ_{A}{}^{\mu})$  
 lead us off -$\hat E$. 
 Recalling
 that 
 $(T_{(0)}{}^{\mu\nu}, J{_{A}}_{(0)}^{\mu}, S_{(0)}^{\mu})$
obey (\ref{Entropy_ne1}), addition of  (\ref{Entropy_ne1}) and (\ref{ad_1})
 yields
the following fundamental relation 
between the primary variables
 $(S^\mu, T^{\lambda\mu}, J_{A}{}^{\mu})$ 
and the parameters 
$(\Theta_{A}, b^{\mu})$ 
 describing the background equilibrium state
  \begin{equation}
	S^\mu = P(\Theta_A, b)b^\mu - \sum_{A=1}^n \Theta_A J_{(A)}^\mu - b_\lambda T^{\lambda \mu} - Q^\mu(\delta J_A^\mu, \delta T^{\lambda \mu}, X_{(i)}^{\mu \nu..}),
\label{Entropy_Off_E}
\end{equation}
where the term $ Q^\mu $ takes
care of the quadratic
and higher order 
contributions 
in the Taylor series expansion of $ S^\mu $ 
around the equilibrium state.
In this formula,
$P(\Theta_{A}, b)$ is the thermodynamical pressure
of the background equilibrium state
and the perturbations
$ \delta J_A^\mu:=J^{\mu}-J_{0}{}^{\mu} $, $ \delta T^{ \mu\nu}:=T^{\mu\nu}-T_{0}{}^{\mu\nu} $
describe
deviations from the equilibrium state.\\
Formula (\ref{Entropy_Off_E}) has similar structure
as the generalized entropy $s_{ge}(\vec x,t)$
 introduced 
 in the development of (EIT) in section $(6)$. 
 Depending upon the structure of 
 the $Q^{\mu}$ term, the entropy vector $S^{\mu}$ may receive contributions
 from terms describing 
 deviations from the background equilibrium state.
 As we shall see further ahead, 
the Eckart \cite{Eck} theory and Landau-Lifshitz \cite{LaL} theory
are generated by taking $Q^{\mu}:=0$,
while second order theories\footnote{The terms ``first order theories'', ``second order theories'' seem to have been 
coined by
 Hiscock and Lindblom in Ref.\cite{His1}.}
postulate that $Q^{\mu}\neq0$.
However before we discuss properties of the resulting theories,
we first point out a few implications
 of the fundamental relation in (\ref{Entropy_Off_E}).\\

As it stands, this relation
combines the primary variables
describing the non equilibrium state and the variables of a reference background equilibrium state
 and this mixing make
difficult to extract
out of 
 (\ref{Entropy_Off_E}) 
  the relevant physics.
 However, one point  worth recognizing
is the non uniqueness property of the reference equilibrium state 
parametrized by $ (\Theta_{A}, b^\mu) $.
Israel and Stewart observe  
 that if the parameters $ (\Theta_{A}, b^\mu)$
 that specify an equilibrium state 
 $(T_{(0)}{}^{\mu\nu}, J{_{A}}_{(0)}^{\mu}, S_{(0)}^{\mu})$ 
  which is near 
to $(T^{\mu\nu}, J_{A}^{\mu}, S^{\mu}, X_{i}^{\mu\nu....})$, 
are displaced 
 to nearby values
 $(\Theta'_{A}=\Theta_{A}+\delta \Theta_{A}, b'_{\mu}=b_{\mu}+\delta b_{\mu}) $,
 then
(\ref{Entropy_Off_E}) can be written in the equivalent form:
 \begin{equation}
 \begin{split}
S^\mu = & P(\Theta'_A- \delta \Theta_{A}, b'_{\nu}-\delta b_{\nu})(b'^\mu-\delta b^{\mu}) - \sum_{A=1}^n 
(\Theta'_A- \delta \Theta_{A})J_{A}^{\mu}
- (b'_\nu-\delta b_{\nu}) T^{\mu\nu} - \\ 
& -Q^\mu(\delta J_A^\mu, \delta T^{\lambda \mu}, X_{(i)}^{\mu \nu..})
= P(\Theta'_A,b'_{\nu})b'^\mu - \sum_{A=1}^n 
\Theta'_A J_{A}^{\mu}
- b'_\nu T^{\mu\nu} - Q'^\mu,
\end{split}
\label{TranS}
\end{equation}
\begin{equation}
Q^{{\mu}'}=Q^{\mu}- \sum_{A=1}^n(J_{A}^{\mu}-J^{\mu}_{(0)A})\delta \Theta_{A}-(T^{\mu\nu}-T_{0}^{\mu\nu})\delta b_{\nu}
\label{TQ}
\end{equation}
where
terms $(\delta \Theta_{A})^{c}$ and $(\delta b_{\mu})^{c}$, with $c\geq2$ are neglected.
In turn, 
(\ref{TranS}) implies 
that if terms quadratic in 
$(\delta \Theta_{A}, \delta b_{\mu}) $,
are neglected, then both
$(\Theta_{A}, b_{\mu})$ 
 and
 $(\Theta'_{A}=\Theta_{A}+\delta \Theta_{A}, b'_{\mu}=b_{\mu}+\delta b_{\mu}) $ 
  can serve as a reference background equilibrium state
 for $(T^{\mu\nu}, J_{A}^{\mu}, S^{\mu}, X_{i}^{\mu\nu....})$.\\
 This freedom in the choice of the reference
equilibrium state 
 simplifies matter
considerably.
For a 
 non equilibrium state\footnote{In order to avoid technicalities,
in the following analysis, we restrict attention to the case of a simple fluid.
The case of fluid mixture can be treated similarly and details can be found in ref.\refcite{Isr2}.}
specified by $(T^{\mu\nu}, J^{\mu}, S^{\mu}, X_{i}^{\mu\nu....}),$
Israel and Stewart 
assign a reference equilibrium state 
in the following manner:
First they choose a four velocity $u^{\mu}$ within the cone of opening angle $\epsilon$ defined by $u_{E}$ and $u_{N}$
(see Appendix A for the definition of this cone).
 Once a choice of this $u^{\mu}$ has been made, 
the primary
variables $(T_{(0)}{}^{\mu\nu}, J{}_{(0)}^{\mu})$ 
identifying a
reference equilibrium state 
that is near to 
$(T^{\mu\nu}, J^{\mu}, S^{\mu}, X_{i}^{\mu\nu....}),$
are chosen so that the following fitting conditions hold:
 \begin{equation}
(J^{\mu}-J_{(0)}^{\mu})u_{\mu}=(T^{\mu\nu}-T_{(0)}{}^{\mu\nu})u_{\mu}u_{\nu}=0,	
\label{FC}
\end{equation}
while the rest of the thermodynamical variables $s(u), T(u)$ and $P(u)$
for this reference state 
are constructed by appealing 
to the equilibrium equation of state 
and the equilibrium Gibbs relation (\ref{GibbsR}).
For the so defined reference equilibrium state,
the fundamental relation
 (\ref{Entropy_Off_E}) in view of 
 (\ref{Entropy_ne1}) yields 
 
$$
-u_{\mu}(S^{\mu}-S_{0}^{\mu})=
-[ P(\Theta_A, b)b^\mu - \sum_{A=1}^n \Theta_A J_{(A)}^\mu - b_\lambda T^{\lambda \mu} - 
Q^\mu(\delta J_A^\mu, \delta T^{\lambda \mu}, X_{(i)}^{\mu \nu..})-$$
$$-Pb^\mu +\sum_{A=1}^n \Theta_AJ^\mu_{(0)A} +b_\lambda T_{(0)}^{\lambda \mu}]u_{\mu}=u_{\mu}Q^{\mu},$$
which implies that the 
 entropy density $s(x)=-u_{\mu}S^{\mu}$ of the actual state  and the 
entropy density $s_{0}(x)=-u_{\mu}S_{0}^{\mu}$ of the 
reference
equilibrium state 
as measured
by the $u$-observer,
obey
\begin{equation}
s_{0}(x)-s(x)=-u_{\mu}Q^{\mu},
\label{FC2}
\end{equation}
i.e.  the two densities agree to first order deviations from the equilibrium
and differences between these densities appear in second order deviations.
Moreover 
(\ref{FC2}) shows that 
under the assumption that the fitting conditions
(\ref{FC}) hold, then  among all states with the same $(\rho(u),n(u))$
then $s(x)=-u_{\mu}S^{\mu}$ attains its maximum value at equilibrium,
if and only if $Q^{\mu}$ is timelike and future directed.\\

Relation (\ref{FC2}) has another important consequence:
the thermodynamical pressure\footnote{Notice that in addition, there exist two inter-related issues that needed to be 
clarified. First how one identifies the thermodynamical pressure $P(u)$ and secondly how one differentiates
$P(u)$ from the bulk 
pressure $\pi(u)$ that appears 
in the formula (\ref{PP1}) bellow. The analysis that follows addresses also these two issues.}
$P{(u)}$
for the off-equilibrium state
as measured by the $u$ observer
and the equilibrium pressure $P(\Theta, b)$ 
that appears in
(\ref{Entropy_Off_E}) satisfy
\begin{equation}
P{(u)}-P(\Theta, b)=O_{2},
\label{PD}
\end{equation}

i.e. the two pressures
agree to first order in deviations from equilibrium
and this property can be seen as follows: For the off-equilibrium state
specified by $(T^{\mu\nu}, J^{\mu}, S^{\mu}, X_{i}^{\mu\nu....})$,
the thermodynamical pressure $P(u)$ 
and the
bulk pressure
$\pi(u)$ 
as perceived by the $u$ observer
satisfy
 \begin{equation}
\frac {1}{3}T_{\mu\nu}\Delta(u)^{\mu\nu}=P(u)+\pi(u), \quad  \Delta(u)^{\mu}{}_{\nu}=\delta^{\mu}{}_{\nu}+u^{\mu}u_{\nu},
\label{PP1}
\end{equation}
where $ \Delta(u)^{\mu}{}_{\nu}$
is the projection tensor associated to the particular four velocity $u$
(relations  (\ref{PP1})
will become clear further below).
Relative to the rest frame specified by $u$ 
and for a simple fluid,
the thermodynamical pressure $P(u)$ can be 
defined either by appealing to the relation 
$s(u) = b\left(\rho(u) + P(u)\right) - \Theta(u) n(u)$
see equacion 
(\ref{ET:15}) in the Appendix C
or by appealing to the non equilibrium Gibbs relation,
see equation (\ref{ET:9}) in the 
same Appendix. The latter equation implies:
\begin{equation}
	d\hat s=T^{-1}\left[ de+P(u)d\left(\frac {1}{n}\right)\right]=
T^{-1}[d(\rho V)+P(u)d(V)],\quad V=n^{-1},
\end{equation}
and thus for any process that maintains the entropy particle $\hat s:=sn^{-1}$ fixed, obeys:
$$d(\rho V)+P(u)d(V)=0,$$
which identifies the thermodynamic pressure $P(u)$ according to
\begin{equation}
	P(u)=-\frac {\partial (\rho V)}{\partial V}\Big\vert_{(sV, n)}.
\label{THERP}
\end{equation}
This formula is util to establish validity of
(\ref{PD}). Indeed we find
\begin{equation}
	P(u)=-\frac {\partial (\rho V)}{\partial V}\Big\vert_{(sV, n)}=-\frac {\partial (\rho_{0} V)}{\partial V}\Big\vert_{[(s_{0}+u_{\mu}Q^{\mu})V,n_{0}]}=P(\Theta, b)+O_{2},
\end{equation}
where the partial derivatives are computed at fixed entropy per particle and particle density
and we passed to the second equality by appealing to relations
(\ref{FC}) and 
(\ref{FC2}).
Therefore the thermodynamical pressure $P(u)$ in 
(\ref{PP1})
and 
the equilibrium pressure $P(\Theta, b)$ 
appearing in (\ref{Entropy_Off_E}),
agree to first order deviations from equilibrium.
Since we are interested only in the 
dynamics of first order deviations from equilibrium,
here after we do not differentiate between the two pressures.
Parenthetically, notice that
(\ref{THERP}) offers the operational means of differentiating\footnote{Although
the derivation of 
(\ref{THERP})
presupposes a simple fluid, a similar formula holds for the case of a fluid mixture.
Details of this can be found in Ref.\cite{F1}.}
 between $P(u)$ and $\pi(u)$.\\

We now consider some implications
arising from the non-uniqueness of the  four velocity $u$
that enters into the theory through the fitting conditions
(\ref{FC}).
Within the framework of the transient 
thermodynamics,
 the four velocity $u$
 that enters into these fitting 
 conditions is arbitrary
except that it is restricted to lie within the cone of opening angle $\epsilon$ specified by $u_{E}$ and $u_{N}$.
Once however, a choice of $u$ has been made,
the primary variables
$T^{\mu\nu}$ and $J^{\mu}$
can be 
decomposed\footnote{In this and the remaining sections, we
write 
$\rho(u), P(u), h(u), \tau(u)^{\mu}{}_{\nu}$ etc. in
order to remind the reader that these 
quantities are 
measured by the $u-$observer, and ought to keep in mind
that in general the
decompositions  in 
(\ref{OD}), (\ref{OD_11}),
(\ref{SD}) and (\ref{PRS})
depend upon the chosen $u$ and thus are frame dependent.} 
according to
\begin{equation}
T^{\mu\nu}=\rho(u)u^{\mu}u^{\nu}+P(u)\Delta(u) ^{\mu\nu}+h(u)^{\mu}u^{\nu}+h(u)^{\nu}u^{\mu}+\tau(u)^{\mu\nu},
\label{OD}
\end{equation}
\begin{equation}
	J^{\mu}=n(u)u^{\mu}+n(u)^{\mu},\quad h(u)^{\mu}u_{\mu}=n(u)^{\mu}u_{\mu}=u_{\mu}\tau(u)^{\mu\nu} = 0,
\label{OD_11}	
\end{equation}
where $h(u)^{\mu}, n(u)^{\mu}$ 
is the energy flow and particle ``drift'' relative to the $u$-frame,
$\Delta ^{\mu\nu}(u)$ is the projection tensor\footnote{In order to avoid proliferation of new symbols, the four velocity $u$
in 
(\ref{OD})
should not be confused with the four velocity $u$ that defines the unique rest frame in 
equilibrium states, (see (\ref{tensor energia momento_1})).}
and the spatial symmetric pressure tensor $\tau(u)^{\mu\nu}$ is 
decomposed according to 
\begin{equation}
\tau(u)^{\mu\nu}=\pi(u)\Delta(u) ^{\mu\nu}+\pi(u)^{\mu\nu},\quad \pi(u)^{\mu}_{\mu}=0,
\label{SD}
\end{equation}
where $\pi(u)$ and $\pi(u)^{\mu\nu}$ stand for the bulk and shear stresses
(as measured by the $u$-observer).
Combining these decompositions
with the fitting conditions 
(\ref{FC}) and the relation
$T(u)s(u)={\rho(u) +P(u)}-
T(u) \Theta(u) n(u)$
which implies:
$P(u)b^{\mu}=s(u)u^{\mu}-\rho(u) b^{\mu}+\Theta(u)[J^{\mu}-n(u)^{\mu}]$,
we find
that the fundamental relation
(\ref{Entropy_Off_E}) takes the equivalent form:
\begin{equation}
S^{\mu}=s(\rho(u), n(u))u^{\mu}-\Theta(u) n(u)^{\mu}+\frac {h(u)^{\mu}}{T(u)}-Q^{\mu}(u),\quad
J^{\mu}=n(u)u^{\mu}+n(u)^{\mu}.
\label{PRS}
\end{equation}
The dependance of the right handside of this relation
 upon the arbitrarily chosen  four velocity
$u$
raises delicate questions regarding the interpretation of the theory
 and the implementation of  
 the second law.\\
 For states near equilibrium,
the variables $\tau^{\mu\nu}(u), \rho(u), n(u), $ $P(u), s(u), T(u), \Theta(u)$ 
in 
(\ref{OD}-\ref{PRS}), except the term $Q^{\mu}$,
are all invariant to first order deviation from equilibrium and thus they are considered as 
been ``frame independent''. To define this later term more precisely,  let
$Z(u)$ stands for an arbitrary thermodynamical variable
measured by the observer $u$,
and let 
a change of 
the rest frame which is implemented by a change: 
$u^{\mu} \to {\hat u}^{\mu}$.
Under 
such change, 
the variation
 $\delta Z:=Z(\hat u)-Z(u)$
can be cast in the form:
$$\delta Z=Z(\hat u)-Z(u)=a_{0}+a_{1}O_{1}+a_{2}O_{2}+..$$
where $(a_{0},a_{1},a_{2}....)$ are well defined functions
and $O_{1}, O_{2},...$ 
 stand for terms of first, second order..... deviations from equilibrium\footnote{Notice
 that once a first order change in the frame is implemented via 
$u^{\mu} \to \hat u^{\mu}=u^{\mu}+\hat \epsilon^{\mu}$
then besides the term $\epsilon:=O_{1}$ is also introduced the smallness parameter
$\hat \epsilon$ (see the formulation of the following lemma)
and the terms $(a_{0},a_{1},a_{2}....)$ may contain this new smallness parameters
$\hat \epsilon$.}.
 Variable characterized by the property 
that $a_{1}:=0$ 
are considered to  be
frame-invariant
and the
following two lemmas 
describe the transformations properties of several thermodynamical variables
under a change of the rest frame (Israel \cite{Isr1}).\\

\begin{lemma}
Let ($ u $, $ \hat{u} $) be two arbitrary (future directed) timelike unit vectors within the cone 
of opening angle $\epsilon$
spanned by $ u_E $ and $ u_N $ and let
\begin{equation}
	\hat u^\mu = (1+\delta^2)^{1/2}u^\mu + \delta^\mu, \qquad \delta^2 = \delta^a \delta_a, \qquad \delta^a u_a = 0,
\label{RFT}
\end{equation}
with $\delta^{\mu}$ subject to:
$\delta^{\mu}=\hat \epsilon^{\mu}:=\hat \epsilon \leq O_{1}$
 so that $ \hat u^{\mu}-u^{\mu}=\hat{\epsilon} \leq O_1 $.
Let 
 the primary variables $ T^{\mu \nu} $ and $ J^\mu $ are decomposed relative to the $u$-frame 
according to 
(\ref{OD}) and 
(\ref{OD_11})
while relative to the 
to the 
$ \hat u$-frame these decompositions 
are written in the form:
\begin{equation}
\begin{split}
	T^{\mu \nu} & = {\rho}(\hat u)\hat{u}^\mu \hat{u}^\nu + {P(\hat u)}\Delta(\hat u)^{\mu \nu} + {h(\hat u)}^\mu u^\nu + \hat{h}(\hat u)^\nu \hat{u}^\mu + \hat{\tau}(\hat u)^{\mu \nu},\\	
	J^\mu & = n(\hat{u})\hat{u}^\mu + \hat{n}^\mu(\hat{u}), \quad h(\hat u)^{\mu}\hat u_{\mu}=n(\hat u)^{\mu}\hat u_{\mu}=\hat u_{\mu}\tau(\hat u)^{\mu\nu}=0.
\end{split}
\label{THAT}
\end{equation}	

Then under 
the transformation
under first-order changes of the rest-frame $u^{\mu}$ 
described by 
$$u^{\mu} \to \hat u^{\mu}=u^{\mu}+\hat \epsilon^{\mu}$$
that follows (\ref{RFT}),
the following relations hold:\\
\begin{equation}
	\delta\rho:=\rho(\hat u)-\rho(u)=\hat \epsilon O_{1},
\label{L11}
\end{equation}
\begin{equation}
	\delta h^{\mu}:=h(\hat u)^{\mu}-h(u)^{\mu}=\hat \epsilon^{\mu}=\hat \epsilon,
\label{L22}
\end{equation}
\begin{equation}
	\delta \tau^{\mu\nu}:=\tau(\hat u)^{\mu\nu}-\tau(u)^{\mu\nu}=\hat \epsilon O_{1}.
\label{L33}
\end{equation}

\end{lemma}
\begin{lemma}
Under the same assumptions as in the previous lemma, the combination
\begin{equation}
q^{\mu}(u)=h(u)^{\mu}-\frac {\rho(u) +P(u)}{n(u)} n(u)^{\mu},
\label{IHF}
\end{equation}
is frame invariant i.e.
\begin{equation}
	\delta q^{\mu}:=q(\hat u)^{\mu}-q^{\mu}(u)=a_{2}O_{2}+.....
\end{equation}

\end{lemma}
The proofs of these  two Lemmas are discussed in the Appendix D (see also derivation in Ref\cite{F1}).
For completeness, we also mention that the 
the variation $\delta$ of the
other thermodynamical variables like $(P(u), n(u), s(u), T(u), \Theta(u))$
under a first order 
change in the frame 
$u^{\mu} \to \hat u^{\mu}=u^{\mu}+\hat \epsilon^{\mu}$
described above,
are
all of order
 $\hat \epsilon O_{1}$
with exception the variation $\delta n^{\mu}$ in the particle drift $n^{\mu}(u)$. Like the energy flux $h(u)^{\mu}$, its variation is
of order $\hat \epsilon $.\\

As a first consequence of these Lemmas,
it is worth noticing that the spatial vector $q^{\mu}(u)$ in 
(\ref{IHF})
can be interpreted as defining the frame independent heat flux vector, since
relative to the Eckart frame this $q^{\mu}(u)$  reduces to the energy flux\footnote{For 
the case of fluid mixture, the invariant heat flux  $q^{\mu}(u)$ is replaced by
$q_{A}{}^{\mu}(u)=h(u)^{\mu}-\frac {\rho(u) +P(u)}{n_{A}(u)} n_{A}(u)^{\mu}$ 
which has the 
property that if $u$ is identified by the one of the 
$u_{A}$ satisfying $J_{A}^{\mu}=n_{A}u_{A}^{\mu}$
then 
$q_{A}{}^{\mu}(u_{A})=h(u_{A})^{\mu}$.
Notice however, that for 
a fluid mixture,
the drifts of the particles taken relative to the energy frame are more
convenient variables.} $h^{\mu}(u_{N})$,
while relative to the energy frame specified by $u_{E}$,
this $q(u_{E})^{\mu}$ is proportional to the particle drift 
$\nu^{\mu}$.\\
In addition these lemmas, permit us to eliminate the explicit dependance of the
various thermodynamical variables upon the arbitrarily chosen $u^{\mu}$.
For instance, if
in (\ref{PRS}),
we eliminate the energy flux $h^{\mu}(u)$ in favor of 
the heat flux $q^{\mu}(u)$, 
then ($S^{\mu} , J^{\mu})$ reduce to
\begin{equation}
S^{\mu}=\frac {s}{n} J^{\mu}+\frac {q^{\mu}}{T}-Q^{\mu}, \quad J^{\mu}=nu^{\mu}+n^{\mu},
\label{ENX}
\end{equation}
and if for the moment we ignore  the $Q^{\mu}$ contribution in $S^{\mu}$,
then the resulting $S^{\mu}$ is manifestly frame invariant
under frame change as long as $O_{2}, O_{3}....$
deviations from equilibrium are neglected.
For this reason in the entropy flux $S^{\mu}$ in (\ref{ENX}),
we write
$s, n$.. 
instead of 
$s(u), n(u)$ etc
and this convention will used further ahead.\\
We now
consider the implementation of the 
second law i.e. impose the inequality $\nabla_{\mu}S^{\mu}\geq0$.
To do so, we form the 
covariant divergence of 
(\ref{Entropy_Off_E}),
which in 
view of 
the conservation laws satisfied by the primary variables, yields:
\begin{equation}
\nabla_{\mu}S^{\mu}=\nabla_{\mu}(Pb^{\mu})-J^{\mu}\nabla_{\mu}\Theta-T^{\mu\nu}\nabla_{\mu}b_{\nu}-\nabla_{\mu}Q^{\mu}.
\label{DEN1}
\end{equation}
Since (\ref{FuG}) implies
\begin{equation}
\nabla_{\mu}(Pb^{\mu})=J_{0}^{\mu}\nabla_{\mu}\Theta+T_{0}^{\mu\nu}\nabla_{\mu}b_{\nu},
\label{DEN2}
\end{equation}
therefore
(\ref{DEN1})
takes the form
\begin{equation}
\nabla_{\mu}S^{\mu}=(J_{0}^{\mu}-J^{\mu})\nabla_{\mu} \Theta
+(T_{0}^{\mu\nu}-T^{\mu\nu})\nabla_{\mu}b_{\nu}
-\nabla_{\mu}Q^{\mu}.
\label{DEN}
\end{equation}

By inserting the decompositions of $ T^{\mu \nu} $ and $ J^\mu $ in (\ref{OD}) 
in view of the special forms of $T_{0}{}^{\mu\nu}$ and $J_{0}{}^{\mu}$,
we find
\begin{equation}
\nabla_{\mu}S^{\mu}=W-\nabla_{\mu}Q^{\mu},
\label{FEN1}
\end{equation}
\begin{equation}
W := h(u)^{\mu}\left[\nabla_{\mu}\left(\frac {1}{T(u)}\right)-\frac {a_{\mu}}{T(u)}\right]-n(u)^{\mu}\nabla_{\mu}\Theta(u) 
-\frac {\tau(u)^{{\mu\nu}}}{T(u)}\nabla _{\mu}u_{\nu},
\label{FEN2}
\end{equation}
where in above and here after $a_{\mu}=u^{\nu}\nabla_{\nu}u_{\mu}$ stands for the four acceleration\footnote{Note 
a misprint in equation $(2.28)$ of Ref.\cite{Isr2}.
The term proportional to the four acceleration
in that equation should appear with a negative sign. It is however quoted correctly in 
\cite{Isr1}, see equations $(72, 73)$ in that reference.}
 of the velocity field\footnote{For the case of a fluid mixture, the term
 $n(u)^{\mu}\nabla_{\mu}\Theta(u)$ in $W$ should be replaced by 
 $\sum\limits_{A} n_{A}(u)^{\mu}\nabla_{\mu}\Theta_{A}(u)$.}   
 $u$.\\
It follows from the last two formulas that
in order to impose and work out consequences of the second law, the term $Q^{\mu}$ needs to be specified.
This amounts to the specification of a particular theory
and for completeness,
we first treat
the class of first order theories
and these  theories 
are generated by assuming
that $Q^{\mu}:=0$ in the fundamental relation
(\ref{Entropy_Off_E}) (or equivalently in (\ref{ENX})).

\section{The Eckart and Landau-Lifshitz first order theories}\label{Eckart}

The choice $Q^{\mu}=0$ in (\ref{ENX}), implies that the entropy vector $S^{\mu}$ reduces to 
\begin{equation}
S^{\mu}=\frac {s}{n} J^{\mu}+\frac {q^{\mu}}{T},\quad J^{\mu}=nu^{\mu}+n^{\mu},
\label{FOT}
\end{equation}
which is the
 entropy vector 
due to the convected motion (first term) followed by an irreversible contribution generated by the heat flux $q^{\mu}$
(second term).
Even though this $S^{\mu}$ has a simple form, we should not loose side of the fact that we are dealing with
fluid states admitting dissipation and heat conduction
and in the right hand side of 
 (\ref{FOT})
is hidden 
the choice of the rest frame $u$. Depending upon the choice of the rest frame, 
 these class of theories
include
as particular case the 
Landau-Lifshitz
and Eckart theories.
Below we shall analyze only
these latter theories (a general treatment of
first order
theories can be found in Ref. \refcite{His2}).\\
For the 
Landau-Lifshitz theory (for a detailed  account of this theory, see for instance \cite{LaL}), one chooses 
the rest frame $u$ to be
the unique timelike eigenvector $u_{E}$ of the
energy momentum tensor $T_{\mu\nu}$. For this choice,
the decompositions\footnote{In the formulas (\ref{LL}) through
(\ref{PCE}), for simplicity of the presentation we write $u$ instead of $u_{E}$.}
in (\ref{OD}-\ref{OD_11}) imply:
\begin{equation}
T^{\mu\nu}=\rho u^{\mu}u^{\nu}+(P+\pi)\Delta ^{\mu\nu}+\pi^{\mu\nu}\quad J^{\mu}=nu^{\mu}+n^{\mu},
\label{LL}
\end{equation}
and upon substituting these decompositions in 
(\ref{DEN}) with $Q^{\mu}=0$, we get 
 
\begin{equation}
\begin{split}
\nabla_{\mu}S^{\mu}=&
-n^{\mu}\nabla_{\mu}\Theta-[\pi \Delta^{\mu\nu}+\pi^{\mu\nu}]\nabla_{\mu}b_{\nu} =\\
= &-n^{\mu}\nabla_{\mu}\Theta-\frac {\pi}{T} \Delta^{\mu\nu}\nabla_{\mu}u_{\nu}-\frac {\pi ^{\mu\nu}}{T}\nabla_{\mu}u_{\nu}=\\
=& -n^{\mu}\nabla_{\mu}\Theta-\frac {\pi}{T}\nabla_{\mu} u^{\mu} 
-\frac {\pi ^{\mu\nu}}{T}\langle \nabla_{\mu}u_{\nu}\rangle,
\label{LL1}
\end{split}
\end{equation}
where in above and here after
$\langle\nabla_{\mu}u_{\nu}\rangle$ signifies the symmetric traceless part of $\nabla_{\mu}u_{\nu}$.
Returning to the
definition of the heat flux vector $q^{\mu}$ defined 
in (\ref{IHF}) and taking $h^{\mu}=0$, we write
$\ref{LL1}$ in the form:
\begin{equation}
T\nabla_{\mu}S^{\mu}=\frac{nq^{\mu}}{(\rho +P)}T\nabla_{\mu}\Theta
-\pi \nabla_{\mu}u^{\mu}-\pi^{\mu\nu}\langle \nabla_{\mu}u_{\nu}\rangle.
\label{LLF}
\end{equation}
Implementation of the second law leads
to the phenomenological relations:
\begin{equation}
q^{\mu}=\kappa \frac {n}{\rho+P}T^{2}\Delta^{\mu\nu}\nabla_{\nu}\Theta,\quad \pi=-\frac {1}{3}\zeta_{v}\nabla_{\mu}u^{\mu},\quad \pi_{\mu\nu}=-2\zeta \langle \nabla_{\mu}u_{\nu}\rangle,
\label{PCE}
\end{equation}
where in a standard notation,
$\kappa$ is the coefficient of thermal conductivity and 
$(\zeta_v$, $\zeta)$ are the coefficients of bulk and shear viscosity.\\
The Eckart theory
 (for an introduction to this theory, see for instance Ref. \refcite{Eck}),
  is generated by 
 choosing the rest frame to be 
 the four velocity
$u_{N}$ obeying  $J^{\mu}=nu_{N}^{\mu}$.
 For this choice, we have\footnote{Here again in the formulas (\ref{EF}-\ref{ECF})
that follow, $u$ stands for $u_{N}$.}

\begin{equation}
T^{\mu\nu}=\rho u^{\mu}u^{\nu}+(P+\pi)\Delta ^{\mu\nu}+h^{\mu}u^{\nu}+h^{\nu}u^{\mu}+\pi^{\mu\nu},\quad J^{\mu}=nu^{\mu},
\label{EF}
\end{equation}
and upon substituting 
these decompositions  in 
(\ref{DEN}) with $Q^{\mu}=0$, a short calculus shows that
\begin{equation}
\nabla_{\mu}S^{\mu}=-h^{\mu} \left[\frac {a_{\mu}}{T}-\nabla_{\mu}\left(\frac {1}{T}\right)\right]-
\frac {\pi}{T} \nabla_{\mu}u^{\mu}-\frac {\pi^{\mu\nu}}{T} \langle \nabla_{\mu}u_{\nu}\rangle.
\label{EF11}
\end{equation}
Relative to this frame,
 formula (\ref{IHF}) shows that the heat flux $q^{\mu}$ is identical to energy 
 flux i.e. $h^{\mu}=q^{\mu}$ and by similar analysis as for the case
Landau-Lifshitz case, we deduce the 
 following phenomenological relations:
\begin{equation}
q^{\mu}=-\kappa T \Delta^{\mu\nu}\left(\frac {\nabla_{\nu}T}{T}+a_{\nu}\right),\quad \pi=-\frac {1}{3}\zeta_{v}\nabla_{\mu}u^{\mu},\quad \pi_{\mu\nu}=-2\zeta \langle \nabla_{\mu}u_{\nu}\rangle,
\label{ECF}
\end{equation}
where the coefficients $(\kappa$, $\zeta_{v}$,
 $\zeta)$ have the same meaning as for the 
Landau-Lifshitz theory.\\
Although the theories
of Landau-Lifshitz
and Eckart are simple theories, nevertheless
as we mentioned in the introduction, they  are pathological theories.
 They exhibit infinite propagation 
 of disturbances, instabilities  of the equilibrium states
 and for a critical analysis of the problems plugging
 them see for instance Refs. \refcite{His2,His1}.

\section{Second Order Theories: The Hydrodynamical Approximation}\label{IST}

In this section, we consider
second order theories
i.e. theories generated by choosing a non vanishing term  $Q^{\mu}$
in the fundamental relation (\ref{Entropy_Off_E}), and
the choice that we discuss below
generates the Israel-Stewart transient thermodynamics.\\
Israel and Stewart, motivated from
relativistic kinetic theory of diluted gases, 
proposed that 
the entropy flux vector
$S^{\mu}$ (and thus $Q^{\mu}$), 
should be independent of the gradients of $J^{\mu}$ and $T^{\mu\nu}$ and 
should be quadratic in the 
deviations from the state of local equilibrium.
To describe these deviations in a practical and intuitive manner, they first choose an admissible four velocity $u$
and make use of the expansions
in (\ref{OD}-\ref{SD}) 
that define
variables $\pi(u), \pi(u)^{\mu\nu}, h(u)^{\mu}$ 
and also use the invariant heat flux  $q(u)^{\mu}$ defined in 
(\ref{IHF}).
Moreover 
Israel and Stewart postulate
that 
of the infinite number of the auxiliary variables  
$X_{(i)}^{\mu\nu\lambda...}$ with $i\in (1,2,....)$,
that would appear in the exact equation of state shown in (\ref{EOS1}), none of them will
appear explicitly in the $Q^{\mu}(u)$ term.\\
Denoting here after by 
 $Q^{\mu}(u)$ 
the term $Q^{\mu}$ evaluated relative to the $u$-frame,
Israel-Stewart proposed that
 for a simple fluid $Q^{\mu}(u)$
should have the form:
\begin{equation}
Q^{\mu}(u)=
\frac {1}{2}u^{\mu}[\beta_{0}\pi ^{2}+\beta_{1}q^{\nu}q_{\nu}+\beta_{2}\pi^{\lambda\nu}\pi_{\lambda\nu}]-
\alpha_{0}\pi q^{\mu}-\alpha_{1}\pi^{\mu\nu}q_{\nu}+R^{\mu}(u),
\label{FQ1}
\end{equation}
where $R^{\mu}(u)$ stands for
\begin{equation}
R^{\mu}(u)=\frac {1}{T(\rho+P)}\left[
\frac {1}{2}u^{\mu}h^{\nu}h_{\nu}+\tau^{\mu\nu}h_{\nu}\right],
\label{FR}
\end{equation}
and $\alpha_{j}$, $j \in (0, 1)$,
$\beta_{i}$, 
$i \in (0, 1, 2)$,
are 
undetermined coefficients (notice that 
the
  $R^{\mu}(u)$ term  is free of arbitrary functions\footnote{An $R^{\mu}(u)$ term of the form
 $T(\rho+P)R^{\mu}(u)=[
\frac {1}{2}\gamma_{1}u^{\mu}h^{\nu}h_{\nu}+\gamma_{2}\tau^{\mu\nu}h_{\nu}+
\gamma_{3}\pi h^{\mu}]$
 combined with 
the requirement that
 the entire $Q^{\mu}(u)$ should be
 frame-independent to order $O_{2}$, 
 demands  $\gamma_{1}=\gamma_{2}=1$ and $\gamma_{3}=0$.
 For these values of the $\gamma_{i}$, $i \in (1,2,3)$
  the term $Q^{\mu}$ 
  in 
   (\ref{FQ1})  
becomes frame independent
 and for this reason
 the explicit dependance of 
 $T$, $\pi$, $\pi^{\mu\nu}$,...etc
 upon $u$
 has been omitted. For further discussion regarding the structure of 
 the $Q^{\mu}(u)$ term in  (\ref{FQ1}), the reader is referred
  to original article of Israel \cite{Isr1} and Israel and Stewart Ref.\cite{Isr2} (see
  also Ref.\refcite{F1}).}).\\
For a fluid mixture,
it is more convenient 
to express $Q^{\mu}(u)$,
in terms
of the $n$-particle drifts
$$ \nu^\mu_{\ A}:=\Delta(u_{E})^{\mu}{}_{\nu}J_{A}^{\nu}=n_{A}{}^{\mu}(u_{E}),\quad A \in (1,...,n) ,$$
where $u_{E}{}^{\mu}$ specifies  the
energy frame.
In terms of 
these drifts 
and in the notation of ref. Ref.\refcite{Isr1},
the term $Q^{\mu}(u)$
reads:
\begin{equation}
	Q^\mu(u) = \pi \sum_A a_{\ 0}^A \nu^\mu_{\ A} + \pi^{\mu}_{\ \lambda} \sum a_1^{\ A}\nu_{A}^{\ \lambda} + \frac{1}{2}u^\mu \left(\beta_0 \pi^2 + \sum_{A,B}\beta_1^{AB}\nu^{\alpha}_A \nu_{B \alpha} + \beta_2 \pi_{\alpha \beta}\pi^{\alpha \beta} \right) + R^\mu,
	\label{SOT:Q1}
\end{equation}
where the summation over the indice $(A,B)$ extends over all $n$-particle species,
and 
the coefficients
$(a_{\ 0}^A, a_{\ 1}^A , \beta_{0},  \beta_1^{AB}, \beta_2)$
with $A, B \in (1,2,...n)$
are undetermined functions, while 
$R^{\mu}$ in
(\ref{SOT:Q1})
has the form as in 
(\ref{FR}).\\
As for the case of first order theories, 
the 
phenomenological equations
for the Israel-Stewart transient thermodynamics
follow by imposing the second law
and for generality purposes, we 
analyze
the implications 
of the second law
for the choice of $Q^{\mu}(u)$ shown in (\ref{SOT:Q1}).\\
A direct substitution of this $Q^{\mu}(u)$ in (\ref{FEN1}) yields a long expression.
To minimize algebra, we use the W term shown in (\ref{FEN2}) and the form of $R$ 
in (\ref{FR})
and compute
$W - \nabla_\mu R^\mu$:

\begin{equation}
\begin{split}
	W - \nabla_\mu R^\mu & = h^\mu(u)\left[\nabla_\mu \left(\frac{1}{T}\right) - \frac{a_\mu}{T}\right] - \frac{\tau^{\lambda \mu}\nabla_\mu u_\lambda}{T} - \sum_A {n}^\mu_A(u) \nabla_\mu \Theta_A - \\
	& - \nabla_\mu \left[\frac{1}{2}\frac{u^\mu h^a h_a}{T(\rho + P)}\right] - \nabla_\mu \left[\frac{\tau^{\mu \nu}h_\nu}{T(\rho + P)}\right],
\end{split}	
\end{equation}
where $a_{\mu}$ is the four acceleration of the four velocity $u$.
Moreover eliminating the drifts
$ {n}^\mu_A(u) $ in favor of the the $n$ particle drifts $\nu_{A}^{\mu}$ relative to the energy frame
using:

\begin{equation}
	n^{\mu}_{A}(u)=\nu^{\mu}_{A}+\frac {n_{A}}{(\rho + P)}h^{\mu}(u)
	+O_{2}
\label{AD}	
\end{equation}
and
by appealing to the identity
(\ref{SL_CM_GE}), written in the form	 

\begin{equation}
	(\rho + P)\nabla_\mu T^{-1} + \frac{\nabla_\mu P}{T} - \sum_{A}n_A \nabla_\mu \Theta_A = 0,
\end{equation}
one concludes that:

\begin{equation}
\begin{split}
	W - \nabla_\mu R^\mu & = -\frac{h^\mu(u)a_\mu}{T} - \frac{\tau^{\lambda \mu}}{T}\nabla_\mu\left(u_\lambda + \frac{h_\lambda}{\rho + P}\right) - \sum_{A}\nu_A^\mu \nabla_\mu \Theta_A - \\
	& - \frac{h^\mu}{(\rho + P)}\left[\frac{\nabla_\mu P}{T} + \nabla_\nu \left(\frac{\tau_{\mu}^{\ \nu}}{T}\right) \right] - \frac{1}{2}\nabla_\mu \left[\frac{u^\mu h^a h_a}{T(\rho + P)}\right].
\end{split}	
\label{Int}
\end{equation}
Taking into account that

\begin{equation}
\begin{split}
	&  \Delta^\nu_{\ \lambda}\nabla^\mu T_{\mu \nu} \\
	& = (\rho + P + \pi)u^b \nabla_b u_\lambda + \nabla_\lambda(P + \pi) + h_\lambda \nabla_b u^b + u^b \nabla_b h_\lambda + h^b \nabla_b u_\lambda + \\
	& + \nabla^b \pi_{b \lambda} + u_\lambda u_b\nabla^b(P+\pi) + u_\lambda u_a u^b \nabla_b h^a + u_\lambda u_a \nabla_b \pi^{ab}=0,
\end{split}	
\end{equation}

and by eliminating the $h^{\mu}a_{\mu}$ from (\ref{Int}) 
we obtain finally

\begin{equation}
	W - \nabla_\mu R^\mu  =  - \frac{\tau^{\lambda \mu}}{T}\nabla_\mu\left(u_\lambda + \frac{h_\lambda}{\rho + P}\right) - \sum_{A}\nu_A^\mu \nabla_\mu \Theta_A  + O_3. \\	
\label{WR}
\end{equation}
With this simplification, we now have

\begin{equation}
\begin{split}
	\nabla_\mu S^\mu & =  -\sum_A \nu^\mu_A \nabla_\mu \Theta_A - \frac{\tau^{\lambda \mu}}{T}\nabla_\mu \left(u_\lambda + \frac{h_\lambda}{p + \rho}\right) - \\  
	& - \nabla_\mu \left[ \frac{1}{2}u^\mu \left(\beta_0 \pi^2 + \sum_{AB}\beta_1^{AB} \nu^{\alpha}_A \nu_{B\alpha} + \beta_2 \pi_{\alpha \beta}\pi^{\alpha \beta} \right) + 
  \pi \sum a_0^{\ A}\nu^{\mu}_A + \pi^{\mu}_{\ \lambda}\sum a^A_1 \nu^{\lambda}_{\ A} \right]
  + O_{3}.
  \end{split}	
\label{SOT:S2}
\end{equation}
Following the approach\cite{Isr2}, we neglect the gradients of the coefficients
$(a_{\ 0}^A, a_{\ 1}^A , \beta_{0},  \beta_1^{AB}, \beta_2) $ and
thus the right hand side
 of 
 (\ref{SOT:S2})
can be written  in the form:
\begin{equation}
	\nabla_\mu S^\mu = \sum \nu^\mu f^A_{\ \mu} + \pi f + \pi^{\mu \nu}f_{\mu \nu},
	\label{SOT:S3}
\end{equation}

where $ f^A_{\ \mu}, f $ and $ f_{\mu \nu} $ are functions to be determined.
Carrying out the differentiation
in the right hand side of (\ref{SOT:S2}) and grouping terms we arrive at

\begin{equation}
\begin{split}
	\nabla_\mu S^\mu = & -\sum_A \nu^{\mu}_{\ A}\nabla_\mu \Theta_A - \frac{\tau^{\lambda \mu}}{T}\nabla_\mu \left(u_\lambda + \frac{h_\lambda}{p + \rho}\right) - \\
	& - \frac{1}{2}\left( 2\beta_0 \pi \dot{\pi} + 2\sum_{AB}\beta_1^{AB} \dot{\nu}^\alpha_{\ A}\nu_{B\alpha} + 2\beta_2 \dot{\pi}_{\alpha \beta}\pi^{\alpha \beta}\right) - \\
	& - (\nabla_\mu \pi) \sum_A a_0^{\ A} \nu^\mu_{\ A} - \pi \sum_A a_0^{\ A} (\nabla_\mu \nu^\mu_{\ A}) - (\nabla_\mu \pi^{\mu}_{\ \lambda}) \sum_A a_1^A \nu^\lambda_A - \\
	& - \pi^{\mu}_{\ \lambda}  \sum_A a_1^A (\nabla_\mu \nu^\lambda_A) +O_{3}.
\end{split}
\label{SOT:S4}
\end{equation}
where in above and here after an overdot signifies differentiation along
$u^{\mu}$ i.e. $\dot \pi=u^{\mu}\nabla_{\mu}\pi$, etc.
Using the decomposition $\tau^{\mu\nu}=\pi \Delta^{\mu\nu}+\pi^{\mu\nu}$,
the right hand side of 
(\ref{SOT:S4}) yields:

\begin{equation}
\begin{split}
	\nabla_\mu S^\mu = & -\sum_A \nu^{\mu}_{\ A}\left[\nabla_\mu \Theta_A + \sum_B \beta_1^{AB} \dot{\nu}^\alpha_{B\mu} + a^A_{\ 0}\nabla_\mu \pi + a^A_{\ 1}\nabla_\lambda \pi^{A}_{\ \mu} \right] - \\
	& - \pi\left[\beta_0 \dot{\pi} + \frac{\Delta^{\lambda \mu}}{T}\nabla_\mu \left(u_\lambda + \frac{h_\lambda}{p + \rho}\right) + \sum a^A_0 \nabla_\mu \nu^{\mu}_{\ A} \right] - \\
	& - \pi^{\lambda \mu}\left[\frac{1}{T}\nabla_\mu \left(u_\lambda + \frac{h_\lambda}{p + \rho}\right) + \beta_2\dot{\pi}_{\lambda \mu} + \sum_A a^A_{\ 1}\nabla_\mu \nu_{A\lambda}  \right].
\end{split}
\label{SOT:S5}
\end{equation}

Since $ h_\lambda u^\lambda = 0 $ and $ u^\mu u_\mu = -1 $,
the following identity holds: 
\begin{equation}
	\Delta^{\lambda \mu} \nabla_\mu \left(u_\lambda + \frac{h_\lambda}{p + \rho}\right) = (g^{\lambda \mu} + u^\lambda u^\mu)\nabla_\mu \left(u_\lambda + \frac{h_\lambda}{p + \rho}\right) = \nabla_\mu u^\mu+
	\frac  {\nabla_\mu h^\mu}{p+\rho},
	\label{SOT:1}
\end{equation}
and this  identity transforms
(\ref{SOT:S5}) into the form:

\begin{equation}
\begin{split}
	\nabla_\mu S^\mu = & -\sum_A \nu^{\mu}_{\ A}\left[\nabla_\mu \Theta_A + \sum_B \beta_1^{AB} \dot{\nu}^\alpha_{B\mu} + a^A_{\ 0}\nabla_\mu \pi + a^A_{\ 1}\nabla_\lambda \pi^{A}_{\ \mu} \right] - \\
	& - \pi\left[\beta_0 \dot{\pi} + 
	\frac{\nabla_{\mu}u^{\mu}}{T} +\frac{\nabla_{\mu}h^{\mu}}{T(p+\rho)}	
	+ \sum a^A_0 \nabla_\mu \nu^{\mu}_{\ A} \right] - \\
	& - \pi^{\lambda \mu}\left[      
	\frac{\nabla_{\mu}u^{\mu}}{T} +\frac{\nabla_{\mu}h^{\mu}}{T(p+\rho)}	
	+ \beta_2\dot{\pi}_{\lambda \mu} + \sum_A a^A_{\ 1}\nabla_\mu \nu_{A\lambda}  \right].
\end{split}
\label{SOT:S6}
\end{equation}
By inspection of the right hand side, we can write down the phenomenological relations between
$ (\pi , \pi^{\mu \nu}, \nu^{\mu}_{A}) , A \in (1,2,....n)$
that enforce the second law.
For this, it is sufficient to assume 
that $ \pi $, $ \pi^{\mu \nu} $ and $ \nu^{\ \mu}_A $ depend \textit{linearly} on the ``strains" i.e. 
$ \nabla_\mu u^\mu $, $ \langle \nabla_\mu u_\nu \rangle $, $ \nabla_\mu \Theta_A $
i.e. to set

\begin{equation}
	 \nu^{\mu}_{\ A}:= -\Delta^{\mu \nu}T^{2}\sum_B k_{AB}\left(\nabla_\nu \Theta_B + \sum_C \beta_1^{BC}\dot{\nu}_{C\nu} + a_0^{\ B}\nabla_\nu \pi + a_1^{\ B}\nabla_p \pi^{p}_{\ \nu} \right)
\label{SOT:NU1}	 
\end{equation}

\begin{equation}
	\pi := -\frac{1}{3}J_\nu \left(\nabla_\alpha u^\alpha + \frac{\nabla_\alpha h^\alpha}{p + \rho} + T\beta_0 \dot{\pi} + T\sum_A a_0^{\ A}\nabla_\mu \nu_A^{\ \mu}\right)
\label{SOT:PI}	
\end{equation}

\begin{equation}
	\pi_{\mu \nu} := -2J\langle \nabla_\nu u_\mu + \frac{\nabla_\nu h_\mu}{p + \rho} + T\beta_2 \dot{\pi}_{\mu \nu} + T\sum_A a_1^{\ A}\nabla_\nu \nu_{A\mu}\rangle,
\label{SOT:PI1}	
\end{equation}
where 
$k_{AB}$ 
$A,B, \in (1,2,...,n)$,
is an $(n\times n)$ a semi-positive matrix with real entries and 
the angular bracket in the last equation signifies:
\begin{equation}
\langle A_{\alpha \beta}\rangle =  \left(\Delta^\mu_{(\alpha}\Delta^\nu_{\ \beta)} - \frac{1}{3}\Delta_{\alpha \beta} \Delta^{\mu \nu}\right)A_{\mu \nu}.
\label{AB}
\end{equation}

Equations
(\ref{SOT:NU1}, 
\ref{SOT:PI},	
\ref{SOT:PI1}) are
the phenomenological relations
that follow by imposing the second law
$\nabla_{\mu}S^{\mu}\geq 0$
for the choice 
$Q^{\mu}$ shown in (\ref{SOT:Q1}).
Originally, they have been derived by Israel in
Ref.\refcite{Isr1}, although in this original article Israel analysis includes the case where
there inter-particle reactions
in a fluid mixture.\\
As they are derived here, they are 
valid for any rest frame specified by a four velocity $u$ 
as long as this $u$ lies	
between 
$(0,
min(\epsilon_{1},\epsilon_{2},....\epsilon_{n}))$
where
$\epsilon_{i}$ 
 is the  opening angle defined by $u_{E}$
and the four velocity $u^{\mu}{}_{I}$
parallel to the corresponding $I-$type particle $J^{\mu}_{I}$ (see comments in the Appendix A).
It might be worth mentioning that even though in the derived equaciones
(\ref{SOT:NU1}-\ref{SOT:PI1}) appear explicitly the $n$-particle drifts
$ \nu^\mu_{\ A}:=n_{A}{}^{\mu}(u_{E}), A \in (1,...,n)$
eventually these drifts can be eliminated by appealing to
(\ref{AD}) which shows that 
(\ref{SOT:NU1}-\ref{SOT:PI1}) 
hold relative to an ''arbitrary'' fram $u$.\\
They
simplify slightly 
whenever they are expressed 
relative to a particular frame like the 
energy frame or the particle frame\footnote{For
 a fluid mixture the particle frame is not well defined, or more precisely
there exist $n$ such frames specified by the corresponding $u_{A}$. Relative to each of these frames
the corresponding species of type $A$ is at rest even though the other species exhibit a drift relative to this 
particular frame. Often, for a fluid mixture, 
the material four velocity $u_{M}$ is introduce via
$u_{M}{}^{\mu}=\sum\limits_{A=1}^{n}\frac {w_{A} J_{A}^{\mu}}{w_{A}n_{A}}$ where $w_{A}$ are arbitrary weights
and $n_{A}=-u^{A}_{\mu}J_{A}^{\mu}$.
Relative to this ''material frame'' a heat flux $q^{\mu}$ is defined via
$q^{\mu}=h^{\mu}(u_{M})$.
We shall not write the phenomenological equations relative to this material frame.
They can be derived by a straightforward 
generalization of the approach of this section.}.
For a simple fluid
and relative to the energy frame,
the phenomenological equations are obtained from 
(\ref{SOT:NU1}, 
\ref{SOT:PI},	
\ref{SOT:PI1}) 
by setting $h^{\mu}=0$ everywhere,
removing the summation symbols and the indices $(A, B)$.
They reduce to the form:
\begin{equation}
	 \nu^{\mu}:= -k\Delta^{\mu \nu}T^{2} \left(\nabla_\nu \Theta+ \beta_{1}\dot{\nu}_{\nu} + a_0\nabla_\nu \pi + a_1\nabla_p \pi^{p}_{\ \nu} \right),
\label{SOT:NU2}	 
\end{equation}

\begin{equation}
	\pi := -\frac{1}{3}J_\nu \left(\nabla_\alpha u^\alpha + T\beta_0 \dot{\pi} + T  a_0\nabla_\mu \nu^{\ \mu}\right),
\label{SOT:PII1}	
\end{equation}

\begin{equation}
	\pi_{\mu \nu} := -2J\langle \nabla_\nu u_\mu + T\beta_{2} \dot{\pi}_{\mu \nu} + T a{_1}\nabla_\nu \nu_{\mu}\rangle.
\label{SOT:PI2}	
\end{equation}
where in these equations $u^{\mu}$ stands for $u_{E}{}^{\mu}$,
$\Delta^{\mu \nu}=\Delta^{\mu \nu}(u_{E})$ etc.\\
For a simple fluid, it is often convenient to 
eliminate the particle drift $\nu^{\mu}$
from
(\ref{SOT:NU2}, 
\ref{SOT:PII1},	
\ref{SOT:PI2}) 
in favor of the frame invariant heat flux vector $q^{\mu}$
defined in (\ref{IHF}). Evaluating 
$q^{\mu}$  in the energy frame, 
yields
$$
\nu^{\mu}=-\frac {n(u_{E})}{\rho(u_{E})+P(u_{E})}q^{\mu}+O_{2}.
$$
Returning to the form of $Q^{\mu}$ in
(\ref{SOT:Q1}), removing the summation symbols,
eliminating $\nu^{\mu}$ in favor of $q^{\mu}$
and
setting $h^{\mu}=0$,
we obtain a ``new'' 
 $Q^{\mu}$ term
for a simple fluid.
After some algebra
the phenomenological equations
resulting from the modified expression $\nabla_{\mu}S^{\mu}\geq0$
in (\ref{SOT:S2})
are:
\begin{equation}
q^\alpha = \kappa T \Delta_E^{\alpha \beta}(\eta^{-1}T \nabla_\beta \Theta -\beta_1 \dot{q}_\beta + \alpha_0 \nabla_\beta \pi + \alpha_1 \nabla_\nu \pi_\beta^\nu),\quad \eta=\frac {\rho+P}{n},
\label{X1}
\end{equation}

\begin{equation}
\pi = -\frac{1}{3}\zeta_\nu(\nabla_\alpha u^\alpha + \beta_0 \dot{\pi} - \alpha_0 \nabla_\alpha q^\alpha),
\label{X2}
\end{equation}

\begin{equation}
\pi_{\alpha \beta} = -2\zeta\langle \nabla_\beta u_\alpha + \beta_2 \dot{\pi}_{\alpha \beta} - \alpha_1 \nabla_\beta q_\alpha \rangle,
\label{X3}
\end{equation}
and these are the equations given in
Israel \cite{Isr1} and Israel and Stewart \cite{Isr2}. It is understood that in
(\ref{X1}-\ref{X3}) 
$u$ stands for $u_{E}$.\\
The structure of the phenomenological equations for a simple fluid relative to the 
Eckart frame, 
can be obtained using 
the term $Q^{\mu}$ 
shown  in
 (\ref{FQ1}, \ref{FR})
 and recalling  that
 for this frame 
 $W - \nabla_\mu R^\mu$ reduces to: 
 \begin{equation}
	W - \nabla_\mu R^\mu  = q^\mu(u)\left[\nabla_\mu \left(\frac{1}{T}\right) - \frac{a_\mu}{T}\right]  - \frac{\tau^{\lambda \mu}}{T}\nabla_\mu\left(u_\lambda + \frac{q_\lambda}{\rho + P}\right)  + O_3. \\	
\label{WREC}
\end{equation}
 The resulting  phenomenological equations 
have the form

\begin{equation}
\pi = -\frac{1}{3}\zeta_\nu(\nabla_\alpha u^\alpha + \beta_0 \dot{\pi} - \bar{\alpha}_0 \nabla_\alpha q^\alpha),
\label{Y1}
\end{equation}

\begin{equation}
q^\alpha = -\kappa T \Delta_N^{\alpha \beta}(T^{-1} \nabla_\beta T +\dot{u}_\beta +\bar{\beta}_1 \dot{q}_\beta - \bar{\alpha}_0 \nabla_\beta \pi + \bar{\alpha}_1 \nabla_\nu \pi_\beta^\nu),
\label{Y2}
\end{equation}

\begin{equation}
\pi_{\alpha \beta} = -2\zeta\langle u_{\alpha \beta} + \beta_2 \dot{\pi}_{\alpha \beta} - \bar{\alpha}_1 \nabla_\beta q_\alpha \rangle,
\label{Y3}
\end{equation}
where in these equations $u$ stands for $u_{N}$.\\
The coefficients $\alpha_{0}, \alpha_{1}, \beta_{1}$ 
in 
(\ref{X1},\ref{X2},\ref{X3})
and their barred versions 
in  (\ref{Y1},\ref{Y2},\ref{Y3})
are related via
\begin{equation}
\bar \alpha_{i}-\alpha_{i}=\beta_{1}-\bar \beta_{1}=[(\rho +P)T)^{-1}],
\label{Z1}
\end{equation}
and these relations are important in proving that
the phenomenological equations  (\ref{X1}-\ref{X3})
and 
 (\ref{Y1}-\ref{Y3})
 are 
 equivalent to first order deviations from equilibrium.
 In order to establish that property,
 we should  keep in mind
 that the four velocities 
 $u_{E}$ and $u_{N}$ are related via 
 \begin{equation}
u^{\mu}_{E}=u^{\mu}_{N}+(\rho +P)^{-1}q^{\mu} +O_{2},
\label{Z2}
\end{equation} 
as well as the following relation:
 \begin{equation}
\Delta^{\mu\nu} \eta^{-1} \nabla_{\beta}\Theta=\Delta^{\mu\nu}[\nabla_{\nu}({\frac {1}{T}})-\frac {1}{(\rho +P)T}
(\dot q_{\nu}+\nabla_{\nu}\pi +\nabla_{\lambda}\pi^{\lambda}_{\nu})]-
{\frac {1}{T}}\dot u_{N}^{\mu},
\label{Z3}
\end{equation} 
discussed in Ref.\refcite{Isr1}.

The sets of 
the phenomenological equations\footnote{As
derived above, these equations are equivalent
to an $O_{1}$ accuracy in the change of frame.
However the stability properties
of these equations need to be accessed.
In particularly the gauge freedom associated with different choices of a rest frame
calls for a thorough investigation of 
their stability properties. It is conceivable that different choices of the four velocity $u$
may lead to different stability properties. For an illuminating discussion of that issue consult
Refs.\refcite{RP1,RP2}.} derived above
yield the evolution equations for
the heat flux, bulk and shear stresses
and these 
equations
combined to the equations arising from the conservation laws
yield a closed system of equations.
Their solutions describe the behavior of near equilibrium  states
whose evolution is compatible with the second law. Although
the implementation of the
second law is welcomed, by itself,  it is not
a sufficient reason for the physical acceptance of the theory.
As it is clear from the 
so far analysis, of central importance 
is the issue
whether the set of dynamical equations
constitutes a hyperbolic system of equations
or the stronger restriction whether they constitute a causal set of dynamical equations
at least for states
near equilibrium.
For the moment 
within the context of tarnsient thermodynamics it is not known
whether the resulting equations satisfy that constraint . However there are strong evidences that the theory rests on
strong foundations.\\
Israel and Stewart in Ref.\refcite{Isr2} 
went outside the realm of phenomenology
and derived the equations of transient thermodynamics
from microphysical considerations.
Starting from the relativistic 
Boltzmann equation for a dilute gas
and within Grads approximation \cite{Gra}  
they were able 
in Ref. \refcite{Isr2} to derive
the equations of transient thermodynamics
by
evaluating the first three moments of the Boltzmann equation\footnote{Specifically in \cite{Isr2} they have shown that the first three moments of the distribution function $ f(x,p) $ i.e.  $ \hat J^\mu(x) = \int f(x,p) p^\mu d\Omega $, $\hat T^{\mu \nu}(x) = \int f(x,p) p^\mu p^\nu d\Omega$, $ \hat A^{\mu \nu \lambda}(x) = \int f(x,p)p^\mu p^\nu p^\lambda d\Omega $, as a consequence of the Boltzmann equation satisfy: $ \nabla_\mu \hat J^\mu = \nabla_\mu \hat T^{\mu \nu} = 0$, $ \nabla_\mu\hat  A^{\mu \nu \lambda} = \hat I^{\nu \lambda} $ with $ \hat I^{\nu \lambda} = \int C[f]p^\nu p^\lambda d\Omega $ is the second moment of the collision term $ C[f]  $ for the Boltzmann equation.\label{footnote_KT} }.
Moreover they were able to evaluate the five coefficients $ (\alpha_i, \beta_i) $ in (\ref{Y1}-\ref{Y3}) and they have shown that these coefficients are purely thermodynamical functions i.e. independent of the cross section that enters in the collision 
term in Boltzmann's equation. 
Subsequently they evaluated the 
wave fronts speeds about the equilibrium state and they found that all of them are finite and causal.
These conclusions shows
that transient thermodynamics at least for the coefficients 
$ (\alpha_i, \beta_i) $ predicted by the relativistic 
Boltzmann equation avoids infinite propagation of disturbances.\\
The same conclusion has been also reached by 
Hiscock and Lindblom in \cite{His1}
who have shown 
causality for transient thermodynamics holds for a wider range of circumstances.
Moreover 
Hiscock and Lindblom in \cite{His1}
addressed 
the stability property of these equilibrium states by examining the behavior
of small perturbations about equilibrium states.
They demonstrated a striking connection between 
 causality and stability: at least within the framework of transient thermodynamics,
 these notions are equivalent in the following sense. If
  the theory possesses stable equilibrium states, then linear perturbations propagate causally
 and conversely i.e. if the characteristic velocities are subluminal
 (and the perturbation equations form a hyperbolic system) then the stability 
 of equilibrium states is guaranteed\footnote{Since  the theory admits a gauge freedom
regarding the choice of the rest frame, issues related to the stability properties
under change of the frame are needed to be clarified.
For a discussion regarding that point and further subtleties see Refs:
 \refcite{RP1,RP2} and further references therein.}.
 The stability of equilibrium states implies that all solutions 
 of the perturbations equations about the equilibrium state with
regular initial data
are bounded functions of time and this property has been established by constructing an appropriate energy functional
quadratic in the perturbations.
 These results suggest
that transient thermodynamics
rectifies
the predictions of
Eckart and Landau-Lifshitz
first order theories.\\
Even though all so far evidences regarding transient thermodynamics are encouraging, signaling
a robust theory, nevertheless more work is needed. As we have already mentioned,
for instance,
it would of interest to investigate
whether
the dynamical equations constitute a symmetric-hyperbolic system of equations,
their causality etc.\\
We finish the treatment of transient thermodynamics by mentioning that various aspects of the theory can be found in 
Refs: \refcite{His3}, \refcite{His4},
\refcite{Bel},
\refcite{RMa1},
\refcite{F1},
\refcite{CMon},
\refcite{JS}.

\section{Relativistic (REIT)}\label{R_REIT}

In this section, we introduce 
an alternative theory of non equilibrium thermodynamics describing states of relativistic fluids and this 
theory developed by Liu, M\"uller and Ruggeri in
Ref. \refcite{Mul6} (and is discussed in more detail in chapters $(6)$ in Ref. \refcite{Mul4}).
This theory can be considered as a relativistic
extension of the classical 
rational extended thermodynamics
(REIT) 
that we briefly introduced at the end of section \ref{EIT}
and for this reason often in the sequel we refer to it as relativistic (REIT).\\

Motivated by the desire to 
place irreversible thermodynamics
of fluids on a solid mathematical foundation,
the authors in Ref. \refcite{Mul6} 
introduced a new theory that deals 
with states of a relativistic  simple fluid.
The distinguishing feature of this theory is
that the dynamical fluid equations, by appropriate restrictions, constitute a symmetric hyperbolic, causal system
and thus finite propagation of disturbances is 
automatically built in.\\
In this theory, the $10$ components of the energy momentum tensor $ T^{\mu \nu} = T^{\nu \mu} $ and the
$4$ components of the particle  current $ J^\mu $ are consider to be the dynamical variables\footnote{In the terminology
of the theory of constitutive relations these variables are refereed as the basic variables.}
describing the fluid.
They 
satisfy the 
equations\footnote{ The structure of eqs. (\ref{RET_1},\ref{RET_2}),
 the trace free property of $I^{\mu\nu}$ and the relation
 $A^{\mu \nu}{}_{\nu}=m^{2}c^{2}J^{\mu}$ 
 are motivated by relativistic kinetic theory of a simple gas, see the comment \ref{footnote_KT}
 in the previous page.}:

\begin{equation}
	\nabla_\mu T^{\mu \nu} = \nabla_\mu J^\mu = 0,
\label{RET_1}	
\end{equation} 
\begin{equation}
	\nabla_\mu A^{\mu \nu \lambda} = I^{\nu \lambda},
\label{RET_2}		
\end{equation}
 where $ A^{\mu \nu \lambda} $ is a completely symmetric tensor field 
while $I^{\mu\nu}$ is symmetric 
and traceless $I^{\mu}{}_{\mu}=0$
 and 
 $A^{\mu \nu}{}_{\nu}=m^{2}c^{2}J^{\mu}$
 so that $\nabla_{\mu}A^{\mu\nu}{}_{\nu}=0$.
Because of these symmetries, the system in (\ref{RET_1}, \ref{RET_2}) 
involves $14$ equations, which is equal to the number of the unknown components in $J^{\mu}$ and $T^{\mu\nu}$.
Thus 
(\ref{RET_1} -\ref{RET_2})
could serve as  the dynamical equations for the $14$ unknown fields, provided
that
\begin{equation}
	A^{\mu \nu \lambda} = A^{\mu \nu \lambda}(J^\mu, T^{\mu \nu}),\quad
	I^{\mu \nu} = I^{\mu \nu}(J^\mu, T^{\mu \nu}),
\label{3}	
\end{equation}
i.e. the fields $A^{\mu\nu\lambda}$ and $I^{\mu\nu}$ 
 are viewed as a set of constitutive relations.\\
 In addition to the balance laws
 (\ref{RET_1}, \ref{RET_2}),  
 the authors complete the system by including
 an 
 entropy vector $ S^\mu $
 which is also 
a constitutive function\footnote{It is of interest to point out here the different philosophy 
that underlies the present theory and the transient theory developed in the previous sections. Whereas for the transient theory the entropy vector 
$S^{\mu}$
(see eq. (\ref{EOS1})), depends upon the primary variables and perhaps an infinite set
of 
auxiliary variables  
$X_{(i)}^{\mu\nu\lambda...}$ with $i\in (1,2,....)$,
for the present theory $S^{\mu}$ depends only upon the primary variables
i.e.
$S^\mu = S^\mu(J^\mu, T^{\mu \nu})$.}
 i.e.
\begin{equation}
	S^\mu = S^\mu(J^\mu, T^{\mu \nu}),
\label{RET_5}	
\end{equation}
and demand
that  for any solution\footnote{
A solution 
 $(J^{\mu}, T^{\mu\nu})$
of the field equations
(\ref{RET_1},\ref{RET_2})
 is refereed in 
 \cite{Mul6,Mul4} as a thermodynamical property.} 
 $(J^{\mu}, T^{\mu\nu})$
of  (\ref{RET_1},\ref{RET_2}) the
entropy four vector $S^{\mu}$ obeys:
\begin{equation}
	\nabla_\mu S^\mu(J^{\mu}, T^{\mu\nu}) \geq 0.
\label{RET_entropy1}	
\end{equation}
  The authors 
place severe constraints upon the form
of the
constitutive functions
$A^{\mu \nu \lambda}(J^\mu, T^{\mu \nu})$,  $I^{\mu \nu} = I^{\mu \nu}(J^\mu, T^{\mu \nu})$
and $S^{\mu}=S^{\mu}(J^\mu, T^{\mu \nu})$ by
appealing to the following three principles:\\

\begin{itemize}
	\item[a)] Entropy Principle\\
	
	\item[b)] Principle of Relativity\\
	
	\item[c)] Hyperbolicity.\\
\end{itemize}
Since the applications of these principles as means to pin down the structure of 
the constitutive
functions
involve long computations,
in this section, we
only highlight the essential steps
and frequently refer the reader to the original article of Ref \refcite{Mul6} and to chapter\footnote{Actually, chapter $ 6 $ in Ref. \refcite{Mul4} provides a detail and systematic treatment of the theory
and the interested reader is recommended to go through this chapter.} $(6)$ in the monograph of Ref.\refcite{Mul4}
for further details.\\
The authors begin by breaking the frame invariance of the theory
and employ
the Eckart\footnote{In principle
one could however
decompose 
 $J^{\mu}$ and $T^{\mu\nu}$
relative
to any arbitrary four velocity $\hat u^{\mu}$
and carry out the analysis of this section
relative to this new frame 
$\hat u^{\mu}$.
This freedom in the choice of the rest frame
 raise the question whether
the resulting theories are physically equivalent.
As long as one restricts attention to states near equilibrium ( a term that needs to be defined precisely within
the theory), one expects
identical situation 
as for the case of the transient 
thermodynamics, although these expectations ought to be checked in details. 
}
frame to perform the calculations.
Accordingly and relative to this frame, they
decompose
the energy momentum tensor $T^{\mu\nu}$ and
the particle current $J^{\mu}$ 
according to:
\begin{equation}
    T^{\mu \nu} = \frac{1}{c^2}\rho u^\mu u^\nu+ (P + \pi)\Delta^{\mu\nu}     
   +\frac{1}{c^2}(u^\mu q^\nu + u^\nu q^\mu)+  
     \pi^{\mu \nu}, \quad J^\mu = nu^\mu, 
\label{LMR}
\end{equation}
where as before
the quantities\footnote{
Since all the subsequent computations are done relative to the Eckart frame
and thus there is no danger of confusion, in this section, we drop the 
dependance of $(\rho, P, \pi, q^{\mu},\pi^{\mu\nu})$ upon $u$ (this dependance introduced in 
the decompositions (\ref{OD}-\ref{OD_11}) as a reminder that 
they are frame dependent).}
$(\rho, P, \pi, q^{\mu},\pi^{\mu\nu}, n)$ have the same
 interpretation as those 
 appearing in the expansions shown in (\ref{OD}-\ref{OD_11}).
Notice that  due to the signature employed
in Ref.\refcite{Mul6}, the components of the projection tensor
$\Delta^{\mu \nu} $
are given by 
$\Delta^{\mu \nu} = \frac{1}{c^2}u^{\mu} u^{\nu}- g^{\mu \nu} $,
while the four velocity $u$ is normalized according to $g(u, u)=c^{2}$.
Moreover for the present section,
the energy flow
$ h^\mu $ in (\ref{OD}) is denoted by $ q^\mu $ and is refereed
as heat flux
while relative to the Eckart frame,
the particle drift $ n^\mu  $
defined in 
(\ref{OD_11})
is vanishing identically.\\

For latter use, notice that
$ \pi^{\alpha \beta} = \left(\Delta^\alpha_{\ \mu}\Delta^\beta_{\ \nu} - \frac{1}{3}\Delta^{\alpha \beta}\Delta_{\mu \nu}\right)T^{\mu \nu}$
while the thermodynamical pressure $P$, bulk pressure $\pi$, energy density $\rho$, heat flux vector $q^{\mu}$ and particle density $n$ are
determined by
\begin{equation}
	P +\pi= \frac{1}{3}\Delta_{\mu \nu}T^{\mu \nu}, \quad
\rho = \frac{1}{c^2}u_\mu u_\nu T^{\mu \nu}, \quad c^{2}n^{2}=J^{\mu}J_{\mu}, \quad
   q^\alpha = -\Delta^\alpha_{\ \mu}u_\nu T^{\mu \nu}.
\label{Split}
\end{equation}
In the 
treatment that follows,
the $14$ components 
of $(J^{\mu}, T^{\mu\nu})$
are traded\footnote{As it will become clear further ahead, this trade off 
make 
the physical content 
of the theory more accessible
to intuition,
although as we have already mentioned 
the freedom in the choice of the frame needs to be treated with care.}
 in favor of the $14$ fields $(\rho, P+\pi, q^{\mu},\pi^{\mu\nu}, n, u^{\mu})$ defined in (\ref{LMR}, \ref{Split}).\\

The principle of relativity, when applied to the constitutive relations $A^{\mu \nu \lambda}(J^\mu, T^{\mu \nu})$,  $I^{\mu \nu} = I^{\mu \nu}(J^\mu, T^{\mu \nu})$
and $S^{\mu}=S^{\mu}(J^\mu, T^{\mu \nu})$, dictates that these objects  should remains form invariant under arbitrary coordinate transformations
which  means that
they behave as isotropic tensor-functions with respect to such transformations.
Although one could study the structure of such tensor fields, the authors avoid that root.
Instead, by
employing the components of $(u^{\mu}, q^{\mu},g^{\mu\nu},\pi^{\mu\nu})$ and scalars, 
they 
constructed the most general form of the tensors $I^{\alpha \beta} $ and $A^{\alpha \beta \mu}$ that are linear in $(\pi$, $\pi^{\alpha \beta}, q^{\alpha})$. They have the form:  
  \begin{equation}
    I^{\alpha \beta} = B^\pi_{\ 1}\pi g^{\alpha \beta} - \frac{4}{c^2}B^\pi_{\ 1}\pi u^\alpha u^\beta + B_3 \pi^{\alpha \beta} +\frac{1}{c^2} B_4(q^\alpha u^\beta + q^\beta u^\alpha),
\label{constitutive 1} 
\end{equation}
\begin{equation}
\begin{split}
    A^{\alpha \beta \mu} & = (C^0_{\ 1} + C^\pi_{\ 1}\pi)u^\alpha u^\beta u^\mu + \frac{c^2}{6}(nm^2 - C^0_{\ 1}- C^\pi_{\ 1}\pi)(g^{\alpha \beta}u^\mu + g^{\beta \mu}u^\alpha + g^{\mu \alpha}u^\beta)\\
    & + C_3(g^{\alpha \beta}q^\mu + g^{\beta \mu}q^\alpha + g^{\mu \alpha}q^\beta) - \frac{6}{c^2}C_3(u^\alpha u^\beta q^\alpha + u^\beta u^\mu q^\alpha + u^\mu u^\alpha q^\beta)  \\
    & + C_5(\pi^{\alpha \beta}u^\mu + \pi^{\beta \mu}u^\alpha + \pi^{\mu \alpha}u^\beta),
\end{split}    
\label{constitutive 2} 
\end{equation}  
where $(B^\pi_{\ 1}, B_3, B_4)$ 
respectively
$(C^0_{\ 1}, C^\pi_{\ 1}, C_3, C_5)$
are functions of $(n,\rho)$ to be determined and $ m $ is  a mass scale. A similar construction shows that the
entropy current 
$S^{\alpha}$ quadratic 
in $ (\pi, \pi^{\alpha \beta}, q^\alpha) $ reads:
  
\begin{equation}
\begin{split}
    S^\alpha & = (ns + A^\pi_{\ 1}\pi + A^{\pi^2}_{\ 1}\pi^2 + A^q_{\ 1}q^\beta q_\beta + A^t_{\ 1}\pi^{\beta \mu}\pi_{\beta \mu})u^\alpha \\
    & + (A^0_{\ 2}+ A^\pi_{\ 2}\pi)q^\alpha + A^0_{\ 3}\pi^{\alpha \beta}q_\beta,
\end{split}   
\label{constitutive 3} 
\end{equation}
where again $ (A^\pi_{\ 1}, A^{\pi^2}_{\ 1}, A^q_{\ 1}, A^t_{\ 1}, A^0_{\ 2}, A^\pi_{\ 2}, A^0_{\ 3}) $ are 
functions to be determined while $ (n, s) $ are the particle number density and entropy per particle as
measured relative to the Eckart frame.
The determination of the $14$ coefficients
$(A,...B,...C,...)$  that appear in (\ref{constitutive 1}-\ref{constitutive 3}) is a major undertaking and 
in this task, the entropy principle plays an important role. 

A clever  
device 
to take a maximal advantage of the entropy principle 
was
put forward by Liu in Ref.
\refcite{Liu}. In the Appendix B, we provide a brief introduction to 
Liu's procedure for the implementation of the entropy principle
 and also discuss a variance of this procedure developed by Ruggeri
and coworkers see \cite{Rug4}. Both of these procedures use Lagrange
multipliers and these multipliers arise via the following considerations.\\
With reference to the system (\ref{RET_1},\ref{RET_2}) coupled to the entropy inequality (\ref{RET_entropy1}),	
let the 
fields
$(\zeta, \zeta_{a}, \zeta_{ab})$ with $\zeta_{ab}=\zeta_{ba}$ and $g^{ab}\zeta_{ab}=0$
acting as Lagrange multipliers
in the following sense:
the inequality
\begin{equation}
\nabla_\alpha S^\alpha + \zeta \nabla_\alpha J^\alpha + \zeta_\beta \nabla_\alpha T^{\beta \alpha} + \zeta_{\beta \mu}(\nabla_\alpha A^{\alpha\beta \mu } - I^{\beta \mu}) \geq 0,
\label{LM1}
\end{equation}
 is required
to hold for 
arbitrary smooth configuration $(T^{\mu\nu}, J^{\mu})$
(and not only
for solutions of
(\ref{RET_1},\ref{RET_2})). 
The existence of the field of the Lagrange multipliers 
$(\zeta, \zeta_{a}, \zeta_{ab})$
that satisfy this inequality
has been addressed 
 in Ref.
\refcite{Liu} (see also Appendix B).
Introducing a new field $S'^{\alpha}$ via
\begin{equation}
S'^{\alpha} = S^\alpha + \zeta J^\alpha + \zeta_\beta T^{\beta \alpha} + \zeta_{\beta \mu}A^{\alpha \beta \mu },
\label{BasR}
\end{equation}
then 
(\ref{LM1}) can be rewritten in the form
\begin{equation}
\nabla_\alpha S'^{\alpha} - J^\alpha \nabla_\alpha \zeta - T^{\alpha \beta}\nabla_\alpha \zeta_\beta - A^{\beta \mu \alpha}\nabla_\alpha \zeta_{\beta \mu} - \zeta_{\beta \mu}I^{\beta \mu} \geq 0.
\label{YY}
\end{equation}
Since $ (J^\mu, T^{\mu \nu}) $ and $ (\zeta, \zeta_\mu, \zeta_{\mu \nu}) $ have the same number
of components, the latter  can be traded for
the former through a non singular  transformation of the form\footnote{This transformation and in 
particularly,
its global nature, is subtle. In the section $(3)$
of Ref.\refcite{Mul4} and for the case of non relativistic fluids, the authors discuss this point in some details.
They show that global invertibility is ensured provide the entropy density is a concave
function of the basic variables. For
the relativistic case matters are not that simple. However as we shall see further ahead
one expects local invertibility to hold (see also discussion in
Appendix B).}
\begin{equation}
	(J^{\mu}, T^{\mu\nu})\mapsto(\zeta, \zeta_{a}, \zeta_{ab}).
\label{RET_transformation}	
\end{equation}
Based in an analysis of Ref.\refcite{Liu},
the authors in 
Refs. \refcite{Mul6}, \refcite{Mul4}
proved that this transformation can be generated from the vector function
$S'^{\alpha}$
defined in 
(\ref{BasR})
considered now as a function 
of 
$ (\zeta, \zeta_\mu, \zeta_{\mu \nu}) $.
Performing the differentiation in
$\nabla_{\mu}S'^{\mu}$, then (\ref{YY}) 
yields
\begin{equation}
\left(\frac{\partial S'^{\alpha}}{\partial \zeta} - J^\alpha\right)\nabla_\alpha \zeta + \left(\frac{\partial S'^{\alpha}}{\partial \zeta_\beta} - T^{\alpha \beta} \right)\nabla_\alpha \zeta_\beta + \left(\frac{\partial S'^{\alpha}}{\partial \zeta_{\beta \mu}} - A^{\beta \mu \alpha}\right)\nabla_\alpha \zeta_{\beta \mu} - \zeta_{\beta \mu}I^{\beta \mu} \geq 0,
\label{RET_lagrange_m_f}
\end{equation}
and since this inequality must hold for arbitrary $ (\zeta, \zeta_\beta, \zeta_{\beta \alpha}) $, 
it implies:

\begin{equation}
J^\alpha = \frac{\partial S'^{\alpha}}{\partial \zeta}, \quad T^{\alpha \beta} = \frac{\partial S'^{\alpha}}{\partial \zeta_\beta}, \quad A^{\langle \beta \mu \rangle \alpha} = \frac{\partial S'^{\alpha}}{\partial \zeta_{\beta \mu}} - \frac{1}{4}g^{\beta \mu}g_{\rho k}\frac{\partial S'^{\alpha}}{\partial \zeta_{\rho k}}.
\label{constititivas de parametros}
\end{equation}
where in above, as before, the angular bracket signifies
taking the symmetric and trace free part.\\
Introducing the entropy production $ \sigma $ via
$\sigma := -\zeta_{\beta \mu}I^{\beta \mu}$, then validity of 
(\ref{RET_lagrange_m_f}) now demands: 
\begin{equation}
\sigma = -\zeta_{\beta \mu}I^{\beta \mu} \geq 0.
\label{EP}
\end{equation}
It follows from
(\ref{constititivas de parametros})
that the components $(J^{\mu}, T^{\mu\nu}, A^{\mu\nu\lambda})$ can be
considered as depending upon
the multipliers $(\zeta, \zeta_{\mu}, \zeta_{\mu\nu})$
and moreover
these components 
can be generated\footnote{\label{Comm_GL} 
As we shall seen in the next section
as long as $T^{\mu\nu}=T^{\nu\mu}$, actually 
the components of $(J^{\mu}, T^{\mu\nu}, A^{\mu\nu\lambda})$ 
can be obtained by differentiating  a 
scalar function and this observation simplifies mater considerably.
We shall discuss this property in the next section.} by differentiating
the ''vector potential" $S'^{\alpha} = S'^{\alpha}(\zeta, \zeta_\mu, \zeta_{\mu \nu})$.\\
Due to the significance of this 
''vector potential",
in Refs, \refcite{Mul6},\refcite{Mul4} construct  
at first an explicit
 representation 
of $S'^{\alpha}  $
as function of
$(\zeta, \zeta_{\mu}, \zeta_{\mu\nu})$
by appealing to the principle of relativity. 
As for the case
of the tensors $I^{\alpha \beta},A^{\alpha \beta \mu}$ and $S^{\alpha}$ treated above,
this principle
requires that 
$ S'^\alpha $ 
to behave as
an isotropic vector function 
under arbitrary coordinate transformations.
Since however 
 $S'^{\alpha} = S'^{\alpha}(\zeta, \zeta_\mu, \zeta_{\mu \nu})$
and the only vectors that one can construct out of the Lagrange multipliers 
that are quadratic in $\zeta_{\mu\nu}$ are
$\zeta_{\mu}, \zeta_{\mu\nu} \zeta^{\nu}, \zeta_{\mu\nu}^{2}\zeta^{\nu}$, it follows that
 $ S'^\alpha $ must be a linear combination of these vectors with suitable scalar coefficients
functions of the Lagrange multipliers.
Introducing the scalars

\begin{equation}
  \begin{split}
    H_0 & = \zeta								\\
    H_2 & = \text{tr}(\zeta^2_{\alpha \beta})	\\
    H_3 & = \text{tr}(\zeta^3_{\alpha \beta})	\\
    H_4 & = \text{tr}(\zeta^4_{\alpha \beta})
  \end{split}
\qquad \qquad
  \begin{split}
	G_0 & = \zeta^\alpha \zeta_\alpha						\\
	G_1 & = \zeta^\alpha \zeta_{\alpha \beta} \zeta^\beta	\\
	G_2 & = \zeta^\alpha \zeta_{\alpha \beta}^2 \zeta^\beta	\\
	G_3 & = \zeta^\alpha \zeta_{\alpha \beta}^3 \zeta^\beta
  \end{split}
\end{equation}
then it was shown in \cite{Mul6,Mul4}
that
the most general form $S'^{\alpha}$ quadratic in $\zeta_{\mu\nu}$ must have the form\footnote{
The representation of $S'_\alpha$ 
in
(\ref{GEN}) is one of the central formulas
and the reader is refereed 
to Refs. \refcite{Mul4}, section $(2.3)$ chapter $ 6 $ for more 
details leading 
to its derivation.}:

\begin{equation}
\begin{split}
	S'_\alpha & = \left[\Gamma_0 + \frac{\partial \Gamma_1}{\partial G_0}G_1 +\frac{1}{2}\frac{\partial^2 \Gamma_2}{\partial G_0^2}G_1^2 + \frac{\partial \Gamma_2}{\partial G_0}G_2 + \frac{1}{4}\Gamma_2 H_2 \right]\zeta_\alpha + \\ 
	& + \left[\Gamma_1 + \frac{\partial \Gamma_2}{\partial G_0}G_1\right] \zeta_{\alpha \beta}\zeta^\beta + \Gamma_2 \zeta^2_{\alpha \beta}\zeta^\beta.
\end{split}	
\label{GEN}
\end{equation}
where $(\Gamma_{0}, \Gamma_{1}, \Gamma_{2})$ are functions
of $\zeta$ and $G_{0}$. In arriving at the above formula, the authors 
took into considerations the symmetries of the fields  
$(T^{\mu\nu}, A^{\mu\nu\lambda})$ and the requirement that
$(J^{\mu}, T^{\mu\nu}, A^{\mu\nu\lambda})$ should be linear functions of $\zeta_{\mu\nu}$.
These conditions fix the functions $(\Gamma_{1}, \Gamma_{2})$ in terms of  the $\Gamma_{0}(\zeta, G_{0})$ and 
two arbitrary 
function of $\zeta$. Therefore, although in the following 
formulas the functions $ \Gamma_1, \Gamma_2 $ appear explicitly, in fact 
they are considered as determined by $ \Gamma_0(\zeta, G_0) $
and two arbitrary functions of $\zeta$.

Using $S'_\alpha$ shown 
in (\ref{GEN})
combined 
with 
(\ref{constititivas de parametros}), it follows that
the components of $ J^\alpha $, $ T^{\alpha \beta} $ and $ A^{\beta \mu \alpha} $ to linear order in $ \zeta_{\alpha \beta} $
 have the form:

\begin{equation}
J^\alpha = \left(\frac{\partial \Gamma_0}{\partial \zeta} + \frac{\partial^2 \Gamma_1}{\partial \zeta \partial G_0}G_1\right)\zeta^\alpha + \frac{\partial \Gamma_1}{\partial \zeta}\zeta^{\alpha \beta}\zeta_\beta,
\label{RET_j_1}
\end{equation}

\begin{equation}
\begin{split}
	T^{\beta \alpha} = & \left(\Gamma_0 + \frac{\partial \Gamma_1}{\partial G_0}G_1\right)g^{\beta \alpha} + \Gamma_1 \zeta^{\beta \alpha} + 2\left(\frac{\partial \Gamma_0}{\partial G_0} + \frac{\partial^2 \Gamma_1}{\partial G_0^2}G_1\right)\zeta^\beta \zeta^\alpha \\
	& + 2\frac{\partial \Gamma_1}{\partial G_0}(\zeta^\beta \zeta^{\alpha \mu}\zeta_\mu + \zeta^\alpha \zeta^{\beta \mu}\zeta_\mu),
\end{split}	
\label{RET_T_1}
\end{equation}

\begin{equation}
\begin{split}
	A^{\beta \mu \alpha}  = & \frac{1}{2}\left(\Gamma_1 + \frac{\partial \Gamma_2}{\partial G_0}G_1\right)(g^{\beta \mu}\zeta^\alpha + g^{\mu \alpha}\zeta^\beta + g^{\alpha \beta}\zeta^\mu) +	\\
	& + \frac{1}{2}\Gamma_2(g^{\beta \mu}\zeta^{\alpha \nu}\zeta_\nu + g^{\mu \alpha}\zeta^{\beta \nu}\zeta_\nu + g^{\alpha \beta}\zeta^{\mu \nu}\zeta_\nu + \zeta^{\beta \mu}\zeta^\alpha + \zeta^{\mu \alpha}\zeta^\beta + \zeta^{\alpha \beta}\zeta^\mu) + \\
	& + \left(\frac{\partial \Gamma_1}{\partial G_0} + \frac{\partial^2 \Gamma_2}{\partial G_0^2}G_1\right)\zeta^\beta \zeta^\mu \zeta^\alpha + \frac{\partial \Gamma_2}{\partial G_0}(\zeta^\beta \zeta^\mu \zeta^{\alpha \nu}\zeta_\nu + \zeta^\mu \zeta^\alpha \zeta^{\beta \nu}\zeta_\nu + \zeta^\alpha \zeta^\beta \zeta^{\mu \nu}\zeta_\nu),
\end{split}
\label{ZZ}
\end{equation}

while the entropy vector  $ S^\alpha $ has the form:

\begin{equation}
\begin{split}
	S^\alpha = & -\left[\zeta \frac{\partial \Gamma_0}{\partial \zeta} + 2\frac{\partial \Gamma_0}{\partial G_0}G_0 + \left(\zeta \frac{\partial^2\Gamma_1}{\partial \zeta \partial G_0} + 2G_0\frac{\partial^2 \Gamma_1}{\partial G_0^2} + 3\frac{\partial \Gamma_1}{\partial G_0}\right)G_1\right]\zeta^\alpha - \\
	& - \left[\Gamma_1 + \zeta \frac{\partial \Gamma_1}{\partial \zeta} + 2G_0 \frac{\partial \Gamma_1}{\partial G_0}\right]\zeta^{\alpha \beta}\zeta_\beta.
\end{split}
\end{equation}

The above representations of $(J^\alpha, T^{\alpha\beta }, A^{\beta \mu \alpha}, S^\alpha)$ 
are formal since they depend
upon the Lagrange multipliers
 $(\zeta, \zeta_\mu, \zeta_{\mu \nu}) $ 
 whose physical significance is not yet clear.
 So the second step
 in the approach of
 Refs.\refcite{Mul4,Mul6} 
is  to express the multipliers in terms of the physically relevant quantities $(\rho, P+\pi, q^{a},\pi^{\mu\nu}, n, u^{\mu})$ 
that appear in (\ref{LMR},\ref{Split}). 
Combining 
(\ref{LMR},\ref{Split}).
with
(\ref{RET_j_1}), (\ref{RET_T_1})
one gets:
\begin{equation}
	nmu^\alpha = \left(\frac{\partial \Gamma_0}{\partial \zeta} + \frac{\partial^2 \Gamma_1}{\partial \zeta \partial G_0}G_1\right)\zeta^\alpha + \frac{\partial \Gamma_1}{\partial \zeta}\zeta^{\alpha \beta}\zeta_\beta,
\label{RET_50}	
\end{equation}

\begin{equation}
\begin{split}
	&   \frac{1}{c^2}\rho u^\alpha u^\beta  + (P + \pi)\Delta^{\alpha\beta} + \frac{1}{c^2}(u^\alpha q^\beta + u^\alpha q\beta)     + \pi^{\alpha \beta} 	
	  = \left(\Gamma_0 + \frac{\partial \Gamma_1}{\partial G_0}G_0\right)g^{\alpha\beta} + \\ 
	&  + \Gamma_1 \zeta^{\alpha \beta} + 2\left(\frac{\partial \Gamma_0}{\partial G_0} + 
	\frac{\partial \Gamma_1}{\partial G_0^2}+G_1\right)\zeta^\alpha\zeta^\beta+ 2\frac{\partial \Gamma_1}{\partial G_0}\left(\zeta^\beta \zeta^{\alpha \mu}\zeta_\mu + \zeta^\alpha \zeta^{\beta \mu}\zeta_\mu \right).
\end{split}	 
\label{RET_51}
\end{equation}
These relations are considered as a system of $ 14 $ 
equations relating  the Lagrange multipliers $ (\zeta, \zeta_\mu, \zeta_{\mu \nu}) $ to
the $14$ fields $ (n, \rho, \pi^{\mu \nu}, \text{etc}) $ appearing in
(\ref{LMR},\ref{Split}).
However, due to their non linear nature,
resolving these equations for the multipliers is not a trivial task.
In 
 \refcite{Mul6},\refcite{Mul4}
settled that problem ''perturbativelly'':
they first established the relationship
 between
  $ (\zeta, \zeta_\mu, \zeta_{\mu \nu}) $
and the physical variables
 $ (n, \rho, \pi^{\mu \nu}, \text{etc}) $ 
 for the particular family of fluid states namely the family of equilibrium states.
The identification of the equilibrium states
within their theory 
needs some considerations. 
Such states
are 
 characterized 
  by the vanishing of the entropy production $\sigma$ (see (\ref{EP}))
  and this condition requires $I^{\mu\nu}=0$. 
However the additional requirement that
  $\sigma$ should be a local minimum in equilibrium demands that
equilibrium states
are characterized by the property: $\zeta_{\mu\nu}=0$ (for details leading to this conclusion see
discussion in  Refs. \refcite{Mul6},\refcite{Mul4}).\\
Accepting this property, and using the subscript $E$ to denote equilibrium values,
one finds from the expression for $S^{\alpha}$ and (\ref{RET_50}, \ref{RET_51})
the following formulas valid only for equilibrium sates:

\begin{equation}
S^\alpha_{E} :=S_{E}u^{\alpha}= \left (-(\zeta \frac{\partial \Gamma_0}{\partial \zeta} + 2\frac{\partial \Gamma_0}{\partial G_0}G_0)\zeta^{\alpha}\right)_{E}
\label{X}
\end{equation}

\begin{equation}
	nmu^\alpha = \left( \frac{\partial \Gamma_0}{\partial \zeta}\zeta^\alpha\right)_{E}
\label{XX}	
\end{equation}

\begin{equation}
P\Delta^{\alpha\beta} + \frac{1}{c^2}\rho u^\alpha u^{\beta} = \left(\Gamma_0 g^{\alpha\beta} + 2\frac{\partial \Gamma_0}{\partial G_0}\zeta^\alpha \zeta^\beta\right)_{E}
\label{ERET_5}
\end{equation}
where 
$S_{E}$
is the equilibrium entropy density  as measured by the $u^{\mu}$ observer i.e.
$S_{E}=S^{\mu}{}_{E}u_{\mu}$.\\
Recalling that $\Gamma_{0}:=\Gamma_{0}(\zeta, G_{0})$
and $G_{0}:=(\zeta^{\alpha}\zeta_{\alpha})$, it was argued in 
\refcite{Mul4}, \refcite{Mul6}
that these three relations becomes consistent 
 by identifying 
$\zeta_{E}$ and $G_{0 E}:=(\zeta^{\alpha}\zeta_{\alpha})_{E}$
with the fugacity $\alpha$  and the absolute temperature $T$
and moreover 
set: 
$\Gamma_{0}(\zeta_{E}, G_{0E})=-P$ with $P = P(n, \rho)$ the thermal equation of state.
In details
\begin{equation}
\frac {1}{T}=-\frac {{(G_{0E})}^{\frac {1}{2}} }{c},\quad 
\zeta_{E}=\frac {1}{T} 
 \left( \frac {{\rho}-T {S_{E}} + {P}}{nm}\right)=-\alpha,\quad \Gamma_{0}(\zeta, G_{0E})=-P(n, \rho).
 \label{RET_52}
 \end{equation}
Moreover the identification:
$T^{-2} c^{2}=G_{0E}=(\zeta^{\mu} \zeta_{\mu})_{E}$,
implies that
$(\zeta_{\mu})_{E}  = -cu_{\mu}{T^{-1}}$
and thus in summary the equilibrium values of the multipliers are given by:
\begin{equation}
	(\zeta)_{E}  = -\alpha, \quad (\zeta_\alpha)_{E}  = -\frac{u_\alpha}{T}, \quad (\zeta_{\alpha \beta})_{E}  = 0.
\label{RET_zetas}
\end{equation} 
Using these equilibrium values 
for the multipliers, subsequently 
they perturb them according to the following scheme: 

\begin{equation}
\begin{split}
	\zeta & = -\alpha + \chi,		\\
	\zeta_\alpha & = -\frac{u_\alpha}{T} + \lambda_\alpha + \lambda u_\alpha ,\\
	\zeta_{\alpha \beta} & = \sigma_{\langle\alpha \beta\rangle} + \sigma \Delta_{\alpha \beta} + \frac{1}{c^2}(u_\alpha \sigma_\beta + u_\beta \sigma_\alpha) + \frac{3}{c^2}\sigma u_\alpha u_\beta, \quad \sigma^\alpha_{\ \alpha} = 0,
\end{split}
\label{RET_zetas}
\end{equation} 

where 
$ (\chi, \lambda, \lambda_\alpha, \sigma, \sigma_\alpha, \sigma_{\alpha \beta}) $ are the non equilibrium parts of the multipliers and for notational simplicity we omitted the subscript $E$ describing the leading equilibrium parts. Inserting (\ref{RET_zetas}) back 
to 
(\ref{RET_50}, \ref{RET_51})
they obtain a linear system for
$(\chi, \lambda, \lambda_\alpha, \sigma, \sigma_\alpha, \sigma_{\alpha \beta}) $
whose solutions 
are linear functions of  $(q^{\mu}, \pi, \pi^{\mu\nu})$ and
below, we provide  a sample of the resulting solutions 

\begin{equation}
	\lambda^\alpha = -\frac{1}{T}\frac{\dot{\Gamma}_1}{D_3}q^\alpha := \lambda_q q^\alpha
\end{equation}

\begin{equation}
	\sigma^\alpha = \frac{\dot{P}}{D_3}q^\alpha := \sigma_q q^\alpha
\end{equation}

\begin{equation}
	 \sigma^{\langle \alpha \beta \rangle} = \frac{1}{\Gamma_1}\pi^{\alpha \beta}
\end{equation}
where an over dot signifies derivative with respect to $\alpha$  while  $D_{3}$ is 
the determinant of a $( 3\times 3 )$ matrix
involving derivatives of the equilibrium equation of state $P(\alpha, T)$
and the derivatives of $\Gamma_{1}(\alpha, T) $.
 The other parts of the Lagrange multipliers 
 appearing in (\ref{RET_zetas})
are longer expressions
that we do not report here but they can be found
in eqs. $(2.39) $ page $ (115) $ of Ref. \refcite{Mul4}.\\
Substituting these linearized representations of the Lagrange multipliers  in (\ref{ZZ})
results a long expression 
for  the components $ A^{\mu \nu \lambda} $ 
(too long to be 
given here) that can be found on page $116$ formula $(2.41)$ in Ref. \refcite{Mul4}.
This representation 
of $ A^{\mu \nu \lambda} $ 
has the same structure
as the one shown  in 
(\ref{constitutive 2})
and a comparison between these two 
yields
a representation for the $4$ coefficients 
$(C^0_{\ 1}, C^\pi_{\ 1}, C_3, C_5)$
in terms of an
equilibrium equation of state $ P = P(\alpha,T)$
and  the functions  $ \Gamma_1(\alpha, T) $ and $ \Gamma_2(\alpha, T) $.
The explicit formulas\footnote{Recall that although
in these formula appear terms $ \Gamma_1(\alpha, T) $ and $ \Gamma_2(\alpha, T) $,
these functions are determined by $ P = P(\alpha,T) $ and two functions $ A_1(\alpha) $, $ A_2(\alpha) $
of the fugacity coefficient $\alpha$ (see eqs. (2.68), (2.69) on page $122$).} 
for the 
coefficients 
$(C^0_{\ 1}, C^\pi_{\ 1}, C_3, C_5)$
can be found in eq. $2.43$ on page $116$.

The determination of the coefficients $ (A^\pi_{\ 1}, A^{\pi^2}_{\ 1}, A^q_{\ 1}, ... , A^0_{\ 3}) $ that appear
in the entropy vector $ S^\mu $  in (\ref{constitutive 3}) are 
calculated in\cite{Mul4,Mul6}, 
by noting
that (\ref{BasR}) combined with 
(\ref{constititivas de parametros})
imply
\begin{equation}
	dS^\mu = -\zeta dJ^\mu - \zeta_\nu dT^{\nu \mu} - \zeta_{\nu \lambda}dA^{\mu \nu \lambda}.
\label{entr_LM}	
\end{equation}
Eliminating the Lagrange multipliers using their linearized versions in (\ref{RET_zetas}) 
and the expansions in  
(\ref{LMR}), the above formula can be
re-writen (\ref{entr_LM}) in the form 
\begin{equation}
	dS^\mu = f^{\mu}d\pi + g^{\mu}{}_\nu dq^\nu + G^{\mu}{}_{\nu \lambda}d\pi^{\nu \lambda},
\end{equation}
so that 
\begin{equation}
	\frac{\partial S^\mu}{\partial \pi} = f^{\mu}, \quad \frac{\partial S^\mu}{\partial q^\nu} = g^{\mu}{}_{\nu}
	, \quad \frac{\partial S^\mu}{\partial \pi^{\nu \lambda}} = G^{\mu}{}_{\nu\lambda},
\end{equation}
where $(f^{\mu}, g^{\mu}{}_{\nu}, G^{\mu}{}_{\nu\lambda})$ are well defined functions.
An integration of these relations
and comparison of the resulting form of $S^{\mu}$
with the one 
shown in (\ref{constitutive 3}), 
fixes the coefficients 
$ (A^\pi_{\ 1}, A^{\pi^2}_{\ 1}, A^q_{\ 1}, A^t_{\ 1}, A^0_{\ 2}, A^\pi_{\ 2}, A^0_{\ 3}) $ and 
their explicit form 
can be found on page $117$ of ref.\refcite{Mul4}.\\

To complete the analysis,  the coefficients
$(B^\pi_{\ 1}, B_3, B_4)$  that appear in
the dissipation tensor $I^{\mu\nu}$ in
(\ref{constitutive 1}) need to be considered. 
However these coefficients appear to be underdetermined except that they are restricted to
obey certain inequalities. These inequalities arise by noting 
that the entropy production 
$ \sigma = -\zeta_{\mu \nu}I^{\mu \nu} $ upon using $ I^{\mu \nu} $
in
(\ref{constitutive 1})
and substituting 
$ \zeta_{\mu \nu} $ for the one in (\ref{RET_zetas}),
yields

\begin{equation}
	\sigma = 12B^\pi_{\ 1}\sigma_\pi \pi^2 - 2\frac{1}{c^2}B_4\sigma_q q^\mu q_\mu  - B_3 \frac{1}{\Gamma_1}\pi^{\langle \beta \mu \rangle}\pi_{\langle \beta \mu \rangle} \geq 0.
\label{ENGEN}
\end{equation}
The non-negativity\footnote{Notice that even though in the right hand side appear coefficients like $ \sigma_\pi $, $ \sigma_q $ and $ \Gamma^{-1}_1 $, these coefficients are considered as known and their form can be found in
page $ 115 $ in \cite{Mul4}.} of the right hand side of this $\sigma$  
puts restrictions in the form of inequalities in the coefficients $ (B^\pi_{\ 1}, B_4, B_3) $. Otherwise
they  remain undetermined and
as we shall see below, these coefficients are related to bulk, shear viscosity and the coefficient of heat conduction
of the fluid.
As a side remark, (\ref{ENGEN})
shows
 that there exist three dissipative mechanism 
 generating entropy in the fluid, namely 
the heat flux and the two stresses.\\
Finally, we mention 
briefly the important issue of 
the symmetric-hyperbolic and
causality property of the dynamical equations
within the relativistic (REIT).
As shown in more details in Refs. \refcite{Mul4,Mul6},
these requirement puts restrictions on the coefficients $ C $ and  $ A $
that appear in $A^{\mu\nu\lambda}$ and $S^{\mu}$
(see (\ref{constitutive 2},\ref{constitutive 3})). 
Amongst these is the restrictions of 
 the form of the equilibrium equation of state $ P = P(\alpha, T) $
 although this is not the only restriction.
 In effect when these restrictions hold, an open vicinity
 of around the equilibrium state the dynamical equations 
 a 
 symmetric-hyperbolic and
causal system.
Although from the physical standpoint causality
is a very important for any theory,
we shall not discuss its implications
upon relativistic (REIT). In the next section,
we shall address this 
issue from a wider angle.\\

This concludes a brief summary
of the Liu-M\"uller-Ruggeri relativistic (REIT).
All of the above described work aimed at
specifying the structure of the constitutive functions
$A^{\mu\nu\lambda}, I^{\mu\nu}$ and  $S^{\mu}$
for states near equilibrium so the set of $14$ dynamical equations
for the auxiliary variables
$(\rho, P+\pi, q^{a},\pi^{\mu\nu}, n, u^{\mu})$ 
can be written down.
Solutions of these equations  describe states near equilibrium
and these equations form a closed system provided
the thermal equilibrium equation of state $  P = P(\alpha, T)  $  
a-priori has been specified\footnote{In that regard, it should be
 pointed out 
 the philosophy underlying the work in Refs.\refcite{Mul4,Mul6}.
 Since for relativistic systems it seems hard to determine
 observationally a thermal equation of state,
 the authors
appeal 
 to relativistic kinetic theory of dilute gases
 to specify a reliable equation of state. The restriction 
 to states of relativistic gases in equilibrium comes from the property that for such media one may specify 
 families of physically reliable equilibrium equations of state by appealing to J\"uttner equilibrium distribution.
 It should mentioned however, that except this technicality
their treatment 
  is quite general.}.\\
The
 dynamical equations for the fields 
 $(n, \rho, P, \pi, q^{\mu},\pi^{\mu\nu})$
 are
 \begin{equation} 
 \nabla_{\alpha}(nu^{\alpha})=0
 \label{B1}
 \end{equation}
  \begin{equation} 
 \nabla_{\alpha}\left( \frac{1}{c^2}\rho u^\alpha u^\beta+ (P + \pi)\Delta^{\alpha\beta }     
   +\frac{1}{c^2}(u^\alpha q^\beta + u^\beta q^\alpha)+  
     \pi^{\alpha \beta } \right)=0, 
 \label{B2}
 \end{equation}
 \begin{equation}
	\left(\Delta_{\mu \alpha}\Delta_{\nu \beta} - \frac{1}{3}\Delta_{\mu \nu}\Delta_{\alpha \beta}\right) \nabla_\gamma A^{\alpha \beta \gamma} = B_3\pi_{\mu \nu},
\label{B3}
\end{equation}  
\begin{equation}
	\Delta_{\mu \alpha}u_\beta \nabla_\gamma A^{\alpha \beta \gamma} = -B_4c^2q_\mu ,
\label{B4}
\end{equation}
\begin{equation}
	u_\alpha u_\beta \nabla_\gamma A^{\alpha \beta \gamma} = -3B^\pi_{\ 1}c^2\pi.
\label{B5}
\end{equation}
where in the last three equations the components of $A^{\alpha\beta\gamma}$ 
are those in 
(\ref{constitutive 2}) with the ''constants'' 
$(C^0_{\ 1}, C^\pi_{\ 1}, C_3, C_5)$ determined by
$  P = P(\alpha, T)  $ (and two arbitrary functions of $\alpha$).
 These equations contain in addition  
the three unspecified functions $ (B^\pi_{\ 1}, B_4, B_3) $ that determine
the entropy production.\\

It is worth noting that by evaluating the left hand-sides (\ref{B3}-\ref{B5}) on equilibrium states and denote
their right hand-sides 
as $(\pi^{1}_{\mu\nu}, \pi^{1}, q^{1}_{\mu})$, one recovers Eckart's phenomenological 
equations that we derived in previous sections. Thus relativistic (REIT)  in an appropriate limit, 
contains Eckart's first order theory.\\
If on the other hand on
the left hand-sides of (\ref{B3}-\ref{B5}) one employs
the form of $A^{\mu\nu\lambda}$ that contains linear contribution 
$(\pi_{\mu\nu}, \pi, q_{\mu})$ 
one recovers phenomenological equations
whose form is structurally 
analogous to the phenomenological equations
 (\ref{Y1},\ref{Y2},\ref{Y3})
 derived in section $11$ within the Israel-Stewart transient thermodynamics.  
However there is a pronounced difference between the two families of phenomenological sets.
Whereas
 the equations 
  (\ref{Y1},\ref{Y2},\ref{Y3}) 
  contain eight unspecified coefficients of two variables
  namely, $(\zeta_{u}, \beta_{0}, \bar \alpha_{0}, k, \bar \beta_{1}, \bar \alpha_{1}, \zeta, \beta_{2})$
  the corresponding equations within the 
  Liu-M\"uller-Ruggeri theory  
  contain
  only three unknown functions
   $ (B^\pi_{\ 1}, B_4, B_3) $ of two variables
   and two unknown functions of a single variable
   (correspondingly one unknown function
   of a single variable for 
    the case of a relativistic gas whose equilibrium thermal equation of state arises from a 
    J\"uttner distribution).
 So from this view point,
 the Liu-M\"uller-Ruggeri theory 
  restrict considerably the number of free functions appearing in the phenomenological equations.\\
  
  Explicit result for viscosities
  and coefficient of heat conductivity 
  for near equilibrium states of a relativistic gas
 are discussed  in sections $(4.1-4.7)$ 
   in Ref. \refcite{Mul4}.  
 They specified equilibrium equations of state for the relativistic gas by employing
 a J\"uttner equilibrium distribution and 
for various degrees of 
degeneracy
they work out
the corresponding (equilibrium) equation of state. We refer the reader to monograph of Ref. \refcite{Mul4}
page $(128)$ for a exhaustive analysis of different
degrees of degeneracies exhibited by a relativistic gas in equilibrium.

\section{Dissipative relativistic fluids of divergence type}\label{DRFDT}

In the previous section, 
we discussed at same length 
the Liu-M\"uller-Ruggeri theory
and we have seen that 
it is a restrictive theory 
in the sense that the
resulting phenomenological equations 
involve just three arbitrary functions of two variables (modulo one or two functions of one variable depending upon the
nature of the medio under study).
However, as we have seen, the implementation of the 
Entropy Principle and the Principle of Relativity on
the constitutive relations 
involves tedious computations
so that the nice features of this theory are not
very transparent.
 Pennisi in Ref.\refcite{Pen}
 and Geroch and Lindblom in Ref. \refcite{Ger1},
motivated by the 
structure of the fields equations in (\ref{RET_1},\ref{RET_2}), 
introduced a class of 
relativistic fluid theories satisfying the following
three properties:\\

a) The variables specifying the fluid are the components of the conserved particle current $J^{\mu}$
and the components of the conserved and symmetric energy momentum tensor $T^{\mu\nu}.$\\

b) The dynamical equations 
are the same as in 
(\ref{RET_1},\ref{RET_2}), but
now the tensor fields $A^{\mu\nu\lambda}$ and $I^{\nu\lambda}$ are symmetric and traceless\footnote{
Thus the treatment of Pennisi and Geroch-Lindblom
generalizes
slightly the 
Liu-M\"uller-Ruggeri theory.
While in the, latter theory
 $A^{\mu\nu\lambda}$ is assumed to be totally symmetric  
 in
the  Pennisi, Geroch-Lindblom
theory,
 this tensor is symmetric only in the indices 
$\nu,\lambda$. Also for the later theory, the constraint $A^{\mu\nu}{}_{\nu}=m^{2}J^{\mu}$, it is not any longer imposed.} in the indices 
$\nu,\lambda$.
  Moreover, as for
 the Liu-M\"uller-Ruggeri theory,
$A^{\mu\nu\lambda}$ and $I^{\nu\lambda}$
are considered to be constitutive functions i.e.
algebraic functions of the basic variables
$J^{\mu}$ and $T^{\mu\nu}$.\\

c) There exists an entropy current 
$S^{\mu}(J, T)$ 
whose dependance 
upon $(J,T)$ is such that 
the entropy principle holds i.e.
for any solution $(J, T)$ of the field equations 
(\ref{RET_1},\ref{RET_2}), 
the current satisfies\footnote{
For the theories
analyzed in 
\cite{Ger1}, the authors imposed two additional restrictions:
 in addition to $\sigma(J, T)\geq 0,$
they assumed the theory to be ``generic" in the sense that the
``dissipation tensor" $I^{\mu\nu}$ and of the derivatives of the generating  function $\chi$ (to be defined below) 
satisfy suitable inequalities
to be introduced further ahead.}:

\begin{equation}
	\nabla_{\mu}S^{\mu}(J,T)=\sigma(J,T)\geq0.
\label{SF}
\end{equation}
where $\sigma(J,T)$ some algebraic function 
of $(J,T)$.

The theories defined by these three conditions 
are refereed
as (relativistic) fluid theories of divergence type
and in this section we shall highlight their
most important properties. \\
Already the
analysis 
of the
relativistic (REIT)
carried out 
in the previous section,
gave us a good insights regarding the structure of
fluids described
by 
relativistic (REIT)\footnote{Ought to be mentioned,
that except for 
a minor point regarding the symmetries of $A^{\mu\nu\lambda}$, 
relativistic (REIT) 
and fluid theories of 
of divergence type
are considered as been ''closely related''. In this work we say 
 that a fluid belongs to the class  of divergence type
 whenever the energy momentum tensor $T^{\mu\nu}$ is symmetric
 and of course obey  
the field equations cited above. As we shall see in this section, these class of 
divergence fluids can be treated via a scalar generating function which simplifies mater considerably. However ought to be stressed
that the powerful treatment of
Liu-M\"uller-Ruggeri theory should not disregarded. There are many physical important settings where a relativistic fluid interacts with external fields or with other fluids 
and in these cases one cannot treat them via a scalar generating function.},
at least for fluid states near equilibrium. However,  Pennisi\cite{Pen}  in $1989$ and independently 
Geroch and Lindblom \cite{Ger1} in $(1990)$
offer an alternative treatment to fluids obeying conditions $a)-c)$ cited above
and in this section we analyze their approach.\\

The Pennisi\cite{Pen} -
Geroch and Lindblom \cite{Ger1} approach 
 introduces again the field of Lagrange multipliers\footnote{For comparison
purposes, we denote these multipliers by the same
symbols as those employed for 
 Liu-M\"uller-Ruggeri theory.}
$ (\zeta, \zeta_\mu, \zeta_{\mu \nu}) $ obeying
 $\zeta_{\mu\nu}=\zeta_{\nu\mu}$ and $g^{\mu\nu}\zeta_{\mu\nu}=0$.
 As for
the case of 
Liu-M\"uller-Ruggeri
treatment, they also introduce 
the vector function 
$S'^{\alpha}$ 
as in (\ref{BasR}) and the transformation
shown
in (\ref{RET_transformation}).
They 
noticed however,
that the symmetries of $ T^{\mu \nu} $ imply 
that this transformation 
may be generated by a some 
scalar function $ \chi(\zeta,\zeta_\mu, \zeta_{\mu \nu})$ 
related to the vector function $S'^{\alpha}$ via:
\begin{equation}
S'^\alpha = \frac{\partial \chi}{\partial \zeta_{\alpha}}
\label{DeGF}
\end{equation}
and this fundamental relation makes the important difference
in the formulation of divergence fluids.
By arguments similar to
those 
in \cite{Liu} and in \cite{Mul6},
(see also discussion in Appendix B and \cite{FT1}),
any fluid theory obeying conditions $a)-c)$ above
is determined from the knowledge of the dissipation tensor $I^{\mu\nu}(\zeta, \zeta_\mu, \zeta_{\mu \nu})$
and from  a
smooth scalar function $ \chi (\zeta, \zeta_\mu, \zeta_{\mu \nu}) $ referred as the generating function. This
 scalar function $ \chi (\zeta, \zeta_\mu, \zeta_{\mu \nu}) $ as a consequence 
 of
 (\ref{DeGF})
 combined with
(\ref{constititivas de parametros})
  satisfies:
\begin{equation}
	J^\mu = \frac{\partial^2 \chi}{\partial \zeta \partial \zeta_\mu}, \quad T^{\mu \nu} = \frac{\partial^2 \chi}{\partial \zeta_\mu \partial \zeta_\nu}, \quad A^{\mu \nu \lambda} = \frac{\partial^2 \chi}{\partial \zeta_\mu \partial \zeta_{\nu \lambda}},
\label{DTT1}
\end{equation}
and generates the entropy current $ S^\mu $ via
\begin{equation}
	S^\mu = \frac{\partial \chi}{\partial \zeta_\mu} - \zeta J^\mu - \zeta_\nu T^{\mu \nu} - \zeta_{\nu \lambda}A^{\mu \nu \lambda},
\label{DTT2}
\end{equation}   
while the  source term $ \sigma $ in 
(\ref{SF})
 is defined 
 from the dissipation 
 tensor $I^{\mu\nu}$ and 
 $\zeta_{\mu\nu}$
via:
\begin{equation}
	\sigma = -\zeta_{\mu \nu}I^{\mu \nu}.
\label{DTT3}
\end{equation}
The relations in 
(\ref{DTT1})
are the fundamental relations in the 
Penissi-Geroch-Lindblom formalism
and show
that the components $(J^{\mu}, T^{\mu\nu} , A^{\mu \nu \lambda})$
can be regarded
as functions of 
the Lagrange multipliers
$(\zeta, \zeta_\mu, \zeta_{\mu \nu})$,
 and moreover the dynamical equations take the form:
 
\begin{equation}
\begin{split}
0 & = \nabla_\mu J^\mu  = \nabla_\mu\left( \frac{\partial^2 \chi}{\partial \zeta \partial \zeta_\mu} \right) = \nabla_\mu\left(\frac{\partial \chi^\mu}{\partial \zeta} \right) \\
& = \frac{\partial^2 \chi^\mu}{\partial \zeta \partial \zeta}\nabla_\mu \zeta + \frac{\partial^2 \chi^\mu}{\partial \zeta \partial \zeta_\nu}\nabla_\mu \zeta_\nu + \frac{\partial^2 \chi^\mu}{\partial \zeta \partial \zeta_{\kappa \lambda}}\nabla_\mu \zeta_{\kappa \lambda},\\[10pt]
\end{split}
\end{equation}

 \begin{equation}
 \begin{split}
 0 & = \nabla_\mu T^{\mu \nu} = \nabla_\mu\left(\frac{\partial^2 \chi}{\partial \zeta_\mu \partial \zeta_\nu}\right) =  \nabla_\mu\left(\frac{\partial \chi^\mu}{\partial \zeta_\nu} \right)\\
 & = \frac{\partial^2 \chi^\mu}{\partial \zeta \partial \zeta_\nu}\nabla_\mu \zeta+ \frac{\partial^2 \chi^\mu}{\partial \zeta_\rho \partial \zeta_\nu}\nabla_\mu \zeta_\rho + \frac{\partial^2 \chi^\mu}{\partial \zeta_\nu \partial \zeta_{\rho \lambda}}\nabla_\mu \zeta_{\rho \lambda},\\[10pt]
 \end{split}
 \end{equation}
 
 \begin{equation}
 \begin{split}
I^{\nu \lambda} & = \nabla_\mu A^{\mu \nu \lambda} = \nabla_\mu\left(\frac{\partial^2 \chi}{\partial \zeta_\mu \partial \zeta_{\nu \lambda}}\right) =  \nabla_\mu\left(\frac{\partial \chi^\mu}{\partial \zeta_{\nu \lambda}} \right)\\
 & = \frac{\partial^2 \chi^\mu}{\partial \zeta \partial \zeta_{\nu \lambda}}\nabla_\mu \zeta+ \frac{\partial^2 \chi^\mu}{\partial \zeta_\rho \partial \zeta_{\nu \lambda}}\nabla_\mu \zeta_\rho + \frac{\partial^2 \chi^\mu}{\partial \zeta_{\nu \lambda} \partial \zeta_{\rho \kappa}}\nabla_\mu \zeta_{\rho \kappa}.
 \end{split}
 \end{equation}

Introducing capital indices $ (A, B, C,...) $
so that
$ (\zeta_A $, $I^{A}) $
stand for $\zeta_A := (\zeta, \zeta_\mu, \zeta_{\mu \nu}) $ respectively
 for
$ I^A := (0,0,I^{\mu \nu}) $,
 the above equations can be written in the compact form:

\begin{equation}
\frac{\partial^2 \chi^\mu}{\partial \zeta_A \partial \zeta_B}\nabla_\mu \zeta_B = M^{AB\mu}\nabla_\mu\zeta_B = I^A,\quad
\mu \in (0,1,2,3),
\label{FEL}
\end{equation}
where the Einstein's summation convention has been extended over
the capital indices as well.\\
Thus in the
Penissi-Geroch-Lindblom formalism,
of prime importance
is the pair
$( \chi (\zeta, \zeta_\mu, \zeta_{\mu \nu}), I^{\mu\nu}(\zeta, \zeta_\mu, \zeta_{\mu \nu})$.
Once this
pair
has been specified,
then (\ref{FEL}) 
is a manifestly
symmetric\footnote{In this section, we follow
the terminology 
and definitions employed 
in Ref.\refcite{Ger1}. Thus the system
$M^{AB\mu}\nabla_{\mu} \zeta_B = I^A$ 
is said to  be symmetric provided: 
 $ M^{AB\mu} = M^{BA\mu} $.
 A symmetric system is 
hyperbolic in an open set of a fluid states, if
 the matrix
  $ M^{AB\mu}w_\mu $ is negative definite for some (possibly state-dependent) future directed timelike $ w_\mu$.
Finally, if 
$M^{AB\mu}\nabla_\mu \zeta_B = I^A$
is symmetric, then it said to be causal in an open set of a fluid states,
(i.e. hyperbolic with no fluid signals propagating faster than light) 
if $ M^{AB\mu}w_\mu $ is negative definite for all future
 directed timelike $ w_\mu$.
In these definitions, the contracted principal symbol
 $ M^{AB\mu}w_{\mu}$
 is evaluated typically on an equilibrium state (these equilibrium states are defined precisely further ahead).
 Via continuity arguments, it follows that
 the principal symbol maintains the same 
sign over an open vicinity of fluid states around the equilibrium state.} system of quasi linear equations
for the dynamical variables $
(\zeta, \zeta_\mu, \zeta_{\mu \nu})$.
However, whether this system 
is hyperbolic or (and) causal
depends upon
the nature of the generating function
$ \chi (\zeta, \zeta_\mu, \zeta_{\mu \nu}). $
Below,
we give a few examples of
physically relevant generating functions
$ \chi (\zeta, \zeta_\mu, \zeta_{\mu \nu})$
and briefly discuss\footnote{Within
the Pennisi, Geroch-Lindblom formalism,
the relativistic (REIT) of Liu-M\"uller-Ruggeri that we discussed in the last section, can be generated 
at least for states near equilibrium,
 by the dissipation function
$I^{\mu\nu}$ shown in
(\ref{constitutive 1}) and by a generating function
$ \chi (\zeta, \zeta_\mu, \zeta_{\mu \nu}) $
that can be obtained by integrating 
$S'^{\alpha}$ with respect to $\zeta^{\mu}$.},
the properties of the resulting fluid theories.\\
The first example consists of the class of 
functions
$ \chi (\zeta, \zeta_\mu, \zeta_{\mu \nu})$
that generate
states of a simple perfect fluid. 
Perfect fluid states are generated by setting the dissipation tensor to zero i.e. $I^{\mu\nu}=0$ 
and choosing 
$\chi$ to be a smooth function only of 
$(\zeta, \zeta^{\mu})$
i.e.
 
\begin{equation}
	\chi= \alpha(\zeta,\mu), \qquad \mu = \zeta^\alpha \zeta_\alpha.
\label{PFGFN}
\end{equation}  

For this choice, it follows from 
(\ref{DTT1}) that $A^{\mu\nu\lambda}:=0$, while a simple calculation shows that
the particle current $J_{PF}^{\mu}$,
the symmetric energy momentum tensor $T_{PF}^{\mu\nu}$ and the entropy current $S_{PF}^{\mu}$
are given by:
\begin{equation}
	J^\mu_{PF} = 2\frac{\partial^2 \alpha}{\partial \mu \partial \zeta}\zeta^\mu, \quad	 T^{\mu \nu}_{PF} = 4\frac{\partial^2 \alpha}{\partial \mu^2}\zeta^\mu \zeta^\nu + 2\frac{\partial \alpha}{\partial \mu}g^{\mu \nu},
\label{PFT}
\end{equation}
\begin{equation}
S^\mu_{PF} = -2\left[\zeta \frac{\partial^2 \alpha}{\partial \mu \partial \zeta} + 2\mu \frac{\partial^2 \alpha}{\partial \mu^2}\right]\zeta^\mu.
\label{PFEN}
\end{equation}
By comparing 
the right hand sides of these expressions to the standard
formulas for the particle current
the energy momentum tensor and entropy current of a  perfect fluid,
shows 
that
(\ref{PFGFN}) generates states
of a simple perfect fluid
whose  particle number density $n$, energy density $\rho$, isotropic pressure $P$,
as measured relative to the rest frame of the flow are given by: 
\begin{equation}
	nT= 2\frac{\partial^2 \alpha}{\partial \mu \partial \zeta} \quad \rho + P = -4\mu\frac{\partial^2\alpha}{\partial \mu^2}, \quad P = 2\frac{\partial \alpha}{\partial \mu},
\label{PFQ}
\end{equation}
while the fields $(\zeta, \zeta_{\mu})$ are related to the thermal potential $\Theta$, four velocity $u^{\mu}$,
 local temperature $T$,
 and entropy per particle $\hat s=sn^{-1}$, via:
\begin{equation}
	\zeta := \Theta=\frac{\rho + P}{nT} - \hat s, \quad  \zeta^\alpha = \frac{u^\alpha}{T},  \quad T^2 = -\frac{1}{\mu}.
\label{FVEL}
\end{equation}
Finally, for completeness, we mention that the 
well known property that the
dynamical equations for a simple
perfect fluid form a symmetric-hyperbolic\footnote{This important property of perfect fluids is discussed
for instance in ref.\cite{Chq1}.}
 respectively
symmetric-causal system,
provided that suitable restrictions 
are imposed upon 
the equation of state
can also established within the 
Geroch-Lindblom formalism.
By examining the contraction
 of the
principal symbol $M^{AB\mu}$ using the generating function  in 
 (\ref{PFGFN})
 it follows that causality holds on any fluid state if and only if
the function 
 $ \alpha(\zeta, \mu) $,
in (\ref{PFGFN}) is restricted so that the 
resulting $ (\rho, P) $ satisfy following
inequalities:

\begin{equation}
	\rho + P > 0, \quad \frac{\partial \rho}{\partial P}\Big\vert_J > \frac{\partial \rho}{\partial P}\Big\vert_S \geq 1,
\label{DTT_ine_1}	
\end{equation}
(for details regarding the derivation of these inequalities see
Refs. \refcite{Ger1,Ger3}).\\
 
 The second example 
discusses the structure of the
generating functions
that
within the class of relativistic fluids of divergence type
generate
 equilibrium states.
Geroch and Lindblom in Ref.\refcite{Ger1},
define a fluid 
state to be 
an equilibrium
state if and only if it is described by a solution $ (J^\mu, T^{\mu \nu}) $
of (\ref{RET_1},\ref{RET_2}) 
having the property 
that 
$ I^{\mu \nu}(J, T) = 0 $.
We denoted by $ (J^\mu_E, T^{\mu \nu}_E) $ 
these equilibrium states
and here after we use the subscript $E$ on any quantity to indicate that the underlying quantity is evaluated on an equilibrium state.
Clearly 
on  such states
$ \sigma_{E}(J_E, T_E)  
=-\zeta_{\mu \nu}I_{E}^{\mu \nu}= 0 $.
Under the additional assumption 
that any solution $ (J^\mu, T^{\mu \nu}) $ of  (\ref{RET_1}, \ref{RET_2})  
satisfies
$ \sigma(J,T) \geq 0 $, it was shown in \cite{Ger1} that on any
equilibrium state  $ \zeta_{\mu \nu} = 0 $
a conclusion that holds also for
the Liu-M\"uller-Ruggeri relativistic (REIT).
Properties of equilibrium states can be determined
by
considering a series expansion of the generating 
function
$ \chi (\zeta, \zeta_\mu, \zeta_{\mu \nu})$
around $(\zeta, \zeta_{\mu},  \zeta_{\mu \nu}=0) $.
As long as considerations are restricted on
tensor fields evaluated on equilibrium states, an expansion of
$ \chi (\zeta, \zeta_\mu, \zeta_{\mu \nu})$
 up and including terms linear in $\zeta^{\mu\nu}$ is sufficient.
Such expansion can be presented in the form:
\begin{equation}
	\chi( \zeta, \zeta_{\mu},  \zeta_{\mu \nu})= \alpha(\zeta,\mu),
+\left(\frac{\partial \chi}{\partial \zeta_{\mu \nu}}\right)_E \zeta^{\mu\nu}+O(\zeta^{\mu\nu})^{2},
\quad \mu = \zeta^\alpha \zeta_\alpha,
\label{PFGF}
\end{equation}  
where $ \alpha(\zeta,\mu) $ is an arbitrary smooth algebraic function of $ (\zeta, \mu) $, while the derivative of $ \chi $ with respect to $ \zeta_{\mu \nu} $ evaluated on equilibrium states is taken in the form: 
\begin{equation}
	\left(\frac{\partial \chi}{\partial \zeta_{\mu \nu}}\right)_E = \beta(\zeta, \mu)\left(\zeta^\mu \zeta^\nu - \frac{1}{4}\mu g^{\mu \nu} \right),\quad 
\beta(\zeta, \mu)=\frac {\partial \alpha(\zeta, \mu)}{\partial \mu}.
\label{PFGFG}
\end{equation}
These properties
of the $\chi$ combined 
with
(\ref{DTT1}),
imply  that the particle current $J_{E}^{\mu}$,
the symmetric energy momentum tensor $T_{E}^{\mu\nu}$ and the entropy current $S_{E}^{\mu}$
have identical forms as those shown in
(\ref{PFT}, 
\ref{PFEN})
while a bit of algebra shows that 
\begin{equation}
	A^{\mu \nu \lambda}_E = \frac{1}{2}\beta(4g^{\mu(\nu}\zeta^{\lambda)}-\zeta^\mu g^{\nu \lambda}) + \frac{1}{2}\frac{\partial \beta}{\partial \mu}\zeta^\mu(4\zeta^\nu \zeta^\lambda - \mu g^{\nu \lambda}).
\label{EPF}
\end{equation}
Therefore the  particle number density $n_{E}$, energy density $\rho_{E}$, isotropic pressure $P_{E}$,
as measured relative to the rest frame of the flow are identical to those in 
(\ref{PFQ})
while the fields $(\zeta, \zeta_{\mu})$ are related to the thermal potential $\Theta$, four velocity $u^{\mu}$
 local temperature $T$,
 and entropy per particle $s$
 are as in 
(\ref{FVEL}).
States in equilibrium 
impose additional restrictions 
upon the functions 
$\alpha(\zeta,\mu)$ 
and
$\beta(\zeta, \mu)$
in (\ref{PFGFG}).
These restrictions arise
from the equations:
\begin{equation}
	\nabla_\mu J^\mu_E = \zeta_{\nu} \nabla_\mu  T_E^{\mu \nu} =\zeta_{\nu}\zeta_{\lambda} \nabla_\mu A^{\mu \nu \lambda}_E = 0
\end{equation} 
which yield (for more details see Ref. \refcite{Ger1})

\begin{equation}
	\frac{\partial^2 \alpha}{\partial \mu \partial \zeta}\nabla_m \zeta^m + \frac{\partial^3 \alpha}{\partial \mu \partial \zeta^2}\zeta^m\nabla_m \zeta + \frac{\partial^3 \alpha}{\partial \mu^2 \partial \zeta} \zeta^m\nabla_m \mu = 0,
\label{EQ1}
\end{equation}  

\begin{equation}
\begin{split}
	2\mu\frac{\partial^2 \alpha}{\partial \mu^2}\nabla_m \zeta^m & +  \left[2\mu \frac{\partial^3 \alpha}{\partial \mu^2 \partial \zeta} + \frac{\partial^2 \alpha}{\partial \mu \partial \zeta}\right]\zeta^m \nabla_m \zeta  + \\
	& + 2\left[\mu \frac{\partial^3 \alpha}{\partial \mu^3} + \frac{\partial^2 \alpha}{\partial \mu^2}\right]\zeta^m\nabla_m \mu = 0,
\end{split}
\label{EQ2}
\end{equation}
\begin{equation}
\begin{split}
	\left[3\mu^2\frac{\partial \beta}{\partial \mu} - \mu \beta \right]\nabla_m \zeta^m &+  3\mu \left[\frac{\partial \beta}{\partial \zeta}  + \mu \frac{\partial^2 \beta}{\partial \mu \partial \zeta}\right]\zeta^m\nabla_m \zeta + \\
	& + \left[3 \mu^2 \frac{\partial^2 \beta}{\partial \mu^2} + 6 \mu \frac{\partial \beta}{\partial \mu} + 2\beta \right] \zeta^m\nabla_m \mu = 0.
\end{split}
\label{EQ3}
\end{equation}
These three equations can 
be written in the 
form: 
\begin{equation}
	A^{\nu}{}_{\mu}X^{\mu}=0, \quad X^{\mu}=(\nabla_m \zeta^m ,\zeta^m \nabla_m \zeta , \zeta^m\nabla_m \mu)
\end{equation}
where the $(3\times 3)$ matrix $A$  
can be read from 
equations (\ref{EQ1}-\ref{EQ3})
and its elements
are formed from the derivatives of
the 
functions
$\alpha(\zeta,\mu), \beta(\zeta, \mu)$ defined in  
(\ref{PFGF}, \ref{PFGFG}).
If  these function 
are chosen so
that their derivatives satisfy
$detA\neq0$, then
one concludes that necessary
\begin{equation}
	\nabla_m \zeta^m = \zeta^m \nabla_m \zeta = \zeta^m\nabla_m \mu = 0.
\end{equation} 
 Furthermore, 
the equation $\nabla_{\mu}A_{E}^{\mu\nu\lambda}=0$
yields \cite{Ger1}
\begin{equation}
	\beta(\nabla_\mu \zeta_\nu + \nabla_\nu \zeta_\mu) = 0
\label{Killing_DTT}	
\end{equation}
and 
under
the assumption
that 
$\beta(\zeta, \mu)$
is non vanishing in $U$
it follows that 
 $ \zeta_\mu $ is a timelike Killing\footnote{It is interesting to mention here that
 for the particular class dissipative fluids exhibiting conformal invariance
one finds that 
$\zeta_{\mu}$
satisfies the conformal Killing equation equation
(see discussion further ahead).}
vector field which means that equilibrium states within the class of relativistic
fluid theories of divergent type
are characterized by 
the presence of a timelike Killing and a uniform i.e a constant $ \zeta $.\\
  The so far analysis shows that the space of relativistic fluids of divergence type
it is not empty.
But do there exist 
relativistic dissipative fluids of divergence type
that are physically relevant in the sense that their equations are of symmetric-hyperbolic type
and their evolution respect causality?

The answer to this question  is in the affirmative, and
below we discuss a
family of relativistic fluids that  share this property. 
This family  is a suitable extension of the Eckart theory that we discussed
in section \ref{Eckart}. As we have seen there,
and for states near equilibrium, the particle current $ J^\mu $, stress tensor $ T^{\mu \nu} $ of Eckart's fluids
are described in (\ref{EF}), while the constitutive relations 
are described in (\ref{ECF}). Within the context of a divergent type theory, it is shown in \cite{Ger1} that 
Eckart's theory  is described by:

\begin{align}
	J^\mu & = nu^\mu 
\label{FD_Eckart_1} \\
	T^{\mu \nu} & = \rho u^\mu u^\nu + (P + \pi)\Delta^{\mu \nu} + 2u^{(\mu}q^{\nu)} + \pi^{\mu \nu} \label{FD_Eckart_2}	\\
	A^{\lambda \mu \nu} & = 2Tu^{(\mu}(g^{\nu)\lambda} + u^{\nu)}u^\lambda)
\label{FD_Eckart_3} \\
	I^{\mu \nu} & = -\frac{T}{\hat{\zeta}}\pi^{\mu \nu} - \frac{2T}{3\hat{\zeta_{v}}}(g^{ab} + 4u^\mu u^\nu)\pi - \frac{2}{\kappa}q^{(\mu}u^{\nu)} \label{FD_Eckart_4}\\
	S^\mu & = snu^\mu + \frac{1}{T}q^\mu \label{FD_Eckart_5}\\
	\sigma & = \frac{\pi^2}{\hat{\zeta_{v}} T}+\frac{q^\mu q_\mu}{\kappa T^2} + \frac{\pi^{\mu \nu}\pi_{\mu \nu}}{2\hat{\zeta} T}, \label{FD_Eckart_6}
\end{align}
where $ I^{\mu \nu} $ stands for the dissipation tensor, $(\hat{\zeta}, \hat{\zeta_{v}}) $ are the coefficients of bulk and shear viscosity defined in (\ref{PCE},\ref{ECF}) (here we used hated version 
for these coefficients to avoid confusion with the Lagrange multiplier $ \zeta $),
$u^{\mu}$ is the fluid four velocity,
$\pi$ the bulk pressure, $\pi^{\mu\nu}$ the shear stress, $ q^\mu $ the heat flux and
$T$ the temperature all of them measured relative to the Eckart frame. 
The entropy vector $S^{\mu}$
in (\ref{FD_Eckart_5})
is the same as in 
(\ref{FOT}) for the choice $n^{\mu}=0$ provided 
that we eliminate  the entropy density
$s$ in favor of the entropy density per particle $ sn $ (we use the same symbol for $ s $).\\

In Ref.\refcite{Ger1},
it is shown
that the theory in (\ref{FD_Eckart_1}-\ref{FD_Eckart_6}) can be obtained from the following 
generating function  $ \chi(\zeta, \zeta_\mu, \zeta_{\mu \nu}) $:

\begin{equation}
\chi =\alpha(\zeta, \mu) - \frac{\zeta^{\alpha \beta}\zeta_\alpha \zeta_\beta}{\mu}, \qquad \mu = \zeta^\alpha \zeta_\alpha,
\label{DTE_0} 
\end{equation}
where
$\alpha(\zeta, \mu)$ is an arbitrary smooth function.
Using this
$ \chi(\zeta, \zeta_\mu, \zeta_{\mu \nu}) $ 
and following
similar procedure as for 
the Liu-M\"uller-Ruggeri theory, the
Lagrange multipliers $ (\zeta, \zeta_\mu, \zeta_{\mu \nu}) $ are related to the observable fields $ (n,\rho, P, u^\mu, T, \pi, \pi^{\mu \nu}) $ 
in (\ref{FD_Eckart_1}-\ref{FD_Eckart_6})   via:

\begin{equation}
\zeta = \frac{\rho + P}{nT} - s,
\label{DTE_1} 
\end{equation}
\begin{equation}
\zeta^\alpha = \frac{u^\alpha}{T},
\label{DTE_2} 
\end{equation}
\begin{equation}
\zeta^{\alpha \beta} = \frac{1}{2T^2}\left[\pi^{\alpha \beta} - 2u^{(\alpha}q^{\beta)} + \frac{\pi}{4}(g^{\alpha \beta} + u^\alpha u^\beta)\right],
\label{DTE_3} 
\end{equation}
while the thermodynamical variables like temperature $ T $, particle density $ n $, thermodynamical pressure $ P $ and density $\rho$ are given by: 

\begin{equation}
T^2 = -\frac{1}{\mu},
\label{DTE_4} 
\end{equation}
\begin{equation}
nT= 2\frac{\partial^2 \alpha }{\partial \mu \partial \zeta},
\label{DTE_5} 
\end{equation}
\begin{equation}
P = 2\frac{\partial \alpha}{\partial \mu},
\label{DTE_6} 
\end{equation}
\begin{equation}
\rho + P = -4\mu\frac{\partial^2\alpha}{\partial \mu^2}.
\label{DTE_7} 
\end{equation}
Geroch and Lindblom in Ref.\refcite{Ger1} pointed out that the Eckart theory is 
an
example of a relativistic dissipative
 fluid theory of divergence type, whose dynamical equations
 around the equilibrium state,
fail to constitute
a causal set and thus it cannot be considered to be a satisfactory physical theory
 (this conclusion was expected due to the results in Refs.\refcite{His2,His1}). 
Indeed
for the generating function
$ \chi(\zeta, \zeta_\mu, \zeta_{\mu \nu}) $ defined in
(\ref{DTE_0}),
 the quadratic form $ M^{AB\mu}w_\mu Z_A Z_B $ with arbitrary $ Z_A = (Z, Z_a, Z_{ab}) $ fails to be negative 
 since  
\begin{equation}
	 \frac{\partial^3 \chi}{\partial \zeta_m \partial \zeta_{ab} \partial \zeta_{cd}} 
\label{DTT_C_1} 
\end{equation} 
vanishes identically\footnote{It should be mentioned that the structure of the contraction $ M^{AB\mu}w_\mu Z_A Z_B $ is rather complicated, 
since $ A, B $ run into the interval $ (1,2,..,14) $. But for the case of the $ \chi $ in (\ref{DTE_0}) the vanishing of (\ref{DTT_C_1}) makes the analysis simpler.}  
 since  for the generating function in  (\ref{DTE_0}) is a linear function of $ \zeta_{\mu \nu} $.\\
The advantage of the Pennisi-Geroch-Lindblom formalism is the flexibility that offers
to construct dissipative fluid theories with tractable physical properties and this
flexibility arises from the freedom in choosing the generating function. In that regard, it 
it was noticed in Ref. \refcite{Ger1} that
 by replacing $ \chi(\zeta, \zeta_\mu, \zeta_{\mu \nu}) $ in (\ref{DTE_0}) by:

\begin{equation}
	\chi = \alpha(\zeta, \mu) + \beta(\zeta, \mu)\zeta^{ab}\zeta_a \zeta_b + \chi_2(\zeta, \zeta_a, \zeta_{ab}), 
\label{DTT_ME1}	
\end{equation} 
where $ \chi_2 $ has the form\footnote{Notice that this
$\chi$ in
(\ref{DTT_ME1}	)
combined with (\ref{DTT_ME2}),
it is not the most general from
of the generating function that contains terms quadratic in $\zeta^{ab}$.
For a discussion regarding a more general form see for instance
Ref.\refcite{Nag1,Cal1,Cal2,Cal3,Reu4}.}:

\begin{equation}
	\chi_2 = \frac{\gamma(\zeta, \mu)}{\mu^2}(\mu g_{ab} -2\zeta_a \zeta_b)(\mu g_{cd} -2\zeta_c \zeta_d)\zeta^{ac}\zeta^{bd},
\label{DTT_ME2}		
\end{equation}
results in new theory.
Geroch and Lindblom in Ref.\refcite{Ger1}, investigate the causality property of this theory
by
analyzing the quadratic form
\begin{equation}
	M^{AB\mu}w_\mu Z_A Z_B\quad A,B \in (1,2,3,....,14),
\end{equation} 
and they find that this form is negative definitive for all fluid states
having $ \zeta_{ab} = 0 $ provided the perfect fluid causality
conditions shown in (\ref{DTT_ine_1}) hold and $ \frac{\partial \gamma}{\partial \mu} $ is 
sufficient large. They conclude that the theory is causal for all
sufficiently small $ \zeta_{ab} $ i.e. in some open neighborhood of the equilibrium states, and this conclusion is welcomed. 
It demonstrates that a class of dissipative 
relativistic fluids of divergent  are characterized
by sensible properties
like the causal propagation of disturbances, well posed initial value 
problem, stability of equilibrium states\footnote{This last property is proven in 
Ref.\refcite{Ger1}.} etc.\\
The relativistic (REIT) introduced by Liu-M\"uller and Ruggeri and the 
family of relativistic dissipative fluid theories of divergence type introduced by
Pennisi and Geroch-Lindblom acted as a stimulus 
for further investigations
regarding the properties of this class of fluid theories. 
A large amount of work
in the  literature is centered on
the nature of the generating function $ \chi(\zeta, \zeta_\mu, \zeta_{\mu \nu}) $ 
and 
how the structure of this
function affects the properties of the resulting fluid theory. 
Since equilibrium states are characterized by $\zeta_{\mu \nu} = 0 $, 
therefore states near equilibrium can be studied by
a generating function
$ \chi(\zeta, \zeta_\mu, \zeta_{\mu \nu}) $ whose Taylor series expansion around states in equilibrium
admit  non vanishing 
quadratic contributions in $ \zeta_{\mu \nu} $.
The inclusions of such quadratic contributions 
in $ \chi$ (compare the form of $\chi$ in (\ref{PFGF}))
allows to study the principal part $M^{AB\mu}$ in the dynamical equations
shown in (\ref{FEL})
and 
evaluated this principal part on equilibrium states and thus address issues of causality
for states near equilibrium.
For an explicit representation of such families of generating functions 
see for instance Refs. \refcite{Nag1,Cal1,Cal2,Cal3,Reu4}.\\
 Another aspect of the class of fluid theories of divergent type that has been addressed 
in the literature concerns  the connection between this
class of theories and
relativistic 
 kinetic theory of dilute gases. 
 As we have remarked before,
 for any dilute relativistic gas,
  the first three moments $ \hat{J}^\mu $, $ \hat{T}^{\mu \nu} $ and $ \hat {A}^{\mu \nu \lambda} $ of any distribution function 
 $f(x, p)$ solution of the relativistic Boltzmann equation satisfy: 
 $ \nabla_\mu  \hat {J}^\mu  = \nabla_\mu \hat {T}^{\mu \nu} = 0 $, 
 $ \nabla_\mu \hat {A}^{\mu \nu \lambda}= \hat {I}^{\nu \lambda} $ 
 and these equations look similar
to the standard equations
 (\ref{RET_1}, \ref{RET_2}) 
for the 
Liu-M\"uller-Ruggeri and
Pennisi-Geroch-Lindblom theories.
 However, this similarity is deceiving and in Refs. \refcite{Nag1,Reu4} that question has been discuss at length.
 In general, the second moment of the collision term
 $\hat{I}^{\nu \lambda} $ contains the distribution function
 and thus has different structure than 
 the dissipation tensor $I^{\mu\nu}$
 which depends only upon $J^{\mu}$ and $T^{\mu\nu}$
 and thus at most depends upon the first two moments of the distribution
 function.
In Refs. \refcite{Nag1,Reu4}, the authors
constructed a particular family
of dissipative fluid theories of divergence type 
having the property that the fields 
$J^{\mu},T^{\mu\nu}$ and $A^{\mu\nu\lambda}$
are expressed as moments of a suitable distribution function. 
The resulting system of hyperbolic partial differential equations is very simple and allows one to identify a subclass of manifestly causal theories.\\
Lately, a class of dissipative fluids of divergent type that 
has been analyzed in the literature concerns 
relativistic dissipative fluids that
exhibits  conformal invariance
(for a discussion of 
the implications of this symmetry upon the structure of such fluids consult\cite{Reu3}
 and further references therein).
Physically these fluids could be though as representing the low energy limit of a conformal quantum field theories
(for a discussion on this connection see \cite{Bha}).
In Ref. \refcite{Reu3}, the generating function $\chi$ for this class of fluids
that includes 
second order deviations from equilibrium states
has been constructed. Due to the underlying conformal symmetry, this
$ \chi$
depend only upon 
$( \zeta_\mu, \zeta_{\mu \nu})$.
Equilibrium states within these theories, are states characterized by
the condition
that $\zeta_{\mu \nu}=0$
and in addition the other Lagrange multipler $\zeta_\mu$ is  a conformal Killing vector field.
It was shown in 
Ref. \refcite{Reu3}, that whenever
such equilibrium states are admitted,
then there exist an open set of fluid states around
equilibrium so that that hyperbolicity holds.

\section{Conclusions}\label{S_conclusiones}

We started this article by pointing out that the recent detection of gravitational waves
by the LIGO observatory, the observational evidence for neutron star mergers, the first photograph of the event horizon of the supermassive black hole at
the center of the $M87$ galaxy by (EHT) and the experimental data on heavy ion collisions 
coming from BNL and CERN laboratories,
brought
the field
of irreversible thermodynamics of relativistic continuous media into the frontiers of current research.
The irreversible behavior exhibited by such media is encountered everywhere: from the subnuclear scale,
to the large astrophysical and cosmological scales
and this realization goes well
with Penrose ideas regarding time asymmetric physics.\\
In the course of this review,
 we have mentioned that
many predictions of 
classical (EIT) and (REIT)
 have had experimental confirmation
 and  moreover these theories find applications
in science and technology. For instance in the  
 treatment of heat transport in micro and nano systems, shock structure of waves propagating on 
 hydrodynamical systems, phenomenological hydrodynamics etc, and the interested reader 
 is refereed Refs. \refcite{JVL,Mul4} 
 for overviews successes (and failures)
of these theories.\\
However, at the relativistic level the situation is different. Despite persistence efforts still
an accepted theory of  non equilibrium thermodynamics
of relativistic dissipative fluids 
or more generally continuous media 
is not forthcoming although
big steps have been taken in the right direction. 
The recent efforts (and progress) in the field 
is shifted towards to the construction of theories 
of relativistic fluids
where the dynamical equations constitute a symmetric-hyperbolic
and preferably causal set of dynamical equations.
This shifting in attitude 
results into 
theories
of irreversible thermodynamics that respect causality
at least in a vicinity of states near equilibrium states
and thus these theories eliminate the
unphysical infinite propagation of disturbances
encountered in 
the theories of Eckart and Landau-Lifshitz.\\
The Israel-Stewart transient thermodynamics
 is a flexible theory and 
with additional efforts targeting the
 mathematical structure of the theory, likely to become a 
useful, practical theory.
Even at this level of development, the theory
has been applied to a number of physical problems.
A tractable system where the effects of transient thermodynamics
can be accounted for involves the spatially homogeneous and spatially isotropic cosmological models. 
As a rule, these cosmologies 
postulate that the cosmic fluid
expands
adiabatically
but
there exist processes in the cosmic evolution where this assumption may be 
questioned. In process like the GUT-phase transitions, reheating after inflation,
decoupling of neutrinos or photons from the matter etc.,
the cosmic fluid may be modeled as a dissipative fluid
where various 
thermodynamical variables
exhibit
steep temporal variations so that transient thermodynamics 
is a more suitable theory to
describe
the underlying physics.
Therefore, the dynamics of a spatially homogeneous and isotropic cosmological model coupled to a 
dissipative fluid has attracted the interest of cosmologist. Since the energy momentum tensor of 
a dissipative fluid
has to respects the symmetries of the background geometry, it follows
that such states 
are characterized 
only by a non vanishing  bulk viscosity 
and this property makes 
the coupled Einstein-dissipative fluid system a tractable system to analyze.
For applications of transient thermodynamics
to problems in Cosmology
the reader is refereed to Refs.
\cite{Pav,His3, His4, Bel, RMa1}
and also chapter $(18)$ of 
Ref. \refcite{JVL}.\\
Gravitational collapse also offers many scenarios where transient thermodynamics
finds fertile ground for applications.
One such scenario
corresponds to the phase during the complete gravitational collapse of a star
where the escaping neutrinos pass from the free streaming 
to the trapped regime. During this transition,
again many thermodynamics variables exhibit
rapid spatial and temporal variations
so that transient thermodynamics is applicable.
We are not aware of any treatment of this problem 
within the Einstein-dissipative fluid system
although some initial attempts in that directions have pursued
in Ref. \refcite{Mar1}. Moreover in Refs. \refcite{Sch1,Sch2} transient
thermodynamics applied to radiation fluids (a mixture of ionized matter and photon)
with emphasis to the behavior of density perturbations
and their implications upon the structure formation.\\ 
 It is of interest to mention that in the relativistic version 
 of the Shakura-Sunyaev geometrically thin, optically thick
accretion disks the accreting material is modeled by a viscous fluid
(for an introduction to the physics of these disks see for instance \cite{Tho1}).
The thermodynamical description of this viscous fluid employs the Eckart frame,
and assumes constitutive relations on the stresses and heat flux 
of the conventional Eckart theory.
It would be of some theoretical interest to examine whether 
the equations of transient thermodynamics admit 
disk like solutions
modeling the accreting matter
and whether such solutions (if exist) have any physical relevance.\\
Finally
we should  mention
the
efforts by high energy physicists 
to simulate analytically the quark-gluon plasma
generated in heavy ion collisions.
Since in this setting, the effects of the spacetime curvature are negligible,
many researchers oriented their efforts to construct
analytical (or semi analytical) solutions of the equations of the transient thermodynamics
on a background Minkowski spacetime.
So far only a few families of such solutions have been obtained
and the reader is refereed to Ref. \refcite{Mar1} for progress and further references.\\
As far as 
the status of the Liu-M\"uller-Ruggeri theory and 
the class of relativistic fluid theories 
in the Pennisi-Geroch-Lindblom formalism
are concerned,
we may add that although by design 
these theories can be made causal theories at least in an open set of fluid states around the equilibrium state,
unfortunately it seems that we lack criteria that
single out a universally accepted theory of relativistic dissipation.
The analysis
of 
Liu-M\"uller-Ruggeri theory in
section $12$ and the discussion
 in the last section,
and in particularly the discussion following
equation (\ref{DTT_ME2}),
suggests that there may exist more than one type of divergent type theories that causality holds
for states around the equilibrium one. 
For instance by 
replacing the generating function
$ \chi(\zeta, \zeta_\mu, \zeta_{\mu \nu}) $ in (\ref{DTT_ME2}) 
by a different combination another sensible causal theory can be generated\footnote{
In that regard, we have mentioned earlier on that there exist more general forms of  
$ \chi(\zeta, \zeta_\mu, \zeta_{\mu \nu}) $ that includes quadratic contributions of equilibrium
(see for instance Ref.\refcite{Nag1}).}.
Which one, if any 
of them, is to be considered as the theory preferred by nature?\\
Besides this uniqueness issue,
there exist another problem related to hyperbolic theories of relativistic
dissipation.
That problem arose as a consequence of  results obtained by Geroch and Lindblom 
in Refs \cite{Ger2,Ger3,Lind1}. In these references,
 a more general class of relativistic fluid theories have been introduced that encompass the class of divergent theories
studied in 
section $(13)$. By 
enlarging 
the number of the dynamical fields 
and the corresponding field equations
in a suitable way
the resulting system is a hyperbolic system
and thus physically acceptable.
 What is however perplexing are the results obtained in
Refs \cite{Ger3,Lind1},
regarding the observability
of the 
additional fields
present in 
these  hyperbolic theories.
It was shown in these references that the dynamics of this general class of hyperbolic fluid theories
of relativistic dissipation
are such that the dynamical fields relax on time scales that is 
characteristic of the particle interaction (typically the mean free time between collisions), to 
field configurations 
that are indistinguishable 
from the fields present in the much simpler Eckart theory\footnote{This property,
for some special configurations, has been established in a mathematically rigorous manner
in Ref.\cite{Kre1}.}.\\
The conclusion in  
Refs. \refcite{Ger3,Lind1} has been contested in Refs \refcite{Pav1,Pav2,Pav3}
who argue
that  the physical content  of hyperbolic theories
 is, in general, much broader than those
of the  parabolic type. It is argued in \cite{Pav1,Pav2,Pav3}
that of relevance is 
the relaxation time which depending upon the nature of the system can be large
and thus hyperbolic dissipation can indeed have observable effects.
While the arguments\cite{Pav1,Pav2,Pav3}
are compelling
and in fact Geroch in 
Ref. \refcite{Ger4}  explicitly states that there may exist substances
manifesting dissipation
where hyperbolic theories can be of importance
for the special case of 
Navier-Stokes fluids
\footnote{
Following the notation of Ref. \refcite{Ger4}, in this discussion the term 
Navier-Stokes fluid stands for a collection
of five fields 
 $(\rho, n, u^{\mu}, q^{\mu}, \tau^{\mu\nu})$
 interpreted in the standard way i.e. as the fluid mass density, particle-number density, velocity, heat flow and stress
 with $u^{\mu}$ defining the Eckart frame.
 Moreover, these fields satisfy
 the standard Eckart equations 
 see equations \ref{EF}, \ref{ECF}
 discussed in section $10$.}
that is not the case.
For Navier-Stokes fluids, real or gedanken,
hyperbolic theories may have a chance to be viable 
 provided the equations satisfied by the heat flux and shear tensor 
 break down
 on some length scale much larger than the length scale on which the meanings of its variables, $(\rho, n, u^{\mu}, q^{\mu}, \tau^{\mu\nu})$
 break down. However, Navier-Stokes fluids do not exhibit this kind of behavior (see discussion in \cite{Ger4}).\\
 In the phase of this argument, combined with the decay property obtained in Refs. \refcite{Ger3,Lind1}, the state of affairs regarding dissipative Navier-Stokes fluids is as follows:
For such a fluid, on large enough length scale where the hydrodynamical description is  valid, it seems
that there is one system of equations the Navier-Stokes system (Eckart system in our terminology) 
 that is appropriate for the description of the physics of the fluid, and a second family of systems the hyperbolic systems, that are appropriate for the mathematics. Geroch in \cite{Ger4} states:  \textit{This splitting of the physics and the mathematics is a novel situation, and it takes some getting used to. But, with a little care, ''theories'' of this type can be applied as effectively as more traditional physical theories}.
 (Here, we ought to have in mind that for numerical simulation, presumably hyperbolic theories
 may be of relevance.)\\
  In sum, the description of dissipation in a relativistic fluid  is a challenging problem.
How eventually that issue will be settled is for the moment unknown.\\\
We finish this paper by stating that this
paper targeted theories of extended thermodynamics describing limited class of Newtonian 
continuous media
and relativistic fluids.
As such, the covered material 
is by no means exhaustive
neither we provided a complete 
list of references.
Due to space and time limitations,
we left out of considerations many
types of interesting classes of 
continuous 
media and their thermodynamical behavior.
In particularly,
 we have not  
discussed 
Carter's theory of relativistic heat conduction from a variational view point
(for an overview of this theory see see Refs.\refcite{Car0},\refcite{CMon})
neither 
the structure of a GENERIC which is a 
general framework
aiming 
to analyze general equations for non-equilibrium reversible-irreversible coupling 
(see for instance \cite{Ott1, Hut1, Grm1}). Also we left out of considerations
 theories of classical or relativistic (REIT) for polyatomic gases or (REIT) for dense gases 
(for references of the last two families of theories, 
consult Ref. \refcite{Rug1}).\\
For an overview of the spectrum of 
the existing theories of non equilibrium thermodynamics, like
nonlocal theories of thermodynamics
of continuous media 
and other approaches,
the reader is recommended to consult
Ref.(\refcite{Isr5})
and other
contributed 
articles in Ref. \refcite{AB}.
Also Ref.\refcite{REZ} discusses extensively thermodynamical aspects of relativistic fluids.

\section{Acknowledgments}\label{ACK}

This work arose after multiple discussions that the authors have had
with the members of the relativity group at the IFM Univ. Michoacana
and our warm thanks to all of them. Special thanks to F. Astorga, O. Sarbach, E. Tejeda and U. Nucamendi
for their interest in this work.
The research of T.Z was supported in part by CONACYT Network Project 280908 Agujeros Negros y Ondas 
Gravitatorias and by CIC Grant from the Univ. Michoacana.
J.F.S thanks CONACYT for a predoctoral fellowship.\\

The penetrating comments and criticisms of a reviewer on an earlier version of this paper
are highly appreciated.

\appendix\markboth{Appendix}{Appendix}
\section{On States near Equilibrium}

Since  by design transient  thermodynamics  
deals with states 
that
are close to thermodynamical equilibrium,
it is of relevance to define precisely 
how this class of states is to be identified.
In section $(8)$ and within the context of transient thermodynamics, we identified a special class of  fluid
states that are 
interpreted as states
 in (global or local)  thermal equilibrium.
In this Appendix, we shall
define a class of 
fluid states that are ''close to states in thermal equilibrium".\\
As we discussed in section-$7$,
an
arbitrary state of a simple fluid is described by a set of primary variables
consisting of the conserved and symmetric
energy momentum tensor $T$, a conserved timelike particle current $J$ and the entropy four vector $S$
obeying $\nabla_{\mu}S^{\mu}\geq 0$.
For classical fluids, the energy momentum tensor $T$
defines a
unique timelike eigenvector $u_{E}$ that determines the Landau-Lifshitz (or energy) frame 
while the particle current $J$ 
via $J^{\mu}=n_{N}u_{N}^{\mu}$, 
  determines the
 timelike vector $u_{N}$ that specifies the Eckart frame (or particle frame).
These two vectors
are in general distinct,
unless the state describes a global or local equilibrium.
At 
 any event within the region occupied by this simple fluid,  
 we may choose
 a local Lorentz frame with a time axis parallel to
 $u_{E}$
 and augment this 
 $u_{E}$ with a triad $e_{i},~i \in (1,2,3)$ of spacelike vectors
 so that 
 $(u_{E}, e_{i}),~i \in (1,2,3)$,  
 constitutes an orthonormal tetrad.
 Since
 the triad $e_{i},~i \in (1,2,3)$ is not uniquely defined, 
 we may, without loss of generality, choose the spatial vectors, so that:
 \begin{equation}
 u_{N}^{\mu}=[1-\frac {v^{2}}{c^{2}}]^{-\frac {1}{2}}u_{E}^{\mu}+\frac {v}{c}[1-\frac {v^{2}}{c^{2}}]^{-\frac {1}{2}}e_{1}^{\mu}=cosh\epsilon~u_{E}^{\mu}+sinh\epsilon~ e_{1}^{\mu},\quad g(u_{E}, e_{1})=0,
 \label{ELL}
 \end{equation}
where $\vec v=ve_{1}$ is the ``relative velocity" of the Eckart 
frame relative to the  
Landau-Lifshitz frame
and introduced the parameter $\epsilon$ as the ``pseudo-angle''
between $u_{E}$ and $ u_{N}$.
 From this relation it follows that  the components 
of the particle current take the form
\begin{equation}
 J^{\mu}=n_{N}u_{N}^{\mu}=n_{N}(cosh\epsilon ~u_{E}^{\mu}+ sinh\epsilon~e_{1}^{\mu})=n_{E}u_{E}^{\mu}+\hat j^{\mu},
  \label{PELL}
 \end{equation}
 which means that the particle densities $n_{N}$ and $n_{E}$ as seen in the two frames are related via 
$n_{E}=n_{N}cosh\epsilon$ while 
the particle drift  
$\hat j^{\mu}$
 perceived in the 
Landau-Lifshitz frame
is given by
$\vec {\hat j}=n_{N} sinh\epsilon ~\vec  e_{1}$.  The pseudo angle 
$\epsilon$ between $u_{E}$ and $u_{N}$ satisfies:
\begin{equation}
	cosh \epsilon=-g(u_{E}, u_{N})=\left[1-\left(\frac {v^{2}}{c^2}\right)\right]^{-\frac {1}{2}},
\label{PSA}
\end{equation} 
and since 
in equilibrium (local or global)
$u_{E}$ and   $u_{N}$ coincide and thus $\epsilon=0$, following Israel \cite{Isr1},
it is natural to define states
of the simple fluid\footnote{For a fluid mixture consisting of $n-$particle currents $(J_{1},J_{2},....,J_{n})$, 
one may define $n$-four velocities $(u_{1},u_{2},....,u_{n})$
and thus introduce $n-$pseudo angles
$(\epsilon_{1},\epsilon_{2},....,\epsilon_{n})$ between $u_{E}$ and the corresponding
$(u_{1},u_{2},....,u_{n})$. A state then is close to equilibrium, whenever
$(\epsilon_{1},\epsilon_{2},....,\epsilon_{n})$ 
obey  $\epsilon_{i}\leq 1$ for all $i \in (1,2,.....,n)$.} 
as been close to equilibrium as those states 
having the property that 
everywhere within the region occupied by the fluid
the 
pseudo-angle $\epsilon$ 
in 
(\ref{PSA})
satisfies:
$\epsilon=\frac {v}{c} <<1$.
For such states, 
(\ref{ELL})
and (\ref{PELL}) 
imply 
 \begin{equation}
u_{N}^{\mu}-u_{E}^{\mu} \simeq \epsilon e_{1}^{\mu} =\frac {v}{c}e_{1}^{\mu}=\frac {\hat j^{\mu}}{n_{N}}=
\frac {\nu^{\mu}(u_{E})}{n_{N}}=
\frac {\nu^{\mu}(u_{E})}{n_{E}}=O_{1}<<1,
  \label{NEC}
 \end{equation}
 where 
 following the notation in Ref.\refcite{Isr1},
 we denote the particle drift 
 $\hat j^{\mu}$
 relative to the energy frame 
by $\nu^{\mu}(u_{E})$ i.e. we  set:
$\nu^{\mu}(u_{E}):=\hat j^{\mu}$,
and the symbol $O_{1}$ in (\ref{NEC})
(and here after)
signify
a term of first order deviations from equilibrium\footnote{We follow closely the notation
 of ref.\refcite{Isr1}. Thus here after 
 $O_{1}, O_{2}, O_{3}.....$ 
 signify terms of first, second, third order... deviations from equilibrium.
 Since
 the phenomenological laws of transient thermodynamics 
are invariant
 under first order change of the rest frame,
  the relation
$n_{E}=n_{N}coshe=n_{N}+(O_{1})^{2}$, 
implies that 
the difference between 
$n_{E}$ and $n_{N}$ is 
second order in deviation from equilibrium
and thus  the distinction between $n_{E}$ and $n_{N}$
(as well as between $\rho_{E}$ and $\rho_{N}$) will be gradually blurred.
That means that  (\ref{NEC}) can be written in the equivalent form
$
u_{N}^{\mu}-u_{E}^{\mu} \simeq \frac {\nu^{\mu}(u_{E})}{n_{E}}=O_{1}<<1
$.} like $\pi(u_{E})^{\mu\nu}, \pi(u_{E}),\nu(u_{E})^{\mu})$ etc.\\
Beyond 
 validity of (\ref{NEC}), states close to equilibrium
are required to satisfy another constrain:
 the viscous stresses, denoted collectively by
  $\tau^{\mu\nu}$,
should be small in the sense
   \begin{equation}
\frac {\tau_{E}^{\mu\nu}}{\rho_{E}}=O_{1}<<1,
 \label{SNEC}
 \end{equation}  
 where  
 $(\tau(u_{E})^{\mu\nu},  \rho(u_{E}))$ are the components
 of the viscous stress and
  the energy density
 as measured relative to the energy frame.
 For the motivation behind this condition, 
 see for instance discussion in \cite{Isr1} as well as the  Appendix C of the present paper.\\ 
The condition (\ref{NEC}) imply that
on the tangent space of each event within the fluid,  an invariant 
``cone" of opening pseudo-angle $\epsilon \simeq \frac {v}{c}=O_{1}<<1$ with $v$ the speed of the Eckart frame
relative to the energy frame can be defined
and this invariant ''cone"  plays an important
role in the description of states near equilibrium.
Any four velocity\footnote{In view of the
comment $(e)$ on page $26$, this four velocity
$u^{\mu}$ could be chosen also as potential four velocity of the fluid.}
$u$ that falls within this cone
can be used as a potential admissible  rest frame 
and observers at rest relative to this frame
determine thermodynamical variables associated with the 
non equilibrium state.
Even though the measured  thermodynamical variables
have a dependance upon 
the four velocity $u$
and 
despite the plurality of these rest frames,
nevertheless a consistent thermodynamical theory can
be developed that is
manifestly invariant under first-order changes of the rest-frame $u$, 
i.e. change in the rest frame described by 
\begin{equation}
	u^{\mu}\mapsto \hat u^{\mu}=u^{\mu}+\delta^{\mu},\quad \delta^{\mu}\leq O_{1}.
\end{equation}
Under this class of frame changes  many thermodynamical variables
that are 
frame dependent,
transform in a well defined manner (see Appendix D, for a derivation of 
such transformation). 
As an example, notice that
relation (\ref{PELL})
implies
that the particle densities
$n_{E}$ and $n_{N}$ obey
$n_{E}-n_{N}=O_{2}$  
and thus to the first order deviations from equilibrium
the measured particle densities are independent
whether the energy frame or the particle frame is employed
(provided we neglect $O_{2}$ and higher order contributions). 
As we shall show in Appendix D, an analogous property holds
for many of the observables measured 
by observers with corresponding four velocities
 $(u, \hat u)$ 
both future pointing and both lying within 
the cone
spanned by $(u_{E},u_{N})$.
In this case the corresponding transformations properties of the thermodynamical variables
are discussed in the Appendix D.\\  
We briefly mention here that the cone of opening angle $\epsilon$ 
offers the means\footnote{In the Appendix C, we introduce another manner to 
identify states  that are ''close'' to an equilibrium state.}
 to identify a reference equilibrium state
 (actually a whole class of such states) specified by $(S^{\mu}_{(0)}, J^{\mu}_{(0)}, T^{\mu\nu}_{(0)})$
which is''close'' to an actual off-equilibrium fluid state described by the primary variables
$(S^{\mu}, J^{\mu}, T^{\mu\nu})$. The details of this identification are
discussed in section $9$.

\section{Implementation of the entropy principle - Liu's and Ruggeri's procedures}
As we have seen in the main sections of the paper, the entropy principle plays an important role in the description of the thermodynamical properties  
of continuous media irrespectively whether such media are treated classically or relativistically.
We recall that the principle states that the dependance of any 
constitutive function upon a set of basic variables
should be such that every solution of 
 the balance laws
should satisfy 
the entropy inequality (or to put matters differently should obey the second law).\\
We also pointed out
the need for procedures that lead into the implementation of this principle.
Furthermore, in the sections $(12), (13)$
the fields of the Lagrange multipliers
appeared as means to implement 
this principle
and in that regard the origin and significance 
of these multipliers were
rather mysterious.\\
This Appendix is dedicated on the problem of implementing the entropy
principle
and in that regard, we 
 discuss two related algorithms for its implementation.
The first one invented by Liu
and is known as Liu's procedure
while the second one introduced by Boillat, Ruggeri and coworkers.
Although
in both of these methods the field of the Lagrange multipliers appers,
the latter approach 
provides a more transparent interpretation of these multipliers.\\
We begin by outlining first 
 Liu's \cite{Liu} procedure
 and in this procedure the following Lemma is of key importance:
\begin{lemma}
Let $A=A_{\Delta \gamma}$,\quad $\Delta \in (1,2,.....p)$,\quad $\gamma \in (1,2,....n)$
be a real $(p\times n)$ matrix and $\vec X=(X^{1}, X^{2}, ....X^{n})\in R^{n}$,\quad $B=
(B^{1}, B^{2}, ....B^{p})\in R^{p}$\quad
$\vec \alpha=(\alpha^{1},\alpha^{2},....\alpha^{n})\in R^{n}$.
Assume that 
$\vec \alpha \neq0$ and $ \beta \in R$ are given
and moreover the following set:
\begin{equation}
S= \left\{ \vec X\in R^{n} ~\Big\vert ~ A_{\Delta\gamma} ~X^{\gamma}+B_{\Delta}=0 \right\},
\label{SET}
\end{equation}
is non empty.
Then the following statements are equivalent:\\

(1) For all $\vec X \in  ~S$ it holds: $\alpha_{\gamma} ~X^{\gamma}+\beta \geq 0$\\

(2) There exist  $\vec \Lambda=(\Lambda^{1}, \Lambda^{2}......\Lambda^{p}) \neq 0\in R^{p}$, refereed as Lagrange multipliers, such that:
$$\alpha_{\gamma} ~X^{\gamma}+\beta-\Lambda^{\Delta}( ~A_{\Delta\gamma} ~X^{\gamma}+B_{\Delta})\geq0 \quad
\forall ~\vec X ~\in R^{n}$$\\

(3) There exist 
 $\vec \Lambda=(\Lambda^{1}, \Lambda^{2}......\Lambda^{p}) \neq 0\in R^{p}$ such that
$$\alpha_{\gamma}-\Lambda^{\Delta} ~ A_{\Delta\gamma}=0,\quad \beta-\Lambda^{\Delta} ~B_{\Delta}\geq 0,\quad
\gamma \in (1,2,....n) $$
\end{lemma}

As it stands, it is not clear how this Lemma is related to the entropy principle. Below, we shall outline briefly this connection
and often refer the reader to the original article
by Liu in Ref.\refcite{Liu} for a detailed discussion, proofs and applications of this Lemma.\\

A key element that connects this Lemma to
the entropy principle
is the notion of  \textit{admissible sets}.
Following Liu's approach, these sets 
are defined first for a system of second order quasilinear partial differential equations of the form:
\begin{equation}
G_{A}(x^{\mu}, U_{\hat a}, U_{\hat a,\mu}, U_{\hat a,\mu\nu})=0,\quad A\in (1,2,......M),\quad \hat a \in (1,2,....N),
\label{L1}
\end{equation}
accompanied by an inequality 
\begin{equation}
I(x^{\mu}, U_{\hat a}, U_{\hat a,\mu}, U_{\hat a,\mu\nu})\geq0,\quad \mu, \nu \in (0,1,2,3).
\label{L2}
\end{equation}
The system is defined in an open set $D$ of  $ R^{4}$,
and $U_{\hat a}$ stand  for $N$- unknown thermodynamical fields
while
$(U_{\hat a,\mu}, U_{\hat a,\mu\nu})$ 
denote their first and second partial derivatives\footnote{It is understood that the $M$-function $G_{A}$ and $I$
in (\ref{L1}, \ref{L2}) are smooth bounded functions of their arguments.
Notice also that in the terminology of thermodynamists any solution $V_{\hat a}$ of (\ref{L1}) is refereed as a thermodynamical process.
For the system 
(\ref{L1}, \ref{L2}), we  may interpret (\ref{L1}) as a set of balance laws while (\ref{L2}) as an ''entropy like inequality''.
In the present context, the entropy principle dictates that
the function $I$ in
(\ref{L2}) should be  chosen so that for
any thermodynamical process
(\ref{L2}) must hold.}
with respect to the local coordinates $x^{\mu}$ covering $D$.\\
 Let now an arbitrary point $p$ in $D$
 and let a set of real constants:
 \begin{equation}
\left\{ A_{\hat a},~ ~  B_{\hat a\mu},~ ~  C_{\hat a\mu\nu},\quad \hat a \in (1, 2,....,N),\quad  \mu, \nu \in (0, 1, 2, 3)
 \right\},
\label{ADMS}
\end{equation} 
chosen so that there exists an open vicinity $O$ of $p$ in $D$
and a process 
$V_{\hat a}(x)$ (i.e solution of (\ref{L1})) such that
 \begin{equation}
\left\{ V_{\hat a}(p)=A_{\hat a}, ~  V_{\hat a , \mu}(p)=B_{\hat a \mu}, ~ V_{\hat a ,\mu\nu}(p)=  C_{\hat a \mu\nu}, ~  \hat a
 \in (1, 2,....,N),  ~ \mu, \nu \in (0, 1, 2, 3) \right\}.
\label{ADMSF}
\end{equation} 
In that event, the set
(\ref{ADMS}), provided non empty, constitutes an admissible set for the system (\ref{L1}) at $p\in D$.
The punch line
in Liu's procedure 
for the implementation of the entropy principle, is the observation that if 
one demands validity of 
the entropy principle, 
i.e. one demands that 
for any process 
$V_{\hat a}(x)$ of 
(\ref{L1}) over $D$,
the inequality
(\ref{L2}) must hold,
then (\ref{L2}) must hold
for any admissible set\footnote{This presupposes
that the system
(\ref{L1})
admits  admissible sets.
Liu in Ref.\refcite{Liu} employed
the Cauchy-Kowalewski theorem (and thus employed analyticity)
to demonstrate 
that
(\ref{L1}) admits  admissible sets.
However, as he also stated, 
this conclusion can be reached 
under much weaker conditions such
as 
the Cauchy problem for (\ref{L1}) 
is well posed.} 
of (\ref{L1}) and at any $p \in D$.
The existence of admissible sets at $p \in D$ 
implies
that for suitably defined matrix $A$ and vector $B$ related to (\ref{L1}, \ref{L2}),
the set $S$ in the above Lemma 
is non empty
which in turn
allows to take advantage of conditions $ (2), (3)$ of that Lemma
and thus demonstrate the existence of the Lagrange multipliers
$\vec \Lambda=(\Lambda^{1}, \Lambda^{2}......\Lambda^{p}) \neq 0$
defined initially over $p \in D$.
These multipliers are subsequently extended as fields
over the open vicinity $O$ of $p$ in $D$.
By appealing to the above Lemma, condition $(1)$ is replaced by conditions $(2)$ and $(3)$
and this replacement sets restrictions upon dependance of the Lagrange multipliers upon
the
$U_{\hat a}$ fields (and possibly their derivatives).
\\
 Liu in 
  Ref.\refcite{Liu}, 
 applied this procedure
 to 
 a simple heat conducting, viscous fluid.
  For this system the basic\footnote{Recall that by the term basic variable, we mean a set of variables
  that have the distinct property that any other constitutive
  function should be depending upon these variables.} variables are considered to be
 the density $\rho(t,\vec x)$, the 
 components $u^{i}(t,\vec x), ~ i\in (1,2,3)$ of the velocity field and the empirical 
 temperature $T(t, \vec x)$.
 The balance laws
of 
mass, momentum and internal energy, 
were obtained in sections $(2-4)$,
and contain the components of the Cauchy stress tensor $\sigma_{ij}$, the heat flux vector $q^{i}$
 and internal energy $e$
 which are 
considered to be constitutive functions
i.e. smooth functions   of 
 $(\rho, u^{i}, T, \dot T, T,_{i})$.
 Here 
$\dot T$ stands for
$$ 
 \dot T:=\frac {\partial T}{\partial t}+u^{i}\frac {\partial T}{\partial x^{i}}:=T,_{t}+u^{i}T,_{i},\quad 
 T,_{i}:=\frac {\partial T}{\partial x^{i}},\quad i \in (1,2,3),
 $$
 and for typographical convenience 
 we have suppressed the dependance 
 of the variables $(\rho, u^{i}, T, \dot T, T,_{i})$
 upon the spacetime coordinates.
  For this fluid, the entropy inequality in (\ref{MCDI}) is written in the equivalent form
  \begin{equation} 
\rho \frac {\partial s}{\partial t}+
\rho u^{i}\frac {\partial s}{\partial x^{i}}
+ \frac {\partial J^{i}}{\partial x^{i}}\geq 0
\label{MCDII}
\end{equation} 
 where both $s$ and $\vec J$ are constitutive functions
 i.e.
 $$
 s=s(\rho, u^{i}, T, \dot T, T,_{i}),\quad J^{i}=J^{i}(\rho, u^{i}, T, \dot T, T,_{i}). 
$$ 
Once the dependancies of
 $(\sigma_{ij},q^{i}, e)$ and $(s, J^{i})$
upon
the
basic variables have been specified,
they are substituted
in the balance laws 
and the entropy inequality.
Liu observes that the resulting system although
rather lengthy, it can be written in the following compact form:
\begin{equation}
A_{\Delta\gamma}(y_{\lambda}) ~X^{\gamma}+B_{\Delta}(y_{\lambda})=0 ,\quad  \alpha_{\gamma}(y_{\lambda}) ~X^{\gamma}+\beta(y_\lambda) \geq 0,\quad \Delta \in (1, 2,...5).
\label{LiuS}
\end{equation}
where 
the nine components $y_{\lambda}$ of $ \vec y$ are
$$y_{\lambda}=(\rho, u^{i}, T,
\frac {\partial T}{\partial t},\frac {\partial T}{\partial x^{i}}),
\quad i\in (1,2,3), \quad \lambda \in (1,2,3,....9),$$
the $5$-components of $B_{\Delta}$ and the constant
$\beta$ are given by
$$
B_{\Delta}=(0,~ -\frac {1}{\rho} \frac {\partial \sigma_{ij}}{\partial T} \frac {\partial T}{\partial x^{j}}~ ,~ 
\rho \frac {\partial \epsilon }{\partial T} \frac {\partial T}{\partial t}+\frac {\partial q^{i}}{\partial T}\frac {\partial T}{\partial x^{i}}),
\quad
\beta=\rho \frac {\partial s}{\partial T}\frac {\partial T}{\partial t}+\frac {\partial J^{i}}{\partial T}\frac {\partial T}{\partial x^{i}},
$$
while the $(5 \times 26)$ dimensional matrix $A_{\Delta \gamma}$
and the $26$ components $X^{\gamma}$ of the vector $\vec X$
in (\ref{LiuS})
are very complicated long expressions whose explicit form can be found in 
Ref.\refcite{Liu}.\\ 
Based on these expressions, Liu demonstrates the existence of admissible sets
and thus at a fixed point $p\in D$ the set $S$ in Liu's Lemma is non empty
and thus he infers the existence of the Lagrange multipliers.
Based on the conditions $(2),(3)$ of this Lemma
he deduces restrictions upon the structure of the Lagrange multipliers
$(\Lambda^{\rho}, \Lambda^{u^{i}},\Lambda^{e})$.
We shall not discuss any further this example (the reader who is interested 
for more details 
is refereed to the original article by Liu in 
 Ref.\refcite{Liu}). Instead, below, we 
 discuss an alternative procedure for implementing
 the entropy principle.\\
  In this alternative procedure,
  the Lagrange multipliers
 re-appear but from the perspective
of the mathematical theory of hyperbolic systems. This 
alternative procedure
invented after the 
 recognition by Boillat in Ref.\refcite{Boi} and Ruggeri-Strumia in Ref.\refcite{RugStr}
 that whenever the Lagrange multipliers 
in Liu's procedure are chosen as thermodynamical variables
(and under certain additional requirement upon the entropy vector like convexity), then the field equations (i.e. the balance laws)
can be turned into 
a symmetric, hyperbolic system.
Various aspects of the Boillat,
 Ruggeri-Strumia 
procedure 
have  developed in Refs.
 \refcite{Boi},
\refcite{RugStr},
 \refcite{RugF1}, \refcite{RugF2},  
\refcite{RugF3}. For completeness purposes, below, we briefly outline this powerful method
and follow closely the treatment in Ref.\refcite{RugStr}.\\
 
 The 
 Boillat-Ruggeri-Strumia procedure pre-assumes a medium that is governed by $n-$balance laws of  the form:
\begin{equation}
\partial_{\alpha} {\vec F}^{\alpha }(\vec u)=\vec f(\vec u),\quad \alpha \in (0,1,2,3), 
\label{FOQL}
\end{equation} 
where $ \vec F^{\alpha}$ for $\alpha \in  (0,1,2,3) $ stand for four column of $n$-components:
$$\vec F^{0}:=(F_{1}^{0},
F_{2}^{0},..........,F_{n}^{0})^{T},\quad 
\vec F^{1}:=(F_{1}^{1},
F_{2}^{1},..........,F_{n}^{1})^{T}, etc$$ 
and 
$$\vec u(t, \vec x):=(u_{1}(t, \vec x),
u_{2}(t, \vec x),..........,u_{n}(t, \vec x))^{T},\quad 
\vec f(\vec u):=(f_{1}(\vec u),
f_{2}(\vec u),..........,f_{n}( \vec u))^{T}$$ 
and in above $T$ signifies transpose.

The $n-$functions 
$\vec u(t, \vec x):=(u_{1}(t, \vec x),
u_{2}(t, \vec x),..........,u_{n}(t, \vec x))^{T}$, 
are assumed to be the basic variables describing the medium under consideration and
thus 
(\ref{FOQL}) stands for a system of $n$- equations for the $n$-unknowns functions
$\vec u(t, \vec x).$\\
The system
(\ref{FOQL}) is accompanied by a 
supplementary equation, interpreted as an entropy law, of the form
\begin{equation}
\partial_{\alpha} {S}^{\alpha }(\vec u)=\Sigma (\vec u),\quad \alpha \in (0,1,2,3), 
\label{SCOL}
\end{equation} 
where $S^{\alpha}$ are the components of a $4$-dim. vector field
and
$\Sigma$ is a scalar, both 
 depending  smoothly upon $\vec u(t, \vec x)$.
The inclusion of this additional equation
implies that (\ref{FOQL}) and
(\ref{SCOL}) becomes an overdetermined system of $(n+1)$ equations for
$n$ unknowns.\\
 Within the field
of the thermodynamics of continuous media,
overdetermined systems of ''conservation laws'' are encountered frequently.
The dynamical equations for the Fourier-Navier-Stokes simple fluid can be cast in the form
 (\ref{FOQL},
\ref{SCOL}) (see for instance   
Refs. \refcite{Rug1}, \refcite{Rug4})
while other physical systems described by systems of the form
(\ref{FOQL},
\ref{SCOL}) are discussed in 
ref. (\refcite{Mul4}). Moreover as we have already seen in sections 
$(12-13)$, the dynamical equations of the 
Liu-M\"uller-Ruggeri and the 
family of relativistic dissipative fluids of divergence type
are particular cases of the 
systems
(\ref{FOQL},
\ref{SCOL}).
For these theories, 
the unknowns $(u_{1}(t, \vec x),
u_{2}(t, \vec x),..........,u_{14}(t, \vec x))$
are identified with the $14$ unknown components  of $J^{\mu}$ and $T^{\mu\nu}$
and these $14$ variables satisfy $15$ equations
(see eqs (\ref{RET_1}, \ref{RET_2}, \ref{RET_5}) in the main text).
Thus  it is worthwhile to briefly describe the features of systems
described by 
(\ref{FOQL}, \ref{SCOL}).\\
In general one does not expect
that overdetermined system of equations to admit solutions.
However under suitable restrictions, Friedrichs in 
Refs. \refcite{Fried0}, \refcite{Fried1} (see also Ref.\refcite{Fried2}) has shown that such systems can be 
turned into symmetric hyperbolic system of equations.
That will be the case 
provided a ''main dependency relation\footnote{In the following, we employ the terminology and notation
of  (\ref{RET_1}, \ref{RET_2}).}'' holds (see equation 
(\ref{FRIE1}))
and in addition a certain quadratic form is positive definite (see relation (\ref{QF})).\\

If one rewrites the system 
(\ref{FOQL},
\ref{SCOL}) in the form:
$$
(\partial_{\alpha} \vec F^{\alpha}, \partial_{\alpha} S^{\alpha})^{T}=(\vec f, \Sigma)^{T},
$$ 
and denote by a dot $ \cdot$
the Euclidean inner product in $R^{n+1}$,
then Friedrich 
''main dependency relation'' 
holds provided
that there exists\footnote{In a practical problem one may appeal to Liu's lemma
to establish the existence of the multipliers fields
$(\Lambda^{1}(\vec u), \Lambda^{2}(\vec u),......,\Lambda^{n}(\vec u))$
and then proceed to check of Friedrich's conditions. For an example see
Ref. (\refcite{FT1}).}

 an $(n+1)$ dimensional vector field 
$\vec Z(\vec u)$ that satisfies
\begin{equation}
\vec Z\cdot (\partial_{\alpha} \vec F^{\alpha}, \partial_{\alpha} S^{\alpha})^{T}=\vec Z \cdot (\vec f, \Sigma)^{T},
\label{FRIE1}
\end{equation}
and this equation should hold for all $\vec u$ and $\partial_{\alpha} u^{\beta}, \alpha \in (0,1,2,3), \beta \in (1,2,...,n)$.
Choosing 
the field $\vec Z$ in the form: $\vec Z =(-\vec \Lambda(\vec u), 1)$
where: $\vec \Lambda (\vec u)=(\Lambda^{1}(\vec u), \Lambda^{2}(\vec u),......,\Lambda^{n}(\vec u))$
with
 the $\Lambda^{i}, i\in (1,2,...n)$
 depending smoothly upon $\vec u(t,\vec x)=
 (u_{1}(t, \vec x),
u_{2}(t, \vec x),..........,u_{n}(t, \vec x))$,  
 it follows that
 (\ref{FRIE1})
 is equivalent to
 \begin{equation}
-\vec \Lambda \cdot \partial_{\alpha} \vec F^{\alpha} +\partial_{\alpha} S^{\alpha}=
-\vec \Lambda \cdot  \vec f+\Sigma,
\label{CC}
\end{equation}
and this relation shows that $\vec \Lambda(\vec u)$ plays the same role of the Lagrange multipliers 
in the Liu's approach.\\
Introducing the $n$-dimensional gradient operator $\hat \nabla$ via
$$
\hat \nabla=(\frac {\partial}{\partial u^{1}},\frac {\partial}{\partial u^{2}},........,\frac {\partial}{\partial u^{n}})
$$
and requiring that  condition (\ref{CC}) to hold for all
$\vec u$ and $\partial_{\alpha} u^{\beta}$
it follows

$$
\vec \Lambda\cdot \hat \nabla {\vec F}^{\alpha}=\hat \nabla S^{\alpha}, \quad \vec \Lambda \cdot \vec f =\Sigma.
$$ 
Making use of the $n$-dimensional exterior differential operator $d$ 
in Euclidean $R^{n}$, these conditions can be written in the equivalent form
\begin{equation}
\vec \Lambda {\cdot} d\vec F^{\alpha}=dS^{\alpha}\quad 
\vec \Lambda {\cdot} \vec f=\Sigma 
\quad\alpha \in (0,1,2,3). 
\label{COM}
\end{equation} 
These two equations are consequence of Friedrich's 
''main dependency relation''  and would be worth to have
an understanding of their deeper significance. We shall not address their
deeper significance in this work, but we shall exploit some of their consequences.\\

Suppose that in
(\ref{FOQL})
it is possible
to choose: $\vec u=\vec F^{0}$.
Then for this choice,
(\ref{COM}) implies that
\begin{equation}
\frac {\partial S^{0}}{\partial {\vec u}}=\vec \Lambda,\quad  \frac {\partial^{2} S^{0}}{{\partial {\vec u}}{\partial {\vec u}}}=
\frac {\partial {\vec \Lambda}}{\partial {\vec u}},
\label{COMM}
\end{equation} 
and this is a set of important relations.\\
Suppose
for this particular choice i.e.
$\vec u=\vec F^{0}$, the entropy density\footnote{This
property is well defined for the case of Newtonian media. However for relativistic media
issues of covariance are hidden in this property.}
$S^{0}(\vec u)$ is a concave function of $\vec u$ i.e.
the matrix  
$\frac {\partial^{2} S^{0}}{{\partial {\vec u}}{\partial {\vec u}}}$ is negative definite,
clearly
from the above relations it follows that the matrix 
$(\frac {\partial {\vec \Lambda}}{\partial {\vec u}})$ is also negative definite.
By appealing to standard theorems on Jacobian matrices, one concludes that the transformation
$\vec u \to \Lambda(\vec u)$ is globally invertible.
 In turn this property allows to view
\begin{equation}
{S}^{\alpha}=S^{\alpha}(\vec \Lambda),\quad \vec F^{\alpha}= \vec F^{\alpha}(\vec \Lambda),\quad \vec f=\vec f(\vec \Lambda).
\label{NDV}
\end{equation}
Combining these relations with
(\ref{COM}) one concludes that 
 exist
four scalars $S'^{\alpha}, \alpha \in (0,1,2,3)$ such that 
\begin{equation}
{\vec F}^{\alpha} \cdot d\Lambda^{\alpha}=dS'^{\alpha},\quad
S'^{\alpha}=\vec \Lambda \cdot \vec F^{\alpha}- S^{\alpha}, 
\quad \Longrightarrow {\vec F}^{\alpha}=\frac {\partial S'^{\alpha}}{\partial \vec \Lambda}.
\label{IMRE}
\end{equation}
Thus the system 
(\ref{FOQL})
 is compatible with the 
complementary law
(\ref{SCOL}), 
provided the components $\vec F^{\alpha}$ are the gradient of $S'^{\alpha}$
with the later fields viewed as functions of the field $\vec \Lambda$.\\
Returning to the system
(\ref{FOQL}) and using 
the representation of
${\vec F}^{\alpha}$ shown in
(\ref{IMRE}),
one gets
\begin{equation}
\frac {\partial ^{2} S'^{\alpha}}{\partial {\vec \Lambda}{\partial {\vec \Lambda}}}\partial_{\alpha}\vec \Lambda=\vec f(\vec \Lambda),\quad\alpha \in (0,1,2,3),
\label{FSH}
\end{equation}
where
 the four $(n\times n)$
matrices: 
$$
 A^{\alpha}{}_{ij}=\frac {\partial ^{2} S'^{\alpha}}{\partial {\Lambda^{i}}{\partial {\Lambda^{j}}}},\quad i, j \in (1,2,....n),
$$ 
are the Hessian matrices of $S'^{\alpha}(\vec \Lambda)$ (with respect to the components of $\vec \Lambda)$
and thus by construction are all symmetric\footnote{ Notice that this property does not hold on general 
whenever the components of the field $\vec u$ are employed as basic variables.}.\\
Beyond the manifestly symmetric property of 
(\ref{FSH}), the system
has another remarkable 
property. As long as the entropy density
$S^{0}$ is a concave function of $\vec u(t,\vec x)$,
the system
(\ref{FSH}) is a symmetric-hyperbolic system. The hyperbolic nature
can be established either by appealing to  
Friedrichs second condition or
checking directly whether 
 the definition of 
a symmetric-hyperbolic system of equations hold true for the system (\ref{FSH}).\\
We recall that Friedrichs second condition 
requires that there exist a co-vector
$\xi=(\xi_{0}, \xi_{1},\xi_{2},\xi_{3})$ so that 
 the quadratic form
\begin{equation}
Q=-\xi_{\alpha}[\vec \Lambda {\cdot} \delta^{2} \vec F^{\alpha}+\delta^{2} S^{a}]>0,
\label{QF}
\end{equation}
is positive definite for all smooth variations $\delta u(t, \vec x)$ of a background $\vec u(\vec x,t)$.
 Validity of this condition combined with
 (\ref{COM}) yields the desired conclusion i.e. the
 symmetric-hyperbolic nature of
 (\ref{FSH}).\\
Notice however, that we  can reach the same conclusion
by appealing to the second alternative
i.e. checking directly whether  Friedrichs definition of a symmetric-hyperbolic system holds true.\\
 Writing the system
(\ref{FSH}) in the form  $A^{\alpha}{}_{ij}\partial_{\alpha}\Lambda ^{i}=f_{j}$, then this is a symmetric-hyperbolic system in 
the sense of Friedrich, provided the four matrices
$A^{\alpha}{}_{ij}$ are symmetric i.e. $A^{\alpha}{}_{ij}=A^{\alpha}{}_{ji}$ for all $\alpha \in (0,1,2,3)$
and in addition a ''positivity condition'' should hold.
This later condition requires that there should exist co-vectors
$\xi=(\xi_{0}, \xi_{1},\xi_{2},\xi_{3})$ so that the matrix $\xi_{\mu}A^{\mu}_{i j}$
should be positive definite. For the system
(\ref{FSH}),
we have already seen that the four matrices 
$A^{\alpha}{}_{ij}$ are all symmetric. To establish positivity, it is sufficient
to choose the co-vector
 $\xi=(\xi_{0}, \xi_{1},\xi_{2},\xi_{3})=(1,0,0,0,0)$. For this choice,  
 positivity holds provided  the matrix
$$ \frac {\partial^{2} S'^{0}}{{\partial {\vec \Lambda}}{\partial {\vec \Lambda}}}$$
is definite. This conclusion however follows by recalling that 
for the choice
$\vec u=\vec F^{0}$
the relation $S'^{\alpha}=\vec \Lambda \cdot \vec F^{\alpha}- S^{\alpha}$ implies that
 $S'^{0}=\vec \Lambda \cdot \vec u- S^{0}$ and thus  $S'^{0}$
is the Legendre transform of
$S^{0}$. This implies that $S'^{0}$ is a concave function of $\vec \Lambda$ since $S^{0}$
has been assumed to be a concave function of $\vec u$. This conclusion proves that the symmetric system
(\ref{FSH}) is hyperbolic.\\
In the terminology adopted by Ruggeri and collaborators, the four fields $S'^{\mu}(\vec \Lambda)$
are refereed as the generator field, while the Lagrange multipliers $\vec \Lambda(t, \vec x)$
constitute the main field. From the so far analysis, it follows that if one would be able to identify 
the generator field $S'^{\mu}(\vec \Lambda)$, then the right constitutive relations i.e.
the functions $F^{\alpha}$ 
compatible to the entropy inequality
(\ref{SCOL}),
subject to $\Sigma (\vec u)\geq 0$,
are those obtained 
via differentiating the generator with respect to the components of the main field.
Here we see a more concrete implementation of the entropy principle. The dependance of the constitutive functions
$F^{\alpha}$ upon the basic variables, 
taken here as the Lagrange multipliers, should be special in order to be compatible with 
the entropy inequality.\\
As far as applications of the above results are concerned, it is suffice to stress
that of relevance here is the construction of the 
generator field i.e. $S'^{a}, \alpha \in (0,1,2,3)$.
For the case of the  Liu, M\"uller and Ruggeri theory,
developed in section $(12)$,
 the fields $S'^{a}, \alpha \in (0,1,2,3)$
are defined in (\ref{BasR}). 
Approximate  expression for  $ S'^\alpha $ 
as a function of  the Lagrange multipliers (see
eq.(\ref{GEN})),
has been constructed
by appealing  on the 
 principle of relativity.
For the case of the 
dissipative relativistic fluids of divergence type developed in section
$(13)$, the approach was different.
There the components 
 of $S'^{a}$
are the gradients of the 
smooth generating scalar function $ \chi (\zeta, \zeta_\mu, \zeta_{\mu \nu}) $
i.e. $S'^\alpha = \frac{\partial \chi}{\partial \zeta_{\alpha}}$ 
and this formula makes clear
the economy in the
Pennisi-Geroch-Lindblom formalism.\\
It should be mentioned here that in both of the above  cited approaches the analysis is local.
The transformation $\Lambda^{i} \to u^{i}$ employed in the
 Liu, M\"uller and Ruggeri theory and also for the fluids of divergence type,
 is only locally invertible and this local invertibility
is a consequence of Friedrichs ''main dependency relation''
and the positive definite character of the quadratic form shown in (\ref{QF}).
Validity of these two conditions lead to a local invertibility
of the  transformation $\Lambda^{i} \to u^{i}$ (see discussion in
Ref. (\refcite{FT1})).\\

\section{On the thermodynamics of relativistic continuous media}

In this Appendix, we discuss 
a few basics aspects of the thermodynamics
of relativistic continuous media.
Because such media exhibit a great diversity and their thermodynamical description
requires a great deal of the theory of constitutive
relations and the entropy principle, therefore
in this Appendix,
we only offer a few remarks concerning the structure 
of the first and the second law
of thermodynamics 
applied to states of such media
and comment on the identification
of their equilibrium states.
For the case of simple fluids,
we provide some details
regarding the thermodynamics of such media.\\

As we have mentioned in the introduction, the laws of thermodynamics,
as applied to 
classical (Newtonian) continuous 
media
have been established long ago,
but their formulation assumed a Newtonian setting.
Upon the arrival of the special theory of relativity (and later of the general theory)  
these laws had to be reformulated
so they become compatible 
with the requirements of the Poincare (or general) covariance. 
That problem 
attracted the 
attention of many 
leading physics of that 
epoch including Einstein, Planck, Pauli amongst others
and these early attempts were 
meet with partial success, disagreements and confusions.
It is sufficient  to recall
 the ``Planck-Ott controversy'' regarding the definition 
and transformation properties of the relativistic temperature,
and the 
``Abraham-Minkowski controversy'' concerning the definitions of stresses and momenta
in polarized media
(a historical account of these early efforts is described
in Pauli's book\cite{Pauli}).\\ 
Modern approaches to
thermodynamics of relativistic media employ
 the fields of local Lorentz frames
 or the proper reference frames associated to a world line 
 of an observer, and in general, these classes of frames  are
 specified by defining a four velocity
field in the spacetime region occupied by the fluid.
Although one can develop
the thermodynamical properties of a medium
relative to an arbitrary chossen velocity field, in practice
one  presuppose that the medium
under consideration defines a preferred  four velocity or families of preferred four velocities\footnote{For the case of a fluid,
a preferred  four velocity $u$ may
be identified as the four velocity that defines  the energy frame or the
four velocity that defines the
 particle frame. More  
 generally
 preferred 
 velocity fields may be restricted to lie within
 the interior of the ''cone'' of opening pseudo-angle $\epsilon$ defined in the Appendix A.}.
In addition to the specification of 
a preferred velocity field, the derivation of the
thermodynamical laws
presented in this Appendix relies on the validity of
the local thermodynamical equilibrium hypothesis ((LTE) in short).
These two ingredients together
lead to 
the deduction of the first law and the formulation of second law.
Based on these laws we comment also
on the nature of relativistic equilibrium for relativistic fluids.\\

We begin by assuming that the medium is 
 propagating on a smooth four dimensions spacetime  $(M,g)$
so that at any event $p$ within the medium 
 is defined a future directed four velocity
$u$ normalized according to $g(u, u)=-1$.
For any such $p$, one may introduce a local Lorentz frame
whose time axis is determined by the four velocity $u$.
Notice however that at the same 
$p$, one may 
introduce the proper reference frame\footnote{For a definition and properties of proper
frames 
see for instance discussion in
ref.
\refcite{MTW}.} 
attached to an observer comoving with $u$ 
. 
Even though the resulting
local coordinate  systems have the same time axis they are not identical.
The local Lorentz frame is a free falling frame and is  best suited for implementing Einstein's principle of equivalence
while the proper frame is in general an accelerating frame.
The later, is a more convenient since it can be extended along the 
trajectory of the fluid element
and this proper frame often is referred as the rest frame.\\

 The medium
is assumed to be described by a set of tensor fields, 
denoted collectively by $\zeta_{A}$, with the index $A\in (1,2,...,N)$ enumerating the various field.
These fields 
 would satisfy a set
 of dynamical equations of tensorial nature and furthermore rules are supplied so that a conserved energy momentum tensor 
$T^{\mu\nu}(\zeta_{A})$
and other conserved currents may be
constructed out of the dynamical equations.\\ 
Moreover for any $p\in (M,g)$, it is assumed that the allowed values of $\zeta_{A}$ define a smooth manifold 
$Y$ describing the possible states of the medium over $p$
and 
these fibers join smoothly
so that they define
a smooth fiber bundle, with base manifold the spacetime
$(M,g)$. 
Geroch in ref. \refcite{Ger3},
(see also ref.\refcite{Ger2})
used this framework to analyze a large class 
of relativistic fluid theories
whose dynamic equations constitute a symmetric-hyperbolic (causal) system
and this framework contains as a special case the class
of relativistic fluids of divergence type discussed in section $(13)$.
We shall not enter on this
mathematically beautiful formalism
but we 
only mention that by appropriate restrictions upon the structure of the field equations, 
for a simple fluid Geroch identified
a class of equilibrium states spanning  a five dimensional submanifold $\hat E$ within each 
fiber $Y$. Moreover, 
he showed that the causality
restriction upon the structure of the field equations
 induces on each fiber $Y$ a positive-definite metric $G_{AB}$
and this metric can be employed to give a precise meaning to the
notion that a fluid state finds itself near equilibrium\footnote{
As we have already seen in Appendix A, transient thermodynamics
identifies states close to thermal equilibrium
by making use of the invariant  cone of opening pseudo-angle $\epsilon$
formed by $u_{E}$ and $u_{N}$. It
would be interesting to investigate
whrther these two methods of defining states close to equilibrium
are equivalent and whether the mathematical setting 
in Geroch approach in defining  ''closeness'' 
offers additional advantages.}.\\
For the purpose of this section, of relevance are the variations $\delta \zeta_{A}$ of the fields 
$\zeta_{A}$ that are defined along the fiber $Y$ over a fixed $p \in (M,g)$.
 These variations, refereed as fibered variations,
will be important in the formulation 
of the first law.
Moreover we shall use the notion of the 
a \textit{thermodynamic temperature}
and temperature 
as defined 
in Ref. \refcite{Ger3}
and these notions will be explained further below.\\

We now assume that the medium under consideration is a simple 
fluid and in the region occupied by the fluid is defined a four velovity $u$
that identifies a family of rest frames. Suppose that
relative to a rest frame defined at $p$ by the four velocity $u$, we consider
 a spatial three volume $V$ residing on the spacelike plane orthogonal to $u(p)$.
 We denote by  $(n, \rho) $ the particle number density, and  the density of mass-energy
as measured by an observer 
at rest relative to this frame,
so that
 $N=nV$ and
$U= \rho V $ 
are the total number of particles and  total mass-energy within this $ V $.
As we have already mentioned, the other key hypothesis 
that will be employed
for the
derivation of the laws of thermodynamics
of continuous 
media
is the validity of local thermodynamical equilibrium hypothesis\footnote{It should be mentioned that
the issue of validity of 
local thermodynamical equilibrium hypothesis (LTE), within the relativistic regime
becomes a subtle issue and this subtleties arise
from the basic principles of relativity that 
neither time intervals nor space intervals by themselves
have an absolute meaning. As a consequence  validity of (LTE) may become observer dependent. Fortunately
within transient thermodynamics and as long as one treats fluid states near equilibrium,
then there is $\epsilon$-''cone'' worth of admissible observers
who have the following property: If relative to the one of these observers (LTE) holds,
then it holds for all 
observers whose four velocity lies within this
 $\epsilon$-''cone''. This property is discussed in more details in
Ref.\refcite{FT2}.}.

Within the present context,
validity of (LTE) 
affirms
that
\textit{the state variables within $V$
satisfy the same thermodynamical relations as if this system was in a state of a global thermodynamical equilibrium}.\\
Let us now assume that the state of the
fluid and the nature of the 
 congruence associated with the four velocity $u$, are such that 
within the three volume $V$, (LTE) holds.
Thus we may introduce the ''entropy density''
$s$  which is a function of 
$(\rho, n)$ i.e. 
 $s=s(\rho, n)$ and this  $s=s(\rho, n)$ is refereed as the equilibrium equation of state
\footnote{For the case where the fluid is a perfect fluid
this 
$s=s(\rho, n)$ is the equilibrium equation of state defined
in section $8$ (see also eqs (\ref{Entro_py},\ref{GibbsR})). For this case 
$S^{\mu}=s(\rho, n)u^{\mu}$ is the Tolman physical entropy flux and $s(\rho, n)=-S^{\mu}u_{\mu}$ is the physical entropy density measured by
the observer comoving the fluid's four velocity. Away from perfect fluids,
 this $s=s(\rho, n)$ is well defined as a consequence of the validity of (LTE) postulate. In this case,
 $s(u)= s(\rho(u), n(u))$ is a purely formal quantity and although refereed as ''entropy density''
 there is no physical reasoning to support this interpretation. Still however
 the mathematical manipulations that follows by employing this $s(u)= s(\rho(u), n(u))$ 
 are well defined.}. Using this 
''entropy density'' $s$
we refer to $S=s V$ is the total ''entropy'' within $V$.\\
Validity of (LTE) allows us to introduce
intensive variables\footnote{It is understood
that all these thermodynamical variables are local and we ought to
indicate their explicitly dependance
upon either $p$ or a set of local coordinates i.e. ought to write either
$n(p), \rho(p),...etc$
or equivalently
$n(x^{\mu}), \rho(x^{\mu}),...etc$.
For simplicity of the presentation we omit
such dependancies.}
 temperature $T$, pressure $P$ and chemical potential
$\mu$ so that $(U, S, N, T, P, \mu)$ etc.
 satisfy
the familiar laws of thermodynamics
for spatially homogeneous equilibrium states. 
In particularly,
the first law takes the form\footnote{The variations $dU$, $dQ$.. etc in this first law
refer to fibered variations over $p$ and these variations defined above.}: 
\begin{equation}
dU=dQ-PdV+\mu dN
\label{F}
\end{equation}
where $dQ$ is that amount of heat entering the volume $V$
and 
$PdV$,
$\mu dN$
have the interpretation
as the work done by expanding (contracting) $V$ or adding
(subtracting) 
particles. For a fluid mixture,
one may introduce $n$-chemical potentials $\mu_{i}$ describing the
$n$ different particle species and denote by $N_{i}$
the total number of particle of type $i$ within 
$V$. For such fluids, the first law
in (\ref{F}) is obtained by replacing $ \mu dN$ by
$ {\sum\limits_{i} \mu_{i} dN_{i}}$. Since the system within $V$ is considered as been 
in equilibrium (thermal, mechanical, chemical) the familiar language  of equilibrium
thermodynamics like reversible, irreversible, quasistatic transformations, extensive, intensive  variables 
etc. apply
to this system but we should keep in mind that these concepts
have local character.\\
If an amount of heat $dQ$ is injected reversibly 
within $V$
then $dQ=TdS$
and for these reversible transformation
(\ref{F}) yields the
 familiar Gibbs relation
\begin{equation}
dU=TdS-PdV+\mu dN.
\label{FS}
\end{equation}
The scaling properties of the extensive variables
$(U, S, N)$
in this formula
leads to
\begin{equation}
U=TS-PV+\mu N
\label{Gibbs}
\end{equation}
which in turn imply that the intensive variables $(T,P, \mu)$
satisfy the Gibbs-Duhem relation:
\begin{equation}
SdT-VdP+Nd\mu=0
\label{GibbsD}
\end{equation}
Dividing
(\ref{Gibbs}) by $V$ yields
\begin{equation}
\rho=Ts-P+\mu n
\label{Chem}
\end{equation}
which via differentiation and in  combination to the  the densitized form of the Gibbs-Duhem relation (\ref{GibbsD}) 
i.e. $sdT-dP+nd\mu=0$,
implies
\begin{equation}
d\rho=Tds +\mu dn
\label{Cibbs2}
\end{equation}
and thus knowledge of the equation of state $\rho=\rho(s, n)$ for this fluid determines
the (local) temperature and (local) chemical potential $\mu$
via differentiation. Relation
(\ref{Cibbs2}) is the form of the first law
for a relativistic fluid
 as perceived by an observer with four velocity $u$ at $p$.\\

Often it is convenient to formulate
this law by employing ``the per particle description''.
 To do this,  we set $U=\rho V=\frac{\rho N}{n}$, 
$PdV=Pd\left(\frac{N}{n}\right)$
and introduce the ''entropy'' per particle $\hat s$ via:
$s=\hat s n$ so that 
 $S=sV=s\frac {N}{n}=\hat sN$.
 In terms of these new variables,
  (\ref{FS}) reduces to
\begin{equation}
	d\left(\frac{\rho N}{n}\right) = -Pd\left(\frac{N}{n}\right) + Td(\hat s N)+\mu dN,
\label{ET:6}	
\end{equation}
and for  any transformations that keeps the total particle number $ N $ within $V$ fixed, the first law
takes the form

\begin{equation}
	d\rho = \frac{\rho + P}{n}dn + nTd\hat s,
\label{ET:7}	 
\end{equation}
from which one concludes:

\begin{equation}
P(n,\hat s)= n\left(\frac{\partial \rho}{\partial n}\right)_{\hat s} - \rho, \quad T(n,\hat s) = \frac{1}{n}\left(\frac{\partial \rho}{\partial 
\hat s}\right)_n.
\label{ET:8}	 
\end{equation}
By introducing the mean internal energy $ e $ per particle via 

\begin{equation}
	\rho = n(m + e),
\label{ET:8}	
\end{equation} 
and returning again to (\ref{ET:6}) keeping $ N $ fixed, we obtain 

\begin{equation}
	de= -Pd\hat V + Td\hat s, \quad \hat V=\frac {1}{n} \quad \Longleftrightarrow\quad
		d\hat s = \frac{1}{T}de + \frac{P}{T}d\hat V.	
	\label{ET:9}	
\end{equation}

Moreover substituting 
(\ref{ET:8})
in  (\ref{Chem}), we find that  the relativistic chemical potential per particle $ \mu $ has the form
\footnote{The classical chemical potential $\mu_{cl}$ is defined via
$\mu_{cl}=e + P \hat V - T\hat s$ so that $\mu=mc^{2}+\mu_{cl}.$}:

\begin{equation}
	\mu= mc^2 + e + P \hat V - T\hat s, \quad \Longleftrightarrow\quad	
	\mu n = \rho c^2 + P - Ts.
\label{ET:11}	
\end{equation}
where  in the formula for $\mu n$ the entropy per particle $\hat s$
and $\hat V$ have been eliminated.
If in (\ref{ET:9}),
we eliminate $\hat s$, $\hat V$ and $e$ 
 in favor $(s, n, \rho)$
and use the above relations we return into
(\ref{Cibbs2}).
\\
If we
 introduce the (relativistic) thermal potential $ \Theta $ and the inverse temperature $ b $ via\footnote{If we
restore for a moment $c$ and Boltzmann' s constant $k$, 
 then the form of the relativistic
 thermal potential $ \Theta $ and the inverse temperature $ b $ 
 read:
 $\Theta= \frac{\mu}{kT}, \quad b = \frac{c^2}{kT}$.}
  via

\begin{equation}
	\Theta= \frac{\mu}{T}, \qquad b = \frac{1}{T},
\label{ET:14}	
\end{equation}
then the formulas $\mu n = \rho + P - Ts$ and 
(\ref{Cibbs2})	
take the equivalent forms:	
\begin{equation}
	s = b\left(\rho + {P}\right) - \Theta n,\quad
	ds =b d\rho - \Theta dn.
\label{ET:15}
\end{equation}
These two relations 
imply the important identity\footnote{Whenever $c \neq1$ the identity reads:
$d(sX) = b d(\rho X) - \Theta d(nX) + b \frac{P}{c^2}dX$}
 (derived first by Israel)
\begin{equation}
	d(sX) = b d(\rho X) - \Theta d(nX) + b {P}dX
\label{ET:16}
\end{equation}
where $X$ stands for an arbitrary smooth function.
Its validity can be seen by 
expanding both sides
of (\ref{ET:16})
yielding:

\begin{equation}
	X[ds - b d\rho - \Theta dn] = \left[-s + b\left(\rho + {P}\right) - \Theta n\right]dX,
\label{ET:17}	
\end{equation}
showing that 
(\ref{ET:16}) is a consequence of (\ref{ET:15}).
We shall employ (\ref{ET:16}) further below but it is worth noticing
that this identity depends only upon the validity of the (local) first law and
the scaling property
as expressed in
(\ref{ET:15})
and is independent of the nature of the background fluid.\\

Finally, it  should be mentioned here that
the form of the first law deduced above
applies to
a relativistic fluid whose equation of state has the form $\rho=\rho(n, s)$ (or equivalently
$s=s(\rho, n)$).
For other type of media, 
and as long as 
the local thermodynamical equilibrium hypothesis is assumed to hold,
one via similar reasoning may formulate a local version of the first law.
Notice however, that
the local 
energy density 
$\rho$ may depend besides the ''entropy'' density $s$, upon 
other fields such as 
the state of strain (for an elastic medium) electric or magnetic fields
and in such event the relation
(\ref{Cibbs2}) (or more precisely
(\ref{F})) would have different structure
(for elastic media the structure of the first law is discussed in refs.
\refcite{Car1},\refcite{Ehl}).\\

Before we enter into the formulation of the second law, we comment  briefly on the
notion of the (local) temperature $T(u)$ that appears in the local version of the first law.
From the above deductions, it follows that as long as
an equilibrium equation of state $s=s(\rho, n)$ (or an equivalent form $\rho=\rho(s, n)$)
is specified, then the local equilibrium hypothesis defines
the (local) temperature $T(u)$ as measured by the $u$-observer,  as the partial derivative of the equation of state
and thus as a function of $(\rho, n)$ or $(s,n)$.
Often however, a given 
equation of state $s=s(\rho, n)$ and given temperature profile $T=T(\rho, n)$,
are specified simultaneously and in such event Maxwell's relations demand that 
these $s=s(\rho, n)$ and $T=T(\rho, n)$
are compatible  with the first law, provided
$T=T(\rho, n)$ satisfies :
$$
n\frac {\partial T}{\partial n}+(\rho+P)\frac {\partial T}{\partial \rho}=
\frac {\partial P}{\partial \rho}T
$$
Geroch in  ref. \refcite{Ger3} 
refers to such 
$T=T(\rho, n)$ as the \textit{thermodynamic temperature}
and points out that 
there is a gauge freedom in the choice
of this 
$T=T(\rho, n)$.
If 
$T=T(\rho, n)$ and 
$\hat T=\hat T(\rho, n)$
satisfy this equation then 
for any constants $(c_{1}, c_{2})$ then
$c_{1}T+c_{2}\hat T$ is also a solution 
of the above equation.
Moreover, if
$T=T(\rho, n)$ and $\hat T=\hat T(\rho, n)$
are solutions 
of thee above equation, then any $\hat {\hat T}$
which 
is a homogeneous function of degree one i.e.
$\hat {\hat T}(\lambda T, \lambda \hat T)) =
\lambda \hat {\hat T}(T, \hat T))$
 is also an admissible thermodynamical temperature.
This freedom in the choice of the
thermodynamical temperature.
raises the question which one of these admissible $T$ should be 
the physically correct temperature.
He argues that may be none of them, but for suitably restricted fluid theories, a meaningful temperature  comes from the description of dissipative effects and in particularly the structure
of the heat flow or energy flow relative to the $u$ observer
(for further discussion on this point the reader is referred to Geroch's article).\\

We now pass to the formulation of the second law 
and the assignment of entropy for states describing continuous relativistic media.
As we have seen above,
the (LTE) postulate allows
to introduce 
the (local) ''entropy'' density scalar $s$ so that $S=sV$ 
is the total ''entropy'' within $V$. However, as we have already mentioned
both $s$ and $S$ are formal quantities and away from the special case 
where the medium is a perfect fluid,
there are no physical arguments to interpret them as the physical entropy density or total entropy within $V$.
Nevertheless, it is suggestive 
to examine the evolution of the total ''entropy''
$S=sV$ as it is transported forward in time 
along the flow lines (more precisely along the integral curves of $u^{\mu}$). It follows at once that 
\begin{equation}
\begin{split}
	\frac{d}{d\tau}S :=	& u^\mu \nabla_\mu (sV) \\
					 =	& (u^\mu \nabla_\mu s)V + s(u^\mu \nabla_\mu V) \\
					 =	& (u^\mu \nabla_\mu s)V + s(\nabla_\mu u^\mu) V \\
					 = & \nabla_\mu(s u^\mu)V\\
					 = & (\nabla_\mu S^\mu)V
\end{split}	
\label{ET:1}
\end{equation}
and this formula suggests that 
as long as 
$(\nabla_\mu S^\mu)\geq 0$
then $S=sV$ does not decrease as it is transported forward in time 
along the flow lines.\\

The expression
$S^{\mu}=su^{\mu}$ 
does represent
the entropy flux for a perfect fluid, provided
$s$ is the equilibrium equation of state
and $u^{\mu}$ is the four 
velocity of the fluid
and in fact as we shall see below for this case
$(\nabla_\mu S^\mu)=\nabla_{\mu}(su^{\mu})=0$. These considerations, 
suggest to 
assign 
 to
any  
 state describing an arbitrary fluid (or a relativistic continuous media) 
 an entropy flux vector $S^{\mu}$ 
which is arbitrary except that is 
subject to obey the constraint: 
 $\nabla_{\mu}S^{\mu} \geq 0.$
 This constraint 
guarantees that for 
 any two non intersecting spacelike hypersurfaces $\Sigma_{t_{1}}, \Sigma_{t_{2}} $
 cutting across an asymptotically flat spacetime,  
with  
 $ \Sigma_{t_{2}} $  to the future of  $\Sigma_{t_{1}}$, the total entropy
 $S(t)=\int_{\Sigma_{t}} S^{\mu}d\Sigma_{\mu}$ satisfies
 $S(t_{2})\geq S({t_{1}})$
 and this inequality implements the second law\footnote{Within the standard equilibrium thermodynamics, 
 the second law deals with the transformation of heat into work
and this law is stated  either in the Kelvin-Planck or Clausius form (see for instance \cite{Huang}).
The Clausius inequality allows to introduce the entropy $S$ as a state function
that has the property that for any isolated system $S$ does not decrease in time.
Within the relativistic domain, the second law is implemented by
introducing the entropy flux $S^{\mu}$ 
and ought to be keept in mind that this primary variable $S^{\mu}$
is distinct to the the ''entropy'' flux $su^{\mu}$ arising via the local equilibrium postulate.} for
states of arbitrary relativistic media.
Moreover, this $S^{\mu}$ has the property that any
observer with four velocity 
 $\hat u^{\mu}$,
measures an entropy density: 
$s(\hat u)=-\hat u_{\mu}S^{\mu}$.\\

The entropy flux vector 
$S^{\mu}$
introduced above does not have any apparent relation 
to the energy momentum tensor $T^{\mu\nu}$ (or on the possible particle currents
$J_{A}^{\mu}$) characterizing the medium. In fact, as has been 
discussed 
in the main sections of the paper, the relation of this
$S^{\mu}$ to variables specifying fluid states is the central issue
of irreversible thermodynamics.
Either via specific postulates of 
 M\"uller - Israel type or by appealing to 
entropy principle which asserts
that 
$S^{\mu}$ is a  constitutive 
function,
the entropy flux vector 
$S^{\mu}$, is required to be specified 
and
the condition 
$\nabla_{\mu}S^{\mu}\geq0$ 
becomes a very restrictive. \\
Just to get some insights into the implications 
and significance
of this $S^{\mu}$, let us consider the particular case of a simple relativistic
fluid
characterized by the property that
$S^{\mu}$
has the Tolman form i.e.
$S^{\mu}=su^{\mu}$
with $s$ the ''entropy'' density as measured by the $u^{\mu}$-observer.
Application of the Gibbs relation
(\ref{ET:15}),
implies 
\begin{equation}
\nabla_{\mu}S^{\mu}=\nabla_{\mu}(su^{\mu})=\left[\frac {1}{T}\dot \rho-\Theta \dot n+s\nabla_{\mu}u^{\mu}\right],
\label{ET:18}	
\end{equation}
where an over dot signifies differentiation along the flow lines (integral curves of $u^{\mu}$).
For a simple perfect fluid, by
taking $u^{\mu}$ to be the uniquely defined
four velocity of the fluid combined with the balance laws 
$\nabla_{\mu}T^{\mu\nu}=
\nabla_{\mu}J^{\mu}=0$
imply that the right hand side of
(\ref{ET:18}) is identically vanishing.
Thus the evolution of perfect fluids states do not generate entropy
and the total entropy
 $S(t)=\int_{\Sigma_{t}} S^{\mu}d\Sigma_{\mu}$ 
 across a spacelike hypersurface 
 $\Sigma_{t}$ remains constant i.e. is independent of 
  $\Sigma_{t}$.
 In  a state of global thermodynamical equilibrium it is expected that the total entropy $S(t)$ to be maximum 
and thus any first order variation $\delta S(t)$ induced by for small perturbations $\delta T^{\mu\nu}$ and $\delta J^{\mu}$ 
of the background perfect fluid configurations $T^{\mu\nu}$ and $J^{\mu}$ 
should  be vanishing to first order.
By appealing to covariant Gibbs relation it follows that will be the case
provided the thermal potential $\Theta$ is uniform and the $b^{\mu}$ is a timelike Killing vector field
conditions that we have already seen in section $8$.\\

Away from perfect fluids,
one may postulate
an entropy flux $S^{\mu}$ 
of the form
\begin{equation}
S^{\mu}=su^{\mu}+\beta {h^{\mu}}+\theta n^{\mu},
\label{EFOT}
\end{equation}
where
$(\beta , \theta)$ are unspecified coefficients, while
$h^{\mu}, n^{\mu}$ 
stand for the energy flow and particle ``drift'' relative to the $u$-frame
(see formulas (\ref{OD}) and 
(\ref{OD_11}) in the main text).
This choice,
defines the class of first order theories that includes the 
Eckart and Landau-Lifshitz theories as a special case
(this class of theories has been introduced in ref. \refcite{His2}).
For this class the phenomenological equations and the nature of equilibrium states i.e. states obeying 
$\nabla_{\mu}S^{\mu}=0,$
have been worked out in
ref. \refcite{His2}.\\
It is interesting
to mention that transient thermodynamics
can be developed
by enlarging the entropy flux
$S^{\mu}$ shown
(\ref{EFOT}),
by the inclusion of 
second order contributions
arising from the energy flux and stresses.
For this one should first define an admissible 
four velocity $u^{\mu}$ i.e. a velocity 
field $u^{\mu}$ lying within the invariant 
''cone" of opening pseudo-angle $\epsilon <<1$
that we have defined in Appendix A
and write down an extended entropy flux vector
$S^{\mu}$ that contains second order deviations from equilibrium.
By demanding that the so extended $S^{\mu}$
satisfies $\nabla_{\mu}S^{\mu}\geq 0$,
coupled to the structure of the balance laws
$\nabla_{\mu}T^{\mu\nu}=
\nabla_{\mu}J^{\mu}=0$
one would obtain
the phenomenological equations of the theory.
However, the issues of the covariance 
of the theory under change of the four velocity $u$ needs to examined
and that is what we have done in the main part of the paper.\\
Equilibrium states within transient are states obeying
$\nabla_{\mu}S^{\mu}=0$. From the structure of the phenomenological equations
it follows that
this condition demands the heat flux and stresses to be vanishing
and this leads to the conclusion
that they are perfect fluid states characterized by a uniform thermal potentials and a four velocity parallel to
a timelike Killing field i.e. states obeying the conditions outlined  in section $8$ (for a derivation of these results
consult Ref.\refcite{His1}).

For the rest of this Appendix, we 
 present a covariant formulation of the 
 thermodynamics for a perfect fluid
 i.e.
 a formulation where  the fundamental thermodynamical relations between two
 fluid sates 
 makes no references to any rest frame nor to quantities defined relative to such frame.\\
In order to do so, let
$u$ be the unique hydrodynamical four velocity defined by the perfect fluid
and let $\Delta^{\mu}{}_{\nu} = \delta^{\mu}{}_{\nu} + u^{\mu}u_{\nu}$
be the associated projection tensor. Clearly, relatively to 
the rest frame $ u^\mu = \delta^\mu_{\ 0} $ and thus  $ \Delta^{\mu}{}_{\nu} = diag(0,1,1,1) $. 
Moreover, 
 relative
to the rest frame
the components
of the energy momentum tensor $ T_{\mu\nu} $ take the form

\begin{equation}
	T_{\mu\nu} = diag(\rho,P,P,P),\quad \Longleftrightarrow\quad T_{\mu\nu} = \rho u_{\mu} u_{\nu} + P\Delta_{\mu\nu},
\end{equation}
while the particle current $J^{a}$ and entropy current $S^{a}$ read

\begin{equation}
	J^a = nu^a, \quad S^a = su^a.
\label{CurEn}
\end{equation}
Returning  to the identity (\ref{ET:16}) and taking $ X = u^\mu $ it follows

\begin{equation}
	dS^\mu = -\Theta dJ^\mu - b_\lambda dT^{\lambda \mu}, 
	\label{C1}
\end{equation}
where we used
$\rho u^\mu = -u_\lambda T^{\lambda \mu}$
and introduced 
$b_\lambda = bu_\lambda = \frac{u_\lambda}{T}$.
Moreover the relation
$s = b\left(\rho + \frac{P}{c^2}\right) - \Theta n$ implies that 
 the entropy vector $S^{\mu}=su^{\mu}$ satisfies
\begin{equation}
	S^\mu = Pb^\mu - \Theta J^\mu - b_\lambda T^{\lambda \mu}.
\label{C2}
\end{equation}
This relation in combination to 
(\ref{C1})
 incorporates the thermodynamics
of a simple perfect fluid. 
They are the covariant versions of 
the two relations in
 (\ref{ET:15}), but	
in sharp contrast 
formulas (\ref{C1}, \ref{C2})
make no reference to any rest frame.
Relation 
(\ref{C1}) is the covariant Gibbs relation
that we have encountered in section$-9$ of the paper.\\

\section{Transformations properties of thermodynamical variables}

In this Appendix, we outline the proofs of the two lemmas $(9.1, 9.2)$ employed in section
(\ref{NES}).
For this, let
 $(u, \hat u)$ 
be two (future pointing) unit timelike vectors lying within 
the ''cone''
of opening pseudo-angle
$\epsilon$ formed
by 
the four velocity
of the energy frame $u_{E}$ and 
the corresponding velocity
 $ u_{N}$
of the particle frame.
The vectors $(u, \hat u)$ define the time axis of the two admissible 
rest frames and 
we complement these vectors by  two triads
of spacelike unit vectors $e_{i}, i \in (1,2,3)$ and
$\hat e_{i}, i \in (1,2,3)$ so that
$(u, e_{i})$ respectively 
$(\hat u, \hat e_{i})$ 
constitute an orthonormal basis at the event under consideration.
For this setting, we have
 
\begin{equation}
	\hat u=\frac {u}{(1-\frac {v^{2}}{c^{2}})^{\frac{1}{2}}}+ \frac {v^{i}e_{i}}{c}\frac {1}{(1-\frac {v^{2}}{c^{2}})^{\frac{1}{2}}},\quad
	v^{2}=v^{i}v_{i}.	
	\label{NVR}
\end{equation} 
 where $v^{i}$ are the components of the three velocity of the frame $u$ relative to the $\hat u$ one.
 Following Israel in Ref. \refcite{Isr1}, we re-express formula
 (\ref{NVR})
into the equivalent form
 \begin{equation}
	\hat u^\mu = (1+\delta^2)^{1/2}u^\mu + \delta^\mu, \qquad \delta^2 = \delta^a \delta_a, \qquad \delta^a u_a = 0.
\label{VR}
\end{equation}
Since both $u$ and $\hat u$ lie within the ''cone''
 of opening 
  pseudo-angle 
$\epsilon \simeq  \frac {v}{c}=O_{1}$
and 
for states near equilibrium
$u_{E}-u_{N}=\epsilon=O_{1}$, 
it follows that  the components $\delta^{\mu}$ of $\delta$ in 
( \ref{VR}) satisfy
 $ \delta^{\mu} = \hat{\epsilon}^{\mu}:=\hat \epsilon \leq \epsilon=O_1 $ so that
 to lowest order
 in $\hat \epsilon$, we have:
 $\hat u^{\mu}-u^{\mu}= \hat{\epsilon} \leq O_1 $. This new smallness parameter 
 $\hat{\epsilon}$ introduced here is a measure of the relative three velocity
 between $(\hat u^{\mu}, u^{\mu})$
 and will be important further below.  
 \\ 
 For a frame change implemented by 
 the transformation
$u^{\mu} \to {\hat u}^{\mu},$
 we want to determine the variation
 induced by this frame change on the 
 various 
  thermodynamical variables as measured relative in these two frames.
\\
To derive these variations, it is convenient to introduce 
the four velocity $u^{\mu}_{E}$ of the energy frame and invoke the expansions analogues to (\ref{VR}):

\begin{equation}
	\hat u^\mu = (1+\hat \delta^2)^{1/2}u^\mu_{E} +\hat  \delta^\mu, \qquad \hat \delta^2 =\hat  \delta^a \hat \delta_a, \qquad \hat \delta_a u^a_E = 0,
\label{VR1}
\end{equation}

\begin{equation}
	u^\mu = (1+ \gamma^2)^{1/2}u^\mu_{E} +\gamma^\mu, \qquad \gamma^2 = \gamma^a \gamma_a, \qquad  
	\gamma_a u^a_E = 0,
\label{VR2}
\end{equation}
where in above
$\hat\delta^{\nu}$ and $\gamma^{\mu}$ are taken to be of order
 $\epsilon=O_{1}$
but relations 
(\ref{VR}, \ref{VR1},  \ref{VR2}) imply\footnote{In order 
to see
the origin of 
$\hat \delta^{\mu}-\gamma^{\mu}=\delta^{\mu}=\hat \epsilon \leq O_{1}$
one notices that (\ref{VR1}, \ref{VR2}) imply to leading order:
$\hat u^{\mu}-u_{E}^{\mu}=\hat \delta^{\mu}$ and
$u^{\mu}-u_{E}^{\mu}=\gamma^{\mu}$
while on the other hand 
(\ref{VR}) implies $\hat u^{\mu}-u^{\mu}=\delta^{\mu}.$ These two estimates lead to 
the relation $\hat \delta^{\mu}-\gamma^{\mu}=\delta^{\mu}$.} that
$\hat \delta^{\mu}-\gamma^{\mu}=\delta^{\mu}=\hat \epsilon  \leq O_{1}$.\\

With these
preliminaries, we now return
to demonstrate 
formulas (\ref{L11}, \ref{L22}, \ref{L33})
in Lemma $(9.1)$ and begin with the proof of 
 (\ref{L11}).\\ 
 Let $ \rho(\hat{u}) $, $ \rho(u) $ be the energy densities
measured by $ \hat{u} $ respectively $  u $, then the following relations hold:

\begin{equation}
\begin{split}
	\delta \rho & = \rho(\hat u) - \rho(u) = T^{\mu \nu}[\hat{u}_\mu \hat{u}_\nu - u_\mu u_\nu] \\
	& = [({\rho}(u_E) + {P}(u_E))u^{\mu}_E u^\nu_E + P(u_E)g^{\mu\nu} + 
\tau^{\mu \nu}(u_E)][\hat{u}_\mu \hat{u}_\nu - u_\mu u_\nu] \\
	& = [\rho(u_E) + P(u_{E})][(\hat{u}^\mu u_{\mu(E)})^2 - (u^\mu u_{\mu(E)})^2] + \tau^{\mu \nu}(u_E)[\hat{u}_\mu \hat{u}_\nu - u_\mu u_\nu] \\
	& =  [\rho(u_E) + P(u_E)](\hat{\delta}^2 - \gamma^2) + \tau^{\mu \nu}(u_E)(\hat{\delta}_\mu \hat{\delta}_\nu - \gamma_\mu \gamma_\nu).
\end{split}
\end{equation}
Recalling that that states close to equilibrium
satisfy 
$\tau (u_{E})^{\mu\nu}=\rho(u_{E})O_{1}$ (see
eq. (\ref{SNEC}))
it follows that 
to leading order 
 $$\tau^{\mu \nu}(u_E)(\hat{\delta}_\mu \hat{\delta}_\nu - \gamma_\mu \gamma_\nu)=\hat \epsilon^{2} O_{1}
+ (O(\hat \epsilon)O_{1})^{2}.$$
On the other hand, if in the first term
$$[\rho(u_E) + P(u_E)](\hat{\delta}^2 - \gamma^2)=
[\rho(u_E) + P(u_E)](\hat{\delta} - \gamma)(\hat{\delta} +\gamma),$$
we eliminate $\hat \delta^\mu$ in favor of  $ \delta^\mu$ using
 $ \hat{\delta}^\mu - \gamma^\mu=\delta^\mu=\hat \epsilon$, 
and
 $ \hat{\delta}^\mu + \gamma^\mu=O_{1}$ 
 we obtain to leading order:
 $$[\rho(u_E) + P(u_E)](\hat{\delta}^2 - \gamma^2)=\hat \epsilon O_{1}$$
 and thus we have the estimate: 
 
 \begin{equation}
	\delta \rho  = \rho(\hat u) - \rho(u) = \hat{\epsilon}O_1 +O_{2}
\end{equation}
In order to prove
(\ref{L22}), we note that the variation $ \delta h^\mu $ of the energy flow $h^{\mu}$
takes the form:
\begin{equation}
\begin{split}
	\delta h^\mu & = h(\hat{u})^\mu - h(u)^\mu \\
	& = -(\Delta(\hat{u})^\mu_{\ a}\hat{u}_b - \Delta(u)^\mu_{\ a}u_b)T^{ab} \\
	& = -[\Delta(\hat{u})^\mu_{\ a}\hat{u}_b - \Delta(u)^\mu_{\ a}u_b][(\rho(u_E) + P(u_E))u^a_E u^b_E + P(u_E)g^{ab} + \tau^{ab}(u_E)]\\
	\end{split}
\end{equation}
However, one notices that the term $P(u_{E})g^{ab}$ 
gives zero contribution, while the term
$\tau^{ab}(u_E)$ supplies an $O_{1}$ contribution to the right hand side of above.
However the lowest order contribution arises from the term:
$$-[\Delta(\hat{u})^\mu_{\ a}\hat{u}_b - \Delta(u)^\mu_{\ a}u_b][(\rho(u_E) + P(u_E))u^a_E u^b_E. $$
Indeed a litle calculus shows that
$$K^{\mu}=
[\Delta(\hat{u})^\mu_{\ a}\hat{u}_b]u^a_E u^b_E=u^{\mu}_{E}(1+\hat \delta^{2})^{\frac {1}{2}} +\hat u^{\mu}
(1+\hat \delta^{2})$$
and by eliminating $\hat u^{\mu}$ using
$\hat u^{\mu}=u^{\mu}_{E}+\hat \delta^{\mu} +O(\delta^{\mu})^{2}$
we finds
$$K^{\mu}=u^{\mu}_{E}(1+(\hat \delta)^{2})^{\frac {1}{2}} +\hat u^{\mu}_{E}
(1+\hat \delta^{2})+\hat \delta^{\mu}(1+\hat \delta^{2})=
\hat u^{\mu} +\hat \delta^{\mu}+O(\hat \delta^{\mu}(\hat \delta))^{2}.$$
This estimate in turn implies that to leading order
$$
\delta h^\mu = h(\hat{u})^\mu - h(u)^\mu =-(\rho(u_E) + P(u_E))(\hat \delta^{\mu}-\gamma^{\mu})=
-(\rho(u_E) + P(u_E))\hat \epsilon ^{\mu}.
$$
\\
Via similar procedure one can easily establish
formula (\ref{L33}) i.e.
$$
\delta \tau^{\mu\nu}:=\tau(\hat u)^{\mu\nu}-\tau(u)^{\mu\nu}= \hat \epsilon O_{1}
$$
and for this proof one starts from
$$
\tau(u)^{\mu\nu}=\Delta(u)^{\mu}{}_{a}\Delta(u)^{\nu}{}_{b}T^{ab}-P(u)\Delta(u)^{\mu\nu}
$$
and applies the same steps as the previous two cases.

The proof of the Lemma 2 follows similar steps
and is omitted. 

We finish this Appendix by mentioning that the thermodynamical variables $(P(u), n(u), s(u), T(u), \Theta(u))$
measured by the $u$-observer,
under a change of the frame described in the above proof,
i.e. $u^{\mu} \to \hat u^{\mu}=u^{\mu}+ \hat{\epsilon}^{\mu}$ with  $\hat{\epsilon}^{\mu}\leq O_1$,
all of them change by $\hat \epsilon O_{1}$
while the particle drift $n^{\mu}(u)$, like the energy flux $h^{\mu}(u)$, change 
by $\hat \epsilon $. The proof of these variations
can be constructed as above or can be found in 
Ref.\refcite{F1}.\\


\end{document}